\documentclass[reqno]{amsart}
\usepackage[foot]{amsaddr}

\usepackage[unicode,psdextra]{hyperref}
\hypersetup{
    colorlinks,
    linkcolor={red!50!black},
    citecolor={blue!50!black},
    urlcolor={blue!90!black}
}

\usepackage{graphicx}
\usepackage{xcolor}
\usepackage{xfrac}

\usepackage{enumitem}

\usepackage[english]{babel}
\usepackage[T1]{fontenc}
\usepackage{libertine}

\usepackage{amssymb}

\newcommand{\pd}{\partial}
\newcommand{\dd}{\mathrm{d}}
\newcommand{\ii}{\mathrm{i}}
\newcommand{\pdv}[2]{\frac{\partial #1}{\partial #2}}
\newcommand{\dv}[2]{\frac{\mathrm{d} #1}{\mathrm{d} #2}}
\DeclareMathOperator*{\Res}{Res} 
\newcommand{\erf}{\;\mathrm{erf}}
\renewcommand{\Re}{\, \mathrm{Re}}
\renewcommand{\Im}{\, \mathrm{Im}}

\newcommand{\abs}[1]{\left| #1 \right|}

\makeatletter
\renewcommand\subsubsection{\@startsection{subsubsection}{3}%
  \z@{.5\linespacing\@plus.7\linespacing}{-.5em}%
  {\normalfont\bfseries}}
\makeatother

\makeatletter
\renewcommand\paragraph{\@startsection{paragraph}{4}%
  \z@{.5\linespacing\@plus.7\linespacing}{-.5em}%
  {\normalfont\bfseries}*}
\makeatother

\usepackage[
    backend=biber,
    style=phys,
    citestyle=numeric-comp,
    sorting=none,
    eprint=true,
    doi=false,
    url=false,
    biblabel=brackets
]{biblatex}
\makeatletter

\begin{document}
\title{First-passage properties of the jump process with a drift. 
       The~general~case}
\author{Ivan N. Burenev}
\address{LPTMS, CNRS, Universit\'e Paris-Saclay, 91405 Orsay, France}
\email{inburenev@gmail.com}

\begin{abstract}
We study the first-passage properties of a jump process with constant drift where jump amplitudes and inter-arrival times follow arbitrary light-tailed distributions with smooth densities. Using a mapping to an effective discrete-time random walk, we identify three regimes determined by the drift strength: survival (weak drift), absorption (strong drift), and critical. We derive explicit expressions for exponential decay rates in the survival and absorption regimes, and characterize algebraic decay at the critical point. We also obtain asymptotic behavior of the mean first-passage time, number of jumps, and their variances for processes starting either close to the origin or far from it. 
\end{abstract}

\maketitle
\vspace{-1cm}

\tableofcontents

\section{Introduction}\label{sec:intro}
\par Imagine a bakery with a single counter where a cashier serves customers who arrive at random times, each requiring a varying amount of time to complete their order. The cashier works at a consistent pace, attending to one customer at a time in the order they arrive, with no overtaking allowed. This setup, that we have so frivolously described, actually represents a fundamental model in queuing theory. Specifically, if both the inter-arrival times and the amount of time it takes to serve a customer are independent and identically distributed random variables, this process is usually  referred to as a G/G/1 queue (see \cite{Bhat2015} for a pedagogical introduction). 

\par 
A simple observation is that if many customers arrive with complex orders, the queue will grow indefinitely. In contrast, if few customers arrive or orders are simple to fulfill, the cashier will remain idle most of the time. So it is natural to wonder whether the queue depletes and, if so, how much time it takes? This is essentially the question we address in the present paper.

\begin{figure}[h!]
    \includegraphics{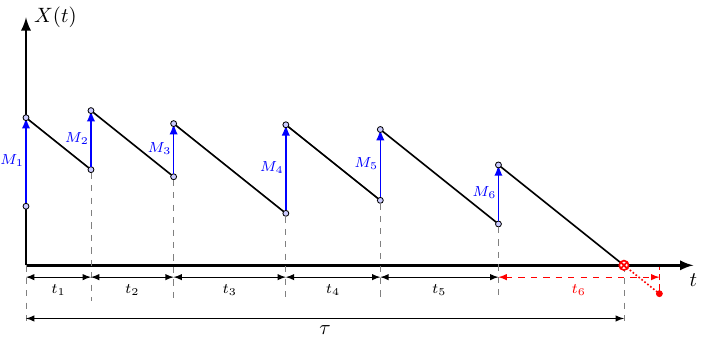}
    \caption{An example of the trajectory. Starting at $X_0$ the process instantaneously undergoes a jump $M_1$, then moves toward the origin with constant velocity $\alpha$ for the time $t_1$ when the next jump $M_2$ occurs. This pattern continues until the process crosses the origin at time $\tau$ after $n$ jumps (here $n=6$). The inter-jump intervals $\{t_1,\ldots,t_6\}$ are \textit{i.i.d.} random variables drawn from $p(t)$, and the jump amplitudes $\{M_1,\ldots,M_6\}$ are also \textit{i.i.d.} random variables following $q(M)$.}\label{fig:Model}
\end{figure}

\par To formulate the problem in a more formal manner, we consider a jump process with a drift, i.e., a stochastic process $X(t)$ evolving on the real line through a combination of deterministic negative linear drift and random positive jumps (see Fig.~\ref{fig:Model}). In the <<bakery>> picture, $X(t)$ is the total amount of requests piled in the queue, the linear decay segments correspond to the service of customers, while the jumps represent the sporadic arrivals of new customers. The dynamics of $X(t)$ unfold as follows:
\begin{enumerate}
    \item The process starts at an initial position $X_0$, and immediately experiences a jump of random amplitude $M_1$. It then drifts toward the origin with the constant velocity $-\alpha$ for a random time interval $t_1$, after which the next jump $M_2$ occurs.
    \item This cycle repeats until the first-passage event, i.e., until the process crosses the origin to the negative side for the first time (the queue is depleted).
    \item The time intervals between successive jumps, $t_i$, are independent and identically distributed (\textit{i.i.d.}), drawn from a distribution with density $p(t)$.
    \item The jump amplitudes $M_i$ are also \textit{i.i.d.}, sampled from a distribution $q(M)$.
\end{enumerate}
The model is controlled by four parameters: the initial position $X_0\ge0$, the drift strength $\alpha>0$, and the two probability distributions $p(t)$ and $q(M)$. Formally, the process can be written as
\begin{equation}\label{eq:X(t)=def}
    X(t) = X_0 - \alpha t + \sum_{j=1}^{n(t)} M_j,
\end{equation}
where $n(t)$ is a renewal process describing the number of jumps that have occurred up to time $t$.

\par 
Our aim is to study the first-passage properties of the process \eqref{eq:X(t)=def}. Specifically, we focus on two observables: the first-passage time $\tau$ and the number of jumps $n$ that occur  before the first-passage event. Clearly, these two quantities are random variables, which are in general correlated. 

{
\par Although we have introduced the process \eqref{eq:X(t)=def} through the lens of queuing theory, it finds applications in many domains where deterministic drift is interrupted by sporadic jumps. For instance, in risk and ruin theory, $X(t)$ represents the capital of a company with constant operating expenses (drift term) and occasional gains (jumps). Examples include pharmaceutical companies where jumps correspond to the net capitalization change due to new discoveries, commission-based businesses such as real estate agencies or brokerage firms, or project-based businesses where jumps correspond to sales. In this context first-passage to zero represents bankruptcy. This setup is known as the (dual) Sparre-Andersen model \cites{GS-05,AGS-07,YS-14}, which reduces to the classical Cramér-Lundberg model when arrivals follow a Poisson process (see \cite{Asmussen2010} for a review).

\par In the mathematical literature, the process \eqref{eq:X(t)=def} is commonly used in its dual version, obtained by reversing the sign of $X(t)$. This dual process belongs to the class of growth-collapse processes \cites{EK-04,BPSZ-06,LS-11} (the original process is sometimes referred to as the decay-surge process \cite{EK-06}). In this formulation, the positive deterministic drift represents continuous growth and the negative jumps correspond to catastrophic collapse events. For instance, such dynamics arise in avalanche phenomena where accumulated stress is suddenly released \cites{LLRM-14,SMR-21,JBMM-25}. Here first-passage to a barrier (equivalent to first-passage to zero in the formulation of this paper) corresponds to the moment at which the system accumulates some given value of stress, which can cause significant structural changes. Alternatively, in population dynamics, the process represents a population that steadily grows but experiences sudden reductions due to catastrophic events such as epidemics or natural disasters, with first-passage to zero corresponding to extinction \cite{H-11,GH-21}.

\par The system can also be viewed within the framework of partial resetting \cites{DCSM-21,TRS-22,HKPS-23,BCHPM-23,H-23,OL-24,BFHMS-24,OG-24,BCGSTP-25,MOPR-25}, a generalization of stochastic resetting \cite{EM-11} (see \cite{EMS-20} for a review). In standard stochastic resetting, a process instantaneously returns to its initial position at random times. In partial resetting, the reset is incomplete~--- the process undergoes jumps of random magnitude rather than full reinitialization. Our drift-jump process provides a natural realization: the deterministic drift toward the origin represents relaxation dynamics, while the jumps $M_i$ act as partial resetting events that interrupt this relaxation. 
}

\par 
Even for seemingly simple one-dimensional processes, obtaining exact results on the first-passage properties presents significant challenges \cite{Redner2001,MajumdarSchehr2024,BMS-13}. The standard approach relies on renewal equations, which lead to integral equations of the Wiener-Hopf type (e.g., \cites{K-10,MV-13,DG-22}). These are notoriously difficult to solve analytically. For the jump process with a drift that we consider, explicit solutions are typically available only when arrivals follow a Poisson process \cites{P-59,Z-91,S-93,DP-97,PSZ-99}, while much less is known for general distributions (see \cite{Kyprianou2014} for a review).

\par 
In \cite{BM-25} we proposed an alternative approach that does not rely on the renewal equation. Specifically, we constructed a mapping of the original continuous-time process onto an effective discrete-time random walk. This mapping allowed us to obtain a closed-form representation of the Laplace transform of the joint probability distribution of $\tau$ and $n$, by applying the Pollaczek-Spitzer formula \cite{P-52,S-56,S-57}, a powerful result in discrete random walk theory. More precisely, we used a generalization of Ivanov's formula \cite{I-94} to the asymmetric random walk \cite{MMS-18}.
This approach was then successfully applied to two exactly solvable cases with Poisson arrivals, where explicit solutions could be obtained and compared against standard renewal equation techniques.

\par 
The present work tackles a significantly more challenging general case, where explicit solutions are no longer available. 
Here, we consider a rather general class of distributions, namely light-tailed $p(t)$ and $q(M)$ with smooth probability densities, thereby moving far beyond the assumption of Poisson arrivals. 
In this general case, the closed-form integral representations of the relevant Laplace transforms are still valid, but extracting the asymptotic behavior  now requires analytic continuation techniques and singularity analysis.
In this work, we provide such analysis and derive several exact  asymptotic results for this general setting.

\par 
The main results concern the behavior of the survival probabilities, i.e., the probability that the process stays positive up to time $\tau$ (or up to the $n$-th jump). We show that there are three distinct regimes determined by the drift strength. In the \textit{survival regime} (weak drift), the process has a finite probability of never crossing the origin, and survival probabilities decay exponentially to a positive constant. In the \textit{absorption regime} (strong drift), the first passage event occurs with certainty, and survival probabilities decay exponentially to zero. We compute these exponential decay rates explicitly for arbitrary light-tailed distributions with smooth densities. At the \textit{critical point} separating these regimes, exponential decay is replaced by algebraic decay. 

\par Beyond the decay rates of the survival probabilities, we obtain exact asymptotic expansions for several quantities as $X_0\to0$ and $X_0\to\infty$. For the \textit{absorption regime}, we derive expansions for the mean and variance of both $\tau$ and $n$. For the \textit{survival regime}, we characterize the asymptotic behavior of the infinite-time survival probability. 
At the \textit{critical point}, we establish the algebraic decay and derive asymptotic expansions for the amplitudes of the power-law behavior.

{
\par  As a continuation of \cite{BM-25}, this paper yields results of similar structure, making the differences worth emphasizing explicitly.
The key novelty is that we treat general inter-arrival and jump distributions where the defining integrals cannot be evaluated explicitly.
Consequently, the asymptotic analysis is significantly more involved and requires fundamentally different methodology. 
Regarding the similarity of the results, this is itself a central contribution of this work: we demonstrate that the regime trichotomy (survival, absorption, critical) holds across all light-tailed distributions with smooth densities. 
While regime boundaries and decay rates depend on $p(t)$ and $q(M)$, the qualitative structure is universal, revealing the two exactly solvable cases to be exemplars of general behavior rather than artifacts of Poisson arrivals.

\par The paper is organized as follows: 
In Section~\ref{sec:model and results}, we formally define the model and recall from \cite{BM-25} the representation for the triple Laplace transform (with respect to $\tau$, $n$, and $X_0$) of the joint probability distribution of the first-passage observables $\tau$ and $n$. Then we briefly summarize the main results derived in this paper.
Section~\ref{sec:example_of_continuation} presents a detailed analysis of a toy example through which we illustrate the analytic continuation techniques used throughout the paper.
In Section~\ref{sec:Analytic structure in lambda-plane}, we study the properties of the Laplace transform in the $\lambda$-plane ($\lambda$ being the Laplace dual of $X_0$) and perform analytic continuation. 
Sections \ref{sec:Survival probability Sinfty}, \ref{sec:Survival probability SnX0}, and \ref{sec:Survival probability StX0} analyze the asymptotic behavior of the survival probabilities: at infinite time, up to the $n$-th jump, and up to time $\tau$, respectively. Section \ref{sec:moments} concerns the asymptotic expansions of the means and the variances of $\tau$ and $n$ in the absorption regime. Finally, we conclude in Section~\ref{sec:conclusion}. 

}

\section{The model and the main results}\label{sec:model and results} 

\par In this paper we consider the process \eqref{eq:X(t)=def} in a rather general setup. That is, we do not specify the distributions $p(t)$ and $q(M)$ and instead only require them to satisfy two conditions:
\begin{enumerate}[nolistsep]
    \item All moments of the distributions $p(t)$ and $q(M)$ are finite (both distributions are light-tailed). 
    \item The densities $p(t)$ and $q(M)$ are continuously differentiable functions, supported on either a finite  interval $[0,\ell]$ or the positive half-line $[0,\infty)$.
\end{enumerate}

\par We start by fixing notation and defining the key quantities. We denote by $\mathbb{P}[\tau,n\,\vert\,X_0]$ the joint probability distribution of the first-passage time $\tau$ and the number of jumps $n$ for the process started at $X_0$. Our analysis focuses on three survival probabilities:
\begin{itemize}
    \item  The probability that the process remains positive until time $\tau$:
    \begin{equation}\label{eq:ST=def}
        S_T(\tau\,\vert\, X_0) \equiv 
        1 - \int_{0}^{\tau}\dd\bar{\tau} 
            \sum_{k=0}^{\infty} \mathbb{P}[\bar{\tau},k\,\vert\,X_0].
    \end{equation}
    \item The probability that the process remains positive  for at least $n$ jumps:
    \begin{equation}\label{eq:SN=def}
        S_N(n\,\vert\,X_0) \equiv
        1 - \int_{0}^{\infty} \dd\bar{\tau} \sum_{k=0}^{n} 
        \mathbb{P}[\bar{\tau},k\,\vert\,X_0].
    \end{equation}
    \item The probability that the process remains positive 
    indefinitely:
    \begin{equation}\label{eq:Sinfty=def}
        S_\infty(X_0) \equiv
        1 - \int_{0}^{\infty} \dd\bar{\tau} \sum_{k=0}^{\infty} 
        \mathbb{P}[\bar{\tau},k\,\vert\,X_0].
    \end{equation}
\end{itemize}

\par Before presenting exact results, let us briefly provide some intuition about the expected behavior. If the drift is strong, it pulls the process toward the origin, so the first-passage event certainly happens and $S_\infty(X_0)=0$. 
If the drift is weak, the jumps dominate and the process can escape to infinity, so there is a finite  probability that first-passage never occurs and $S_\infty(X_0)>0$. 

\par
\begin{figure}[h!]
    \includegraphics{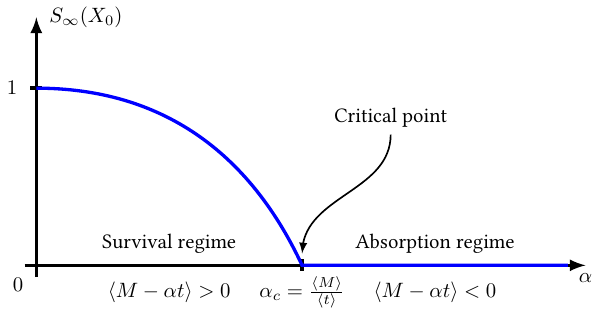}
    \caption{Schematic representation of the survival probability $S_\infty(X_0)$ as a function of the drift strength $\alpha$ for fixed initial position $X_0$. Below the critical value $\alpha_c$, the system is in the \textit{survival regime} with $S_\infty(X_0) > 0$. The \textit{critical point} occurs at $\alpha = \alpha_c$. Above $\alpha_c$, the system is in the \textit{absorption regime} with $S_\infty(X_0) = 0$. }\label{fig:regimes}
\end{figure}

 The transition between these regimes can be understood by examining the average displacement per cycle. Each cycle consists of a jump of size $M$ followed by drift of magnitude $\alpha t$ toward the origin. The sign of the average displacement determines the regime:
\begin{equation}\label{eq:two_regimes}
\begin{aligned}
    \langle M-\alpha t\rangle > 0 &: \qquad S_\infty(X_0) > 0,\\
    \langle M-\alpha t\rangle < 0 &: \qquad S_\infty(X_0) = 0,
\end{aligned}
\end{equation}
where $\langle \cdots\rangle$ denotes averaging with respect to $p(t)$ and $q(M)$. This simple observation gives the critical drift velocity:
\begin{equation}\label{eq:alpha_c_def}
     \alpha_c = \frac{\langle M\rangle}{\langle t\rangle}.
\end{equation}
We treat the survival probability at infinite time as an <<order parameter>> separating the \textit{survival regime}, where $S_\infty(X_0)>0$, from the \textit{absorption regime}, where $S_\infty(X_0) = 0$, with $\alpha$ being the <<control parameter>>. The case $\alpha=\alpha_c$ we refer to as the \textit{critical point} (see Fig.~\ref{fig:regimes} for a schematic representation).

\par The key ingredient in the analysis is the closed-form representation of the Laplace transform $\hat{Q}(\rho,s\,\vert\,\lambda)$ of the joint probability distribution $\mathbb{P}[\tau,n\,\vert\,X_0]$,
\begin{equation}\label{eq:hat(Q)=def}
    \hat{Q}(\rho,s\,\vert\,\lambda) \equiv
    \int_{0}^{\infty} \dd X_0\, e^{ - \lambda X_0} 
    \int_{0}^{\infty}
    \dd \tau\, 
        e^{-\rho \tau} \, 
    \sum_{n=0}^{\infty} 
        s^n\,
        \mathbb{P}[\tau,n\,\vert\, X_0].
\end{equation}
In \cite{BM-25}, we constructed a mapping from the original process to an effective discrete-time random walk via a trick similar to that used in \cites{MDMS-20,MDMS-20b} for the run-and-tumble particle in $d$-dimensions and in \cite{MMV-24} for the cost of excursions (see also \cite{SBEM-25}).
Then, by utilizing the generalized Pollaczek-Spitzer formula, we showed that the Laplace transform \eqref{eq:hat(Q)=def} can be represented as
\begin{equation}\label{eq:LTQ=generalIvanov}
    \hat{Q}(\rho,s\,\vert\,\lambda) = 
    \frac{1}{\lambda+\frac{\rho}{\alpha}}
    - 
    \frac{1 - s \, c(\rho)}{\lambda+\frac{\rho}{\alpha}}
    \;
    \phi^{-}\!\left( \lambda + \frac{\rho}{\alpha}; \rho,s \right)
    \phi^{+}\!\left( 0; \rho,s\right),
\end{equation}
where 
\begin{equation}\label{eq:phi^pm=def}
    \phi^{\pm}(\lambda;\rho,s) \equiv 
    \exp\left[
        -\frac{1}{2\pi} \int_{-\infty}^{\infty}
        \dd k\; \frac{1}{\lambda\pm \ii k}
        \log\left[1 - s\, c(\rho) F(k;\rho)\right] 
    \right],
\end{equation}
and 
\begin{equation}\label{eq:c(rho)=def}
    c(\rho) \equiv \int_0^{\infty} \dd M \, 
    e^{-\rho\frac{M}{\alpha}} q(M).
\end{equation}
The function $F(k;\rho)$ appearing in \eqref{eq:phi^pm=def} is a Fourier transform of the probability distribution of the steps in the effective random walk. In terms of the original probability distributions $p(t)$ and $q(M)$ it reads
\begin{equation}\label{eq:F(k;rho)=def}
    F(k;\rho) \equiv \frac{1}{c(\rho)} 
    \int_{0}^{\infty} \dd M\, e^{ - \rho \frac{M}{\alpha} + \ii k M} q(M)
    \int_{0}^{\infty} \dd t\, e^{-\ii \alpha k t} p(t),
\end{equation}
or equivalently
\begin{equation}\label{eq:F(k;rho)=def_<>}
    F(k;\rho) = \frac{1}{c(\rho)} 
    \left\langle e^{ - \rho \frac{M}{\alpha} + \ii k (M-\alpha t)}
    \right\rangle.
\end{equation}

\par
The joint probability distribution $\mathbb{P}[\tau,n\,\vert\,X_0]$ could, in principle, be found by inverting the transform in \eqref{eq:hat(Q)=def}. 
This is, however, a very thorny path and there is little reason to hope that it can be traversed for arbitrary distributions. 
Instead, we study the analytic structure of the triple Laplace transform $\hat{Q}(\rho,s\,\vert\,\lambda)$ and obtain asymptotic results. 
The main caveat is that the parameters $\lambda$, $\rho$, and $s$ originate from the Laplace transforms; hence the representation \eqref{eq:phi^pm=def} is valid for $\lambda>0$, $\rho>0$, and $s\in(0,1)$. 
However, extracting the asymptotic behavior of the probabilities requires extending this representation to larger domains through proper analytic continuation.

\par 
In the rest of the paper, we analyze the representation \eqref{eq:LTQ=generalIvanov}. 
A well-developed theory exists for studying the analog of this representation for symmetric random walks \cites{CM-05,MCZ-05,ZMC-07,MMS-13,MMS-14,MMS-17,BMS-21b,KVB-23} (see \cite{M-09} for a pedagogical introduction) and symmetric random walks with additional constant drift \cites{MSV-12,MMS-18,BMS-21}. In our case, however, the effective random walk is genuinely asymmetric, which makes the analysis significantly more involved and thus provides one motivation for the present paper.

\par Finally, two remarks are in order. 
First, alternative representations for survival probabilities of discrete-time random walks have recently been obtained using Riemann-Hilbert techniques \cite{RC-25}, providing a complementary approach to the Pollaczek-Spitzer formula. Second, one of the two exactly solvable cases in \cite{BM-25} takes $q(M)$ to be a delta function; hence, the smoothness requirement is not fulfilled, but the detailed computation using \eqref{eq:LTQ=generalIvanov} can still be performed. The reason is that this requirement is mostly technical and can be lifted entirely for a significant portion of the derivations. In addition, the results can readily be extended to cases where only the first three moments of $M$ and $t$ are finite. 
However, tracking exactly which assumptions are needed for each result would require a substantial amount of bookkeeping without offering much additional insight.
For this reason, we choose to sacrifice a small degree of generality in favor of clarity, and  throughout the paper we always assume that both distributions $p(t)$ and $q(M)$ have smooth densities and that all their moments are finite.

\subsection{The main results}\label{sec:main results}
By leveraging the representation \eqref{eq:LTQ=generalIvanov} of the  triple Laplace transform, we derive asymptotic results valid  for arbitrary probability distributions $p(t)$ and $q(M)$. The  key advantage is that all expressions are written in terms of  the Fourier transform $F(k;\rho)$ of the effective random walk  \eqref{eq:F(k;rho)=def}.

\paragraph{Survival regime} 
When the drift is weak ($\alpha<\alpha_c$), there is a finite probability that the process remains positive indefinitely, $S_\infty(X_0) > 0$. The key result here are explicit expressions for the exponential decay rates that govern the approach to this asymptotic value.

\par Both survival probabilities $S_N(n\,\vert\,X_0)$ and $S_T(\tau\,\vert\,X_0)$ approach their infinite-time limit $S_\infty(X_0)$ exponentially fast:
\begin{equation}\label{eq:SN_ST_decays}
    \alpha<\alpha_c:\quad 
    \left\{\begin{aligned}
        S_N(n\,\vert\,X_0) - S_\infty(X_0) 
            &\underset{n\to\infty}{\asymp} 
        \exp\left[ - \frac{n}{\xi_n(\alpha)} \right], \\
        S_T(\tau\,\vert\,X_0) - S_\infty(X_0) 
            &\underset{\tau\to\infty}{\asymp} 
        \exp\left[ - \frac{\tau}{\xi_\tau(\alpha)} \right]. 
    \end{aligned}\right.
\end{equation}
Crucially, the decay rates are given by explicit, computable expressions:
\begin{equation}\label{eq:xi_n xi_tau}
    \frac{1}{\xi_n(\alpha)} = 
    \log s_*,  
    \qquad
    \frac{1}{\xi_\tau(\alpha)} = -\rho_*, 
\end{equation}
where
\begin{equation}\label{eq:s*=result}
    s_* = \left[ 
    \min_{k_I \in (\zeta_{-}, \zeta_+)} 
    F(\ii k_I;0) \right]^{-1},
\end{equation}
and  $\rho_* < 0$ is determined by the condition
\begin{equation}\label{eq:rho_star}
    \frac{1}{c(\rho_*)} = \min_{k_I \in (\zeta_{-}, \zeta_+)}  F(\ii k_I;\rho_*).
\end{equation}
The minimization is performed over purely imaginary arguments within the region where ${\Im(k) \in (\zeta_-, \zeta_+)}$, which is the maximal strip in the complex $k$-plane where the Fourier transform $F(k;\rho)$, defined in \eqref{eq:F(k;rho)=def}, is analytic (see Section~\ref{sec:properties_of_phi} for details).

\par Additionally, we obtain the asymptotic behavior of $S_\infty(X_0)$ itself. If the process starts close to the origin, we obtain the expansion
\begin{equation}\label{eq:S_infty_small_X0}
    S_\infty(X_0) \underset{X_0\to0}{\sim}
    \varsigma_{0} + \varsigma_{1} X_0,
\end{equation}
where the coefficients $\varsigma_{0,1}$ are given in \eqref{eq:S_infty(X)=expansion}.  
For initial positions far from the origin, we instead find exponential behavior
\begin{equation}
    1 - S_\infty(X_0) \underset{X_0\to\infty}{\asymp} e^{- R X_0},
\end{equation} 
where $R > 0$ is the smallest positive solution of $F(\ii R;0)=1$, analogous to the Lundberg exponent in ruin theory.

\paragraph{Critical point}
As the drift approaches the critical value, the rates \eqref{eq:xi_n xi_tau} diverge:
\begin{equation}\label{eq:xiT xin=divergence_res}
    \xi_n(\alpha) \underset{\alpha\to\alpha_c}{\sim}
    \frac{2}{\langle t\rangle^2}
    \frac{\langle (M-\alpha_c t)^2\rangle}{(\alpha-\alpha_c)^2}
    ,
    \qquad 
    \xi_\tau(\alpha) \underset{\alpha\to\alpha_c}{\sim}
    \frac{2}{\langle t\rangle}
    \frac{\langle (M-\alpha_c t)^2\rangle}{(\alpha-\alpha_c)^2}
    .
\end{equation}
This leads to an algebraic decay of the survival probabilities at the critical point $\alpha=\alpha_c$:
\begin{equation}\label{eq:SN_ST_decays crit}
    \alpha=\alpha_c:\quad 
    \left\{\begin{aligned}
        &S_N(n\,\vert\,X_0) \underset{n\to\infty}{\sim} 
        \sqrt{\frac{1}{\pi n}} U(X_0),  
        \\
        & S_T(\tau\,\vert\,X_0) \underset{\tau\to\infty}{\sim} 
        \sqrt{ \frac{\langle t\rangle}{\pi \tau}} U(X_0), 
    \end{aligned}\right.
\end{equation}
where the function $U(X_0)$ is defined through its Laplace transform in \eqref{eq:U(X0)=LT}. This function admits the following asymptotic expansions:
\begin{equation}
    U(X_0)\underset{X_0\to\infty}{\sim}
    U_1 X_0 + U_0,
    \qquad 
    U(X_0)\underset{X_0\to0}{\sim}
    u_0 + u_1 X_0,
\end{equation}
with explicit coefficients $U_{0,1}$ and $u_{0,1}$ given in \eqref{eq:Un(X0->inf)} and \eqref{eq:Un(X0->0)}.

\par In addition, we show that at criticality, there exists a Brownian scaling limit.  When $n\to\infty$ and $X_0\to\infty$ with $X_0/\sqrt{n}$ fixed, we find the scaling behavior
\begin{equation}\label{eq:Sn_erf_scaling}
    S_N(n\,\vert\,X_0) 
    \underset{n\to\infty}{\sim}
    \text{erf}\left(\frac{X_0}{\sqrt{2n\langle (M-\alpha_c t)^2\rangle}}\right),
    \quad
    \frac{X_0}{\sqrt{n}}\text{~--- fixed}.
\end{equation}
Similarly, when $\tau\to\infty$ and $X_0\to\infty$ with $X_0/\sqrt{\tau}$ fixed, we have:
\begin{equation}\label{eq:ST_erf_scaling}
    S_T(\tau\,\vert\,X_0)
    \underset{\tau\to\infty}{\sim}
    \text{erf}\left(
        \frac{X_0}{\sqrt{2\frac{\tau}{\langle t\rangle} \langle (M-\alpha_ct)^2\rangle}}
    \right),\quad
    \frac{X_0}{\sqrt{\tau}}\text{~--- fixed}.
\end{equation}

\paragraph{Absorption regime} 
When the drift is strong ($\alpha>\alpha_c$), the process eventually crosses the origin with certainty. The survival probabilities again decay exponentially:
\begin{equation}\label{eq:SN_ST_decays abs}
    \alpha>\alpha_c:\quad 
    \left\{\begin{aligned}
        S_N(n\,\vert\,X_0)
            &\underset{n\to\infty}{\asymp} 
        \exp\left[ - \frac{n}{\xi_n(\alpha)} \right], \\
        S_T(\tau\,\vert\,X_0)
            &\underset{\tau\to\infty}{\asymp} 
        \exp\left[ - \frac{\tau}{\xi_\tau(\alpha)} \right]. 
    \end{aligned}\right.
\end{equation}
The rates are given by the same expressions as in the survival regime \eqref{eq:xi_n xi_tau}. 

\par In this regime, the moments of $\tau$ and $n$ are finite, and their Laplace transform with respect to $X_0$ can be found from \eqref{eq:LTQ=generalIvanov}. Here we obtain the first two terms in the asymptotic expansions as $X_0\to\infty$ and $X_0\to0$ for the mean and variance of both $n$ and $\tau$:
\begin{gather}\label{eq:E[n]-X0 res}
    \mathbb{E}[n\,\vert\, X_0] 
    \underset{X_0\to\infty}{\sim}
    X_0 A_{1}
    + A_0,
    \qquad
    \mathbb{E}[n\,\vert\, X_0] 
    \underset{X_0\to0}{\sim}
    a_0
    + a_1 X_0,
    \\
    \mathbb{E}[\tau\,\vert\, X_0] 
    \underset{X_0\to\infty}{\sim}
    X_0 \tilde{A}_{1}
    + \tilde{A}_0,
    \qquad
    \mathbb{E}[\tau\,\vert\, X_0] 
    \underset{X_0\to0}{\sim}
    \tilde{a}_0
    + \tilde{a}_1 X_0,
    \\
    \label{eq:Var[n]-X0 res}
    \mathrm{Var}[n\,\vert\,X_0] \underset{X_0\to\infty}{\sim}
    X_0 B_{1} + B_{0},
    \qquad
    \mathrm{Var}[n\,\vert\,X_0] \underset{X_0\to0}{\sim}
    b_0 + b_1 X_0,
    \\
    \label{eq:Var[t]-X0 res}
    \mathrm{Var}[\tau\,\vert\,X_0] \underset{X_0\to\infty}{\sim}
    X_0 C_{1} + C_{0},
    \qquad
    \mathrm{Var}[\tau\,\vert\,X_0] \underset{X_0\to0}{\sim}
    c_0 + c_1 X_0.
\end{gather}
All coefficients are expressed as integrals involving $F(k;\rho)$, with complete expressions provided in Section~\ref{sec:moments}.

\par 
Let us conclude the presentation of the results by  providing some motivation for constructing expansions beyond the leading order.
While subleading coefficients might appear to be technical details, in some cases they carry significant physical meaning. 
A notable example is the constant $U_0$ in the critical regime expansion \eqref{eq:Un(X0->inf)}, which in the symmetric random walk case is closely related to the Milne extrapolation length in radiative transport theory \cite{M-1921} and the trapping problem for Rayleigh flights \cite{Ziff-91} (see \cite{MMS-17} for a more comprehensive overview). 
Similar constants also find applications in packing problems in computer science \cite{CS-93,CFFH-98}. Since these corrections are nontrivial to obtain, we believe it is worthwhile to present here a systematic derivation based on Mellin transform techniques, which provides a robust framework for extracting such asymptotic expansions.

\section{A toy example}\label{sec:example_of_continuation}
{
\par The parameters $\lambda$, $s$, and $\rho$ in \eqref{eq:phi^pm=def} arise from Laplace transforms. Hence, the representation is initially valid for $\lambda>0$, $s\in(0,1)$, and $\rho>0$. The asymptotic analysis of first-passage quantities, however, requires an analytic continuation of $\phi^\pm(\lambda;\rho,s)$ to broader domains. As we will shortly show, this continuation is not as simple as extending \eqref{eq:phi^pm=def} to, say, $\lambda<0$; it demands much more careful consideration. 

\par Before analyzing $\phi^\pm(\lambda;\rho,s)$, it is instructive to consider a simplified toy problem through which to introduce the techniques we shall use and provide the reader with the intuition behind them. Specifically, we define two auxiliary functions
\begin{equation}\label{eq:psi_ex=def}
    \psi^{\pm}(\lambda;A,B) =  \exp\left[
        - \frac{1}{2\pi} \int_{-\infty}^{\infty} \dd k\, 
        \frac{1}{\lambda\pm\ii k} \log \big[ ( k - \ii A) (k + \ii B)\big]
    \right],
\end{equation}
where $A$ and $B$ are positive real parameters and the integral is understood as a principal value. In the complex $k$-plane, the analytic structure of the integrands entering \eqref{eq:psi_ex=def} is apparent: they have a pole at $k=\mp \ii \lambda$ and two logarithmic cuts going from $\ii A$ to $\ii\infty$ and from $-\ii B$ to $-\ii\infty$. In Section~\ref{sec:Analytic structure in lambda-plane}, we demonstrate that $\phi^\pm(\lambda;\rho,s)$ each has a pole and two branch cuts~--- though the branch points have more intricate expressions. In subsequent sections, we show that these are exactly the singularities responsible for the asymptotic behavior.

\par Since the functions $\psi^\pm(\lambda;A,B)$ resemble $\phi^\pm(\lambda;\rho,s)$ in structure, but are sufficiently simple to allow explicit evaluation for all complex $\lambda$, they offer an illustrative example exposing the difference between a naive extension of \eqref{eq:psi_ex=def} and a proper analytic continuation.
 Moreover, there is a pragmatic motive for this detour: similar integrals will reappear in the analysis of first-passage properties, and it is prudent to compute them here, killing two birds with one well-aimed stone.

\par Recall that the full analysis of $\phi^\pm(\lambda;\rho,s)$ requires continuation not only in the $\lambda$-plane, but also in the $s$- and $\rho$-planes. To illustrate these additional continuations, we now introduce $s$-dependence by hand. Specifically, rather than treating $A$ and $B$ as fixed constants, we parametrize them as functions of an auxiliary
variable $s$:
\begin{equation}\label{eq:A,B=A(s), B(s)}
    A \equiv A(s) = k_* + \sqrt{s_* - s}, \quad 
    B \equiv B(s) = -k_* + \sqrt{s_* - s}, \quad
    s\in (0, s_*-k_*^2),
\end{equation}
where $k_*>0$ and $s_*>1$ are parameters satisfying $k_*^2<s_*$. At this point such parametrization may appear arbitrary, but it will be justified in Section~\ref{sec:Analytic structure in lambda-plane}. 
We choose to denote the parameter by $s$, since, as demonstrated in Section~\ref{sec:Survival probability SnX0}, it plays the same role as the Laplace conjugate of $n$ in
\eqref{eq:hat(Q)=def}.

\par Below we first evaluate $\psi^\pm(\lambda;A,B)$ explicitly for all complex $\lambda$ and fixed positive values of $A$ and $B$. Once the exact representations are found, performing analytic continuation in the $s$- and $\lambda$-planes is straightforward. In particular, we shall see that the analytic continuations indeed differ from the naive extensions of \eqref{eq:psi_ex=def}. Our goal here, however, is not to construct such continuations from the explicit results, but rather to perform the continuation directly at the level of the integral representation \eqref{eq:psi_ex=def}. 

\par In short, the continuation in the $\lambda$-plane requires tracking the movement of the pole at $k=\mp\ii\lambda$, and the continuation in the $s$-plane necessitates analyzing the movement of the branch points at $\ii A(s)$ and $-\ii B(s)$ accompanied by appropriate deformations of the integration contour. This is the same procedure used in Section~\ref{sec:Analytic structure in lambda-plane} and Section~\ref{sec:Survival probability SnX0} for the continuation of the original functions $\phi^\pm(\lambda;\rho,s)$ in the $\lambda$- and $s$-planes, respectively. Regarding the continuation in the $\rho$-plane, the procedure is essentially the same as in the $s$-plane, and we thus do not illustrate it with the toy example.
}

\paragraph{Explicit evaluation}
We begin by evaluating $\psi^{\pm}(\lambda;A,B)$ for all complex $\lambda$ with fixed positive $A$ and $B$. The standard method for evaluating the integrals of the form \eqref{eq:psi_ex=def} involves extending them into the complex $k$-plane and closing the contour either in the lower or in the upper half-plane. If the contribution from the semicircle vanishes at infinity, the integral is determined by the singularities inside the contour. However, in the form \eqref{eq:psi_ex=def} it is not apparent why the contributions from the arcs decay sufficiently fast (faster than $\abs{k}^{-1}$). For this reason, we modify this representation so that the standard procedure applies. 

\par 
When integrating \eqref{eq:psi_ex=def} over the real line, the odd part of the integrand cancels due to symmetry, and only the even part remains. Specifically, we can write
\begin{equation}\label{eq:psi(lambda)=int symmetric}
    \psi^{\pm}(\lambda) =
    \exp\left[ 
        - \frac{1}{4\pi}
        \int_{-\infty}^{\infty} \dd k \left\{ 
            \frac{\log\left[
                (k-\ii A)
                (k+\ii B)
            \right]
            }{\lambda \pm \ii k}
            +
            \frac{\log\left[
                (k+\ii A)
                (k-\ii B)
            \right]
            }{\lambda \mp \ii k}
        \right\}
    \right],
\end{equation}
where, to simplify notation, we temporarily suppress the explicit dependence on the parameters $A$ and $B$. Rearranging the terms in the integrand we obtain
\begin{multline}\label{eq:psi(lambda)=two terms}
    \log \left[ \psi^{\pm}(\lambda) \right] =
        - \frac{1}{4\pi}
        \int_{-\infty}^{\infty} \dd k 
            \frac{\lambda }{\lambda^2 + k^2}
            \log\left[
                \left(k^2+A^2\right)
                \left(k^2+B^2\right)
            \right]
        \\ \mp 
        \frac{1}{4\pi}
        \int_{-\infty}^{\infty} \dd k 
            \frac{\ii k }{\lambda^2 + k^2}
            \log\left[
                \frac{(k+\ii A)(k-\ii B)}
                     {(k-\ii A)(k+\ii B)}
            \right]
    .
\end{multline}
Finally, integrating the second term by parts yields a decomposition of $\log \psi^\pm(\lambda)$ into four integrals of essentially identical form:
\begin{multline}\label{eq:log psi(lambda)=sum int}
    \log \left[ \psi^{\pm}(\lambda) \right] =
        - \frac{1}{4\pi}
        \int_{-\infty}^{\infty} \dd k 
            \left( 
            \frac{\lambda }{\lambda^2 + k^2}
            \log\left[
                k^2+A^2
            \right]
            + 
            \frac{\lambda }{\lambda^2 + k^2}
            \log\left[ 
                k^2+B^2
            \right]
            \right)
        \\ \pm 
        \frac{1}{4\pi}
        \int_{-\infty}^{\infty} \dd k 
            \left(
                \frac{A}{ A^2 + k^2 }
                \log\left[k^2+\lambda^2\right]
                -
                \frac{B}{B^2 + k^2}
                \log\left[k^2+\lambda^2\right]
            \right)        
    .
\end{multline}
Each integral in \eqref{eq:log psi(lambda)=sum int} can now be evaluated using
standard complex analysis techniques, by closing the contour in the
complex $k$-plane. In particular, all integrals reduce to the application of the identity
\begin{equation}\label{eq:int log=sign b identity}
    \frac{1}{\pi}\int_{-\infty}^{\infty} \frac{b}{b^2 + k^2} \log(k^2 + a^2)\dd k 
    = 
        \frac{b}{\sqrt{b^2}} \log\frac{b+a}{b-a}
        +
        \frac{a}{\sqrt{a^2}} \log \left[ b^2 - a^2\right],
\end{equation}
where $a$ and $b$ are arbitrary complex parameters, and the integral is taken as a principal value.

\par Applying \eqref{eq:int log=sign b identity} to the integrals in \eqref{eq:log psi(lambda)=sum int} after some algebraic manipulations, we find that
\begin{equation}\label{eq:psi=cases}
    \psi^{+}(\lambda) = 
    \left\{
    \begin{aligned}
        &\frac{1}{B+\lambda}, 
            && \Re(\lambda)>0, \\
        &\sqrt{\frac{A-\lambda}{B+\lambda}},
            && \Re(\lambda) =0, \\
        &A-\lambda,
            && \Re(\lambda)<0 \vphantom{\frac{A}{B}},
    \end{aligned}\right.
    \qquad
    \psi^{-}(\lambda) = 
    \left\{
    \begin{aligned}
        &\frac{1}{A+\lambda}, 
            && \Re(\lambda)>0, \\
        &\sqrt{\frac{B-\lambda}{A+\lambda}},
            && \Re(\lambda) =0, \\
        &B - \lambda, \vphantom{\frac{A}{B}}
            && \Re(\lambda)<0.
    \end{aligned}\right.
\end{equation}
From \eqref{eq:psi=cases} it is evident that the analytic continuations $\psi_\text{a.c.}^\pm(\lambda)$ of the functions $\psi^\pm(\lambda)$ from the positive half of the real line $\lambda\in(0,\infty)$ to the full $\lambda$-plane are given by:
\begin{equation}\label{eq:psi_ac=exact}
    \psi^{+}_\text{a.c.}(\lambda) = \frac{1}{B+\lambda},\qquad   
    \psi^{-}_\text{a.c.}(\lambda) = \frac{1}{A+\lambda},\qquad
    \lambda\in\mathbb{C}.
\end{equation}
Not only do the functions $\psi^\pm(\lambda)$, computed in \eqref{eq:psi=cases} via the original integral expression \eqref{eq:psi_ex=def} for arbitrary $\lambda$, differ from their analytic continuations \eqref{eq:psi_ac=exact}, they are in fact discontinuous at $\Re(\lambda) = 0$.
With this explicit computation in hand, we now show how to perform the analytic continuation at the level of the integral expression \eqref{eq:psi_ex=def}.

\paragraph{Continuation in the $\lambda$-plane}
We first focus on the function $\psi^+(\lambda)$.
The analytic structure of the integrand for $\psi^+(\lambda)$ in the $k$-plane is relatively simple: a branch cut from the logarithm and a pole at $k = \mathrm{i} \lambda$.
When $\Re(\lambda)\to0^{+}$, this pole approaches the real axis from above, falling on the integration contour when $\Re(\lambda)=0$ and crossing it to the lower half-plane for $\Re(\lambda)<0$. This is exactly the mechanism responsible for the discontinuity of $\psi^+(\lambda)$.

\begin{figure}[h!]
    \includegraphics{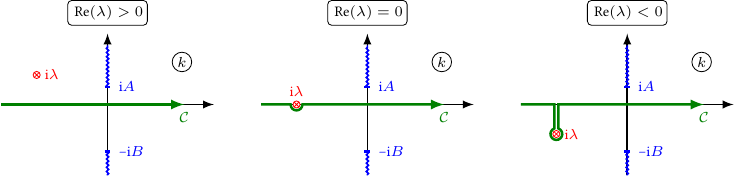}
    \caption{
    Structure of the integrand for $\psi^+(\lambda)$ in \eqref{eq:psi_ex=def} and the deformed contour $\mathcal{C}$ used for the analytic continuation.
    For $\Re(\lambda) > 0$, the contour coincides with the real line.
    For $\Re(\lambda) = 0$, the contour circumvents the pole as a semicircle.
    For $\Re(\lambda) < 0$, the contour encircles the pole in the lower half-plane.
    }
    \label{fig:pole_movement_psi}
\end{figure}

\par 
In order to analytically continue $\psi^+(\lambda)$, we deform the contour of integration to circumvent the pole as shown in Fig.~\ref{fig:pole_movement_psi} and compute the integral explicitly.  
The analytic continuation of the other function $\psi^-(\lambda)$ is performed in essentially the same way. The only difference is that  the pole is located at $k = -\mathrm{i} \lambda$, and when $\Re(\lambda)$ decreases from positive to negative, it crosses the real line in the opposite direction.

\par Let us now compute the integral over the deformed contours for $\Re(\lambda)\le0$.  When $\Re(\lambda) = 0$, the pole lies precisely on the integration contour and the principal value integral differs from the analytic continuation by half the residue at $k = \pm \mathrm{i} \lambda$,
\begin{multline}\label{eq:psi-ac-re0}
\Re(\lambda)=0 : \quad
    \log\left[\psi_\text{a.c.}^\pm(\lambda) \right]
    = \log\left[\psi^\pm(\lambda)\right] 
    \\
    \mp \pi\ii \Res_{k=\pm\ii\lambda}\left[
    \frac{1}{2\pi} 
    \frac{\log\left[(k-\ii A)(k+\ii B) \right]}
         {\lambda\pm\ii k}
    \right].
\end{multline}
For $\Re(\lambda) < 0$, the contour consists of the real line, two vertical segments, and a circle around the pole at $k = \pm \mathrm{i} \lambda$. The integrals over the vertical segments cancel one another, and the circle integral is equal to the residue at $k=\pm\mathrm{i}\lambda$. Thus, the analytic continuation differs from the real-line integral by the full residue, namely, 
\begin{multline}\label{eq:psi-ac-reneg}
\Re(\lambda)<0 : \quad
    \log\left[\psi_\text{a.c.}^\pm(\lambda) \right]
    = \log\left[\psi^\pm(\lambda)\right] 
    \\
    \mp 2\pi\ii \Res_{k=\pm\ii\lambda}\left[
    \frac{1}{2\pi} 
    \frac{\log\left[(k-\ii A)(k+\ii B) \right]}
         {\lambda\pm\ii k}
    \right].
\end{multline}
Computing the residues explicitly, we obtain the analytic continuations $\psi_\text{a.c.}^\pm(\lambda)$ of $\psi^\pm(\lambda)$, 
\begin{equation}\label{eq:psi_ac=cases}
    \psi_\text{a.c.}^\pm(\lambda)
    = 
    \left\{\begin{aligned}
        & \psi^\pm(\lambda), \quad && \Re(\lambda) > 0,\\
        & \frac{\psi^\pm(\lambda)}{\sqrt{(A\mp\lambda)(B\pm\lambda) }},\quad &&\Re(\lambda)=0,\\
        & \frac{\psi^\pm(\lambda)}{(A\mp\lambda)(B\pm\lambda)},\quad 
        && \Re(\lambda)<0.\\
    \end{aligned}\right.
\end{equation}
Comparing the analytic continuation \eqref{eq:psi_ac=cases}, derived directly from the integral representation \eqref{eq:psi_ex=def}, with the continuation \eqref{eq:psi_ac=exact} obtained after evaluating the integrals in \eqref{eq:psi=cases}, we see that the two match exactly. This confirms the validity of the contour deformation procedure and the consistency of the analytic continuation.

\par 
An important remark here is that the above procedure implicitly assumes that the deformed contour $\mathcal{C}$ does not cross the branch cut of the logarithm. This assumption is, however, easily lifted.
Consider a strip $\Im(k) \in (-a,a)$, where $a =\min(A,B)$. Within this strip, the logarithmic term in the integrand \eqref{eq:psi_ex=def} is analytic.
If $\abs{\Re(\lambda)}<a$, then the deformed contour lies within $\Im(k)\in(-a,a)$ where there are no singularities. Hence the analytic continuation \eqref{eq:psi_ac=cases} is valid in this strip. To extend the continuation to the full region $\Re(\lambda)<0$, we combine \eqref{eq:psi_ac=cases} with \eqref{eq:psi_ex=def} to obtain the functional relation
\begin{equation}\label{eq:psi_ac=functional}
    \Re(\lambda)\in(0,a) : \quad 
    \psi^\mp_\text{a.c.}(\lambda) \psi^\pm_\text{a.c.}(-\lambda) 
    = \frac{\psi^\mp(\lambda)\psi^\pm(-\lambda)}{(A\mp\lambda)(B\pm\lambda)}
    = \frac{1}{(A\mp\lambda)(B\pm\lambda)}.
\end{equation}
Now, using  \eqref{eq:psi_ac=functional} as the definition of the analytic continuation for all $\lambda$, we obtain
\begin{equation}\label{eq:psi_ac=cases_2}
    \psi_\text{a.c.}^\pm(\lambda)
    = 
    \left\{\begin{aligned}
        & \psi^\pm(\lambda), \quad && \Re(\lambda) > 0,\\
        & \frac{\psi^\pm(\lambda)}{\sqrt{(A\mp\lambda)(B\pm\lambda) }},\quad &&\Re(\lambda)=0,\\
        & \frac{1}{(A\mp\lambda)(B\pm\lambda)} \frac{1}{\psi^\mp(-\lambda)},\quad 
        && \Re(\lambda)<0.\\
    \end{aligned}\right.
\end{equation}
Clearly \eqref{eq:psi_ac=cases_2} and \eqref{eq:psi_ac=cases} coincide.

{

\paragraph{Continuation in the $s$-plane}
We now demonstrate how to perform analytic continuation in the $s$-plane if the parameters $A$ and $B$ are not fixed but are instead $s$-dependent and given by \eqref{eq:A,B=A(s), B(s)}. From \eqref{eq:psi_ac=exact}, the analytic continuation is straightforward, 
\begin{equation}\label{eq:psi+=ac s-plane}
    \psi^{\pm}_{\text{a.c.}}(\lambda; s) \equiv 
    \psi^\pm_{\text{a.c.}}(\lambda; A(s), B(s)) 
    = 
    \frac{1}{\lambda \mp k_{*} + \sqrt{s_*-s}}.
\end{equation}
Let us now reproduce this result from the integral representation \eqref{eq:psi_ex=def}. 

\par The key observation here is that as $s$ increases, the branch points at $k=\ii A(s)$ and $k=-\ii B(s)$ move toward the origin. Once $s=s_*-k_*^2$, we have $B(s)=0$, and the branch point falls precisely on the integration contour. For $s>s_*-k_*^2$ the contour would cross the branch cut, invalidating the integral representation \eqref{eq:psi_ex=def}. To circumvent this issue, we shift the integration contour from the real axis to a horizontal line $\mathcal{C}_* = \left\{k : \Im(k) = k_* \right \}$ (see Fig.~\ref{fig:psi_lambda_analytic_A}). This yields
\begin{equation}\label{eq:psi(lambda;A,B)=int C_1}
    \psi^{\pm}_\text{a.c.}(\lambda;s) =  \exp\left[
        - \frac{1}{2\pi} \int_{\mathcal{C}_*} \dd k\, 
        \frac{1}{\lambda\pm\ii k} \log \big[ ( k - \ii A) (k + \ii B)\big]
    \right], \quad \Re(\lambda)>k_*.
\end{equation}
Note that if $\Re(\lambda)<k_*$, then the pole at $k=\ii\lambda$ appearing in $\psi^+(\lambda;s)$ lies between the original real-axis contour and the deformed contour $\mathcal{C}_*$. In this case, the deformation would capture the residue at the pole, introducing an additional factor in \eqref{eq:psi(lambda;A,B)=int C_1}. We omit these technical details here, as they are not essential for understanding the general procedure, and for simplicity assume that $\Re(\lambda)>k_*$. In Section~\ref{sec:Survival probability SnX0}, while performing the continuations for $\phi^\pm(\lambda;\rho,s)$ in the $s$-plane, we account for this properly.

\par The representation \eqref{eq:psi(lambda;A,B)=int C_1} provides an analytic continuation in $s$ valid up to the critical value $s=s_*$, where the branch points coincide, trapping the contour of integration so that no deformation can avoid them. This creates a pinch singularity which in the $s$-plane manifests as a branch point at $s_*$. As a side note, such behavior is well-known in the context of analytic structure of the S-matrix in high-energy physics \cite{Eden1966}. 

\par To verify that \eqref{eq:psi(lambda;A,B)=int C_1} coincides with \eqref{eq:psi+=ac s-plane}, we perform a simple change of variables $k\mapsto k + \ii k_*$ in \eqref{eq:psi(lambda;A,B)=int C_1}. This immediately recovers \eqref{eq:psi_ex=def} with $A\mapsto A-k_*$, $B\mapsto B+k_{*}$ and $\lambda\mapsto\lambda\mp k_*$. Computing the integrals as in \eqref{eq:psi_ac=cases} reproduces \eqref{eq:psi_ac=exact}, thereby confirming the consistency of the contour deformation.

\begin{figure}[h!]
    \includegraphics{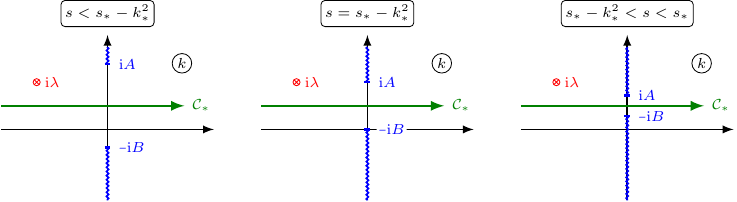}
    \caption{Contour deformation in the $k$-plane. As $s$ increases, the branch point at $k=\ii A(s)$ moves downward and the branch point at $k=-\ii B(s)$ moves upward. At $s=s_*-k_*^2$, the branch point at $k=-\ii B(s)$ falls onto the real line and the integral representation \eqref{eq:psi_ex=def} becomes invalid.  Shifting the integration contour to $\mathcal{C}_*$ at $\Im(k)=k_*$ keeps it between the branch points, allowing analytic continuation \eqref{eq:psi(lambda;A,B)=int C_1} for $s<s_*$.}
    \label{fig:psi_lambda_analytic_A}
\end{figure}

\paragraph{An overview}
Since the technical core of the rest of the paper consists of analytic continuations of the functions $\phi^\pm(\lambda;\rho,s)$, we believe it might be helpful to provide a brief overview of the underlying procedure in light of the toy example described above. The continuation strategy is fundamentally the same, but with considerably more technical details. It is thus useful to keep the simplified picture from the toy example in mind while reading Sections~\ref{sec:Analytic structure in lambda-plane},
\ref{sec:Survival probability Sinfty},
\ref{sec:Survival probability SnX0}, and
\ref{sec:Survival probability StX0} (Section~\ref{sec:moments} is conceptually different and is not illustrated by this example).

\par In Section~\ref{sec:Analytic structure in lambda-plane}, we analyze the analytic structure in the $k$-plane of the integrands defining the functions $\phi^\pm(\lambda;\rho,s)$ in \eqref{eq:phi^pm=def}. We demonstrate that the structure closely resembles that of \eqref{eq:psi_ex=def}, and the singularities governing the asymptotic behavior are: (i) poles at $k=\mp\ii\lambda$ and (ii) two branch cuts. We then perform the analytic continuation in the $\lambda$-plane by tracking the movement of the poles, following the same procedure as in the toy example.

\par In Section~\ref{sec:Survival probability Sinfty}, we study the probability $S_\infty(X_0)$ which, according to \eqref{eq:Sinfty=def} and \eqref{eq:hat(Q)=def}, is obtained by taking the limit $\rho\to0$ and $s\to1$ in \eqref{eq:LTQ=generalIvanov}. In terms of the analytic structure, this limit corresponds to one of the branch points falling onto the real line, as in the toy example at $s=s_*-k_*^2$. This analysis requires careful consideration of the behavior of the integrands in the vicinity of $k=0$.

\par In Section~\ref{sec:Survival probability SnX0}, we perform the continuation in the $s$-plane. Specifically, we show that the asymptotic decay of the survival probability $S_{N}(n\,\vert\,X_0)$ is governed by the point $s_*$ at which two branch points coincide. This leads to the decay rate \eqref{eq:xi_n xi_tau}. In Section~\ref{sec:Survival probability StX0}, we perform a similar procedure for the $\rho$-plane and show that the branch points coincide at $\rho_*$, yielding the corresponding decay rate in \eqref{eq:xi_n xi_tau}.

\par Having outlined this strategy, we now proceed to the computations. We encourage the reader to keep this overview and the toy example in mind throughout the technical analysis that follows.

}

\section{Analytic structure of 
    \texorpdfstring{$\phi^\pm(\lambda;\rho,s)$}{ϕ±(λ; ρ,s)} in the 
    \texorpdfstring{$\lambda$-plane}{λ-plane}}
\label{sec:Analytic structure in lambda-plane}
{
\par In this section, we treat $\phi^\pm(\lambda;\rho,s)$ as functions of $\lambda$ and extend \eqref{eq:phi^pm=def} to the complex $\lambda$-plane. We begin by establishing several key properties of the Fourier transform $F(k;\rho)$ given by \eqref{eq:F(k;rho)=def}, which serves as an elementary building block. In particular, we demonstrate that the analytic structure of the integrands defining $\phi^\pm(\lambda;\rho,s)$ closely resembles that of the toy example discussed in Section~\ref{sec:example_of_continuation}: a pole at $k=\mp\ii\lambda$ and two branch cuts. The analytic continuations $\phi^\pm_\text{a.c.}(\lambda;\rho,s)$ to the $\lambda$-plane are thus constructed following the same approach, namely, by tracking the location of the $\lambda$-dependent pole. In this section, we assume that $s<1$ and $\rho>0$; continuations in the $s$- and $\rho$-planes are addressed in the next sections.
}

\subsection{Properties of \texorpdfstring{$F(k;\rho)$}{F(k; ρ)}}
\label{sec:properties_of_phi} 
\par The key ingredient in $\phi^\pm(\lambda;\rho,s)$ is the Fourier transform $F(k;\rho)$ of the effective random walk. Our analysis therefore relies on several properties of $F(k;\rho)$. 
In particular, we use its asymptotic expansions for real $k$ in the limits $k \to 0$ and $k \to \infty$, as well as its analytic structure in the complex $k$-plane.

\paragraph{Expansions for real $k$} The expansion of $F(k;\rho)$ as $k \to 0$ is straightforward.  Since all moments of $p(t)$ and $q(M)$ are finite, so are the moments of the effective random walk, and thus the expansion reads
\begin{equation}\label{eq:F(k;rho) expansion k->0}
    F(k;\rho) \underset{k\to0}{=} 1 + \ii \mu_1(\rho) k - \mu_2(\rho) \frac{k^2}{2}
    -  \mu_3(\rho) \frac{\ii k^3}{6} + O\left(k^4\right),
\end{equation}
where $\mu_\ell(\rho)$ denotes the $\ell$-th moment of the effective random walk,
\begin{equation}\label{eq:mu(rho)=def}
    \mu_\ell(\rho)  
    \equiv 
    \frac{1}{c(\rho)}
    \int_{0}^{\infty}
    \dd t\, 
    \int_{0}^{\infty}
    \dd M\, 
    e^{-\rho \frac{M}{\alpha}} 
    \left( M- \alpha t \right)^{\ell} 
    q(M)\, p(t),
\end{equation}
or, equivalently, 
\begin{equation}\label{eq:mu(rho)=def<>}
    \mu_\ell(\rho)  
    =
    \frac{1}{c(\rho)}
    \left\langle 
    e^{-\rho \frac{M}{\alpha}} 
    \left( M- \alpha t \right)^{\ell} 
    \right\rangle.
\end{equation}

\par 
For the expansion as $k \to \infty$, we note that $F(k;\rho)$ is a product of two highly oscillatory integrals, each decaying at large $k$. Consider the second integral, assuming that $p(t)$ is supported on $[0,\infty)$. Integration by parts yields
\begin{equation}
    \int_{0}^{\infty}\dd t\,  p(t) e^{-\ii k \alpha t}
    =  \left. - p(t) 
        \frac{e^{-\ii k \alpha t}}
             {\ii k \alpha}  
        \right|_{0}^{\infty}
    +\frac{1}{\ii k \alpha}
     \int_{0}^{\infty} \dd t\, p'(t) e^{-\ii \alpha k t},
\end{equation}
where the boundary term at infinity vanishes due to the decay of $p(t)$. The leading-order behavior is thus determined by the contribution at $t=0$. Since $p(t)$ is continuously differentiable and light-tailed, the remaining integral is subdominant. Hence
\begin{equation}
    \int_{0}^{\infty} \dd t \, p(t) e^{-\ii k \alpha t} 
    \underset{k\to\infty}{=} O\left(\frac{1}{k}\right)  .
\end{equation}
If the probability density is supported on a finite interval instead of the positive real axis, the result includes oscillatory terms, but the absolute value remains of order $1/k$. Repeating the same argument for the integral over $M$ in \eqref{eq:F(k;rho)=def}, we find that it is also $O(1/k)$ as $k \to \infty$. Thus,
\begin{equation}\label{eq:F(k;rho)=k->infty}
    F(k;\rho) \underset{k\to\infty}{=} O\left(\frac{1}{k^2}\right) .
\end{equation}

\begin{figure}[h!]
    \includegraphics{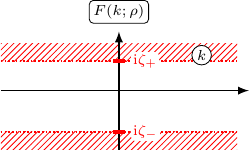}
    \caption{
    The structure of the Fourier transform $F(k;\rho)$ in the complex $k$-plane. The function is analytic in the horizontal strip $\Im(k) \in (\zeta_-, \zeta_+)$, where $\zeta_\pm$ are constants determined by the decay rates of the probability densities $p(t)$ and $q(M)$, as given in \eqref{eq:zeat_pm=def}.
    }
    \label{fig:F_strip}
\end{figure}

\paragraph{Analytic properties in the $k$-plane}
The next step is to analytically continue $F(k;\rho)$ into the complex $k$-plane. Since the distributions $p(t)$ and $q(M)$ are light-tailed, the function defined by \eqref{eq:F(k;rho)=def} is analytic in the complex strip $\zeta_-(\rho) < \Im(k) < \zeta_+$ (see Fig.~\ref{fig:F_strip}). The bounds of this strip are determined by the decay rates of $p(t)$ and $q(M)$ as $t\to\infty$ and $M\to\infty$. Specifically, $\zeta_{+}>0$ and $\zeta_-(\rho)<0$ are $k$-independent quantities  such that
\begin{equation}\label{eq:zeat_pm=def}
    p(t) \underset{t\to\infty}{\lesssim} \exp\left[  - \zeta_+ \, \alpha t \right],
    \qquad 
    q(M) \underset{M\to\infty}{\lesssim} \exp\left[\left(\zeta_-(\rho) + \frac{\rho}{\alpha} \right)M\right].
\end{equation}
The bounds in \eqref{eq:zeat_pm=def} may be equivalently defined as the edges of the region where the integral \eqref{eq:F(k;rho)=def} converges absolutely.

\par In what follows, we write $\zeta_- \equiv \zeta_-(\rho)$, omitting the explicit $\rho$-dependence to simplify notation. In cases where the probability distributions decay faster than exponential (e.g., if $p(t)$ or $q(M)$ is half-Gaussian), we set $\zeta_{\pm} = \pm \infty$. Furthermore, when referring to $F(k;\rho)$, we always assume that $k$ lies within the strip of convergence, i.e., $k = k_R + \ii k_I$ with $k_R \in \mathbb{R}$ and $k_I \in (\zeta_-, \zeta_+)$.

\par 
The first observation about $F(k;\rho)$ is that within the strip of convergence, we have
\begin{equation}\label{eq:|F|<|F(imaginary line)|}
    \abs{F(k_R + \ii k_I; \rho)} < \abs{F(\ii k_I; \rho)}, 
    \qquad k_R \ne 0, \quad k_I \in (\zeta_-, \zeta_+).
\end{equation}
This inequality arises because the real part $k_R$ introduces oscillations in the integral \eqref{eq:F(k;rho)=def}, reducing its absolute value compared to the case where $k$ is purely imaginary.

\par
\begin{figure}[h!]
    \includegraphics[width = \linewidth]{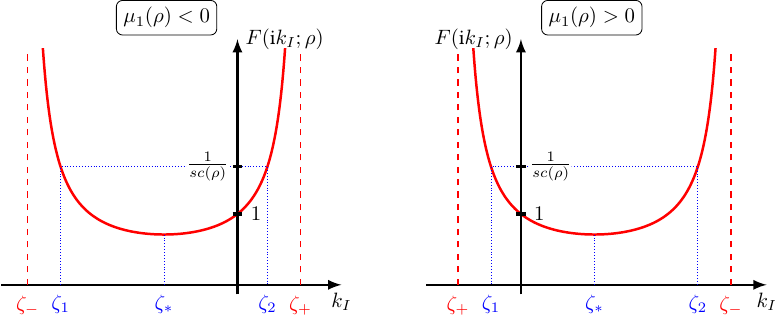}
    \caption{Schematic plot of $F(k; \rho)$ along the imaginary axis.
    The function is convex with the minimum at $\zeta_*\equiv\zeta_*(\rho)$ which is negative if $\mu_1(\rho)<0$ and positive if $\mu_1(\rho)>0$. 
    If $s \, c(\rho)<1$, then there are two real solutions $\zeta_{1,2}\equiv\zeta_{1,2}(\rho,s)$ to \eqref{eq:zeta12=def} located on opposite sides of the origin. 
    }
    \label{fig:F_imaginary_line}
\end{figure}

 
Since the absolute value of $F(k;\rho)$ is bounded by its values along the imaginary axis, let us examine $F(\ii k_I;\rho)$ more closely. From \eqref{eq:F(k;rho)=def}, it immediately follows that
\begin{equation}\label{eq:F(ik;rho)=}
    F(\ii k_I;\rho) = 
    \frac{1}{c(\rho)} 
    \int_{0}^{\infty} \dd M\, q(M)
    \int_{0}^{\infty} \dd t\, p(t)\; 
    e^{ - \rho \frac{M}{\alpha} - k_I (M-\alpha t)},
\end{equation}
and hence 
\begin{equation}
    F(\ii k_I;\rho) \Big|_{k_I=0} = 1, \qquad
    \pdv{}{k_I} F(\ii k_I;\rho) \Big|_{k_I=0} 
    = - \mu_1(\rho).
\end{equation}
Furthermore, for the second derivative we have
\begin{equation}
    \pdv{^2}{k_I^2} F(\ii k_I; \rho) = \frac{1}{c(\rho)} \int_{0}^{\infty} \dd M \, q(M) \int_{0}^{\infty} \dd t \, p(t) \, e^{ - \rho \frac{M}{\alpha} - k_I (M - \alpha t)} (M - \alpha t)^2.
\end{equation}
All factors in the integrand are positive; hence $F(\ii k_I; \rho)$ is strictly convex. This implies a unique minimum, located at some point $k_I = \zeta_*(\rho)$. 
The  position of $\zeta_*(\rho)$ depends on the sign of  $\mu_1(\rho)$: it lies to the left of the origin if  $\mu_1(\rho) > 0$ and to the right if $\mu_1(\rho) < 0$ (see  Fig.~\ref{fig:F_imaginary_line} for a schematic representation).

\par Moreover, we can conclude that, if $s c(\rho)<1$, which is indeed satisfied  for $s\in(0,1)$ and $\rho>0$,  the equation
\begin{equation}\label{eq:zeta12=def}
    1 - s \, c(\rho)\, F(\ii \zeta; \rho) = 0
\end{equation}
admits two real solutions: a negative one $\zeta_1(\rho,s)$ and a positive one $\zeta_2(\rho,s)$. To simplify notation, we will often omit the explicit $s$- and $\rho$-dependence and write $\zeta_{1,2} \equiv \zeta_{1,2}(\rho,s)$ and $\zeta_* \equiv \zeta_*(\rho)$.

\par
\begin{figure}[h!]
    \includegraphics{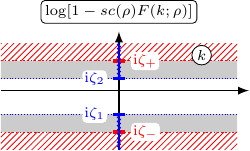}
    \caption{The structure of $\log[1-s c(\rho)F(k;\rho)]$ in the complex $k$-plane.
    The function is analytic in the horizontal strip $\Im(k) \in (\zeta_1, \zeta_2)$ with branch points at $k=\ii \zeta_{1,2}$. The edges of the strip $\zeta_{1,2}$ are two real solutions of \eqref{eq:zeta12=def} as shown in Fig.~\ref{fig:F_imaginary_line}. Note that \eqref{eq:zeta12=def} may have other non-real solutions, leading to more singularities in the logarithmic term, but due to \eqref{eq:|scF|<1 logStrip} these singularities lie in the shaded region. 
    }
    \label{fig:logF_strip}
\end{figure}

 The location of $\zeta_{1,2}$ defines the strip in the $k$-plane, where the logarithmic term  in the integrand \eqref{eq:phi^pm=def} is analytic. Indeed, due to \eqref{eq:|F|<|F(imaginary line)|} we have
\begin{equation}\label{eq:|scF|<1 logStrip}
    \abs{s\, c(\rho)\, F(k;\rho)} < 1 , \quad \Im(k) \in (\zeta_1, \zeta_2).
\end{equation}
Note that \eqref{eq:|scF|<1 logStrip} implies that there are two branch points of the logarithm at $k = \ii \zeta_{1,2}$ (see Fig.~\ref{fig:logF_strip}). In principle, there may be other singularities in the logarithmic term arising from the non-real solutions of \eqref{eq:zeta12=def}, but they all lie outside the horizontal strip $\Im(k)\in(\zeta_1,\zeta_2)$. 

{
\par As a final remark, we note that due to \eqref{eq:|F|<|F(imaginary line)|}, the function $F(k;\rho)$ has a saddle point at $k=\ii\zeta_*$. In its vicinity, we can therefore approximate this function by a quadratic form, which leads to a structure very similar to that of the toy example \eqref{eq:psi_ex=def}. With this observation, we conclude the list of properties of $F(k;\rho)$ that are important for the analysis to come.
}

\subsection{Continuation in the \texorpdfstring{$\lambda$-plane}{λ-plane}}
\label{sec:sub:Analytic continuation of phi in lambda-plane}
Now we proceed to the analytic continuation of $\phi^\pm(\lambda;\rho,s)$ to the $\lambda$-plane, assuming that $s\in(0,1)$ and $\rho>0$. Recall that according to \eqref{eq:phi^pm=def} the functions $\phi^\pm(\lambda;\rho,s)$ are given by
\begin{equation}\label{eq:phi^pm=def1}
    \phi^{\pm}(\lambda;\rho,s) = 
    \exp\left[
        -\frac{1}{2\pi} \int_{-\infty}^{\infty}
        \dd k\; \frac{1}{\lambda\pm \ii k}
        \log\left[1 - s\, c(\rho) F(k;\rho)\right]
    \right].
\end{equation}
The similarity with the example \eqref{eq:psi_ex=def} is apparent. There is only one $\lambda$-dependent singularity, i.e., a pole at $k=\pm\ii\lambda$. 
The analytic structure of the logarithm, on the other hand, depends on the exact form of the function $F(k;\rho)$ and can be rather involved. Fortunately, there is no need to analyze it in detail, and we can focus on the strip $\Im(k)\in(\zeta_1,\zeta_2)$.

\par  
The analytic continuation of $\phi^\pm(\lambda;\rho,s)$ is performed in three steps. The first step is to continue $\phi^\pm(\lambda;\rho,s)$ to the region $\Re(\lambda)>0$ by simply extending the definition~\eqref{eq:phi^pm=def1}. The second step is to consider a strip $\Im(k)\in(\zeta_1,\zeta_2)$ where the logarithmic term has no singularities (see Fig.~\ref{fig:logF_strip}). Within this strip we can freely deform the contour of integration to encircle the pole in the same way as was done in the previous example (see Fig.~\ref{fig:pole_movement_psi}). Then, separating the contribution from the pole and computing the residue we obtain the functional relation 
\begin{equation}\label{eq:phi_ac=functional relation}
    \phi^{\pm}_\text{a.c.}(\lambda;\rho,s)\;
    \phi_\text{a.c.}^{\mp}(-\lambda;\rho,s)
    = \frac{1}{1-s \, c(\rho)\, F(\pm\ii\lambda;\rho)}.
\end{equation}
The third and final step is to use \eqref{eq:phi_ac=functional relation} as the definition of $\phi^\pm_\text{a.c.}(\lambda;\rho,s)$ for all $\Re(\lambda)<0$. This procedure results in the analytic continuation 
\begin{equation}\label{eq:phi_ac=cases}
    \phi_\text{a.c.}^\pm(\lambda;\rho,s)
    \equiv 
    \left\{\begin{aligned}
        & \phi^\pm(\lambda;\rho,s), \quad && \Re(\lambda) > 0,
        \\
        & \frac{\phi^\pm(\lambda;\rho,s)}{\sqrt{ 1 - s\, c(\rho)\, F(\pm \ii \lambda;\rho)}},\quad &&\Re(\lambda)=0,\\
        & \frac{1}
               {1 - s\, c(\rho)\, F(\pm \ii\lambda;\rho)}\,
               \phi^{\pm}(\lambda;\rho,s),\quad 
        && \pm\zeta_{\mp}<\Re(\lambda)<0.\\
    \end{aligned}\right.
\end{equation}
Recall that $F(k;\rho)$ is, in general, defined for $\Im(k)\in(\zeta_-,\zeta_+)$, hence the lower bound for $\Re(\lambda)$ in \eqref{eq:phi_ac=cases}. 

\par 
The functions $\phi^\pm(\lambda;\rho,s)$ are regular when $\Re(\lambda) \ne 0$. Consequently their analytic continuations have no singularities in the right half-plane $\Re(\lambda)>0$ and all singularities that may appear in the left half-plane $\Re(\lambda)<0$ are those of the prefactor. As for the imaginary line $\Re(\lambda)=0$, the analytic continuations are regular there by construction.
\begin{figure}[h!]
    \includegraphics{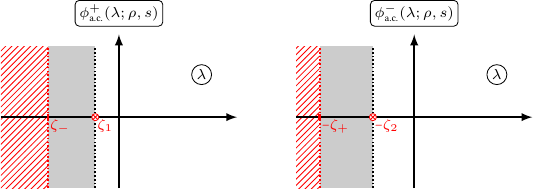}
    \caption{
    Analytic structure of $\phi^\pm_\text{a.c.}(\lambda;\rho,s)$ in the $\lambda$-plane. 
    Left panel: \eqref{eq:phi_ac=cases} defines the function $\phi^+_\text{a.c.}(\lambda;\rho,s)$ in the region $\Re(\lambda)>\zeta_-$; the rightmost singularity is a pole at $\lambda=\zeta_1(\rho,s)$. 
    Right panel: \eqref{eq:phi_ac=cases} defines the function $\phi^-_\text{a.c.}(\lambda;\rho,s)$ in the region $\Re(\lambda)>-\zeta_+$; the rightmost singularity is a pole at $\lambda=-\zeta_2(\rho,s)$. 
    }
    \label{fig:phi_lambda_anal}
\end{figure}

\par 
In other words, the functions $\phi^\pm_\text{a.c.}(\lambda;\rho,s)$ have singularities at $\lambda^\pm_*$ in the left half-plane, which are solutions of
\begin{equation}\label{eq:lambda*=def_general}  
    1 - s\,c(\rho)\,F(\pm \ii\lambda^\pm_*;\rho) = 0, \quad \Re(\lambda^\pm_*)<0,  
\end{equation} 
This is the same equation (up to the sign of the argument) as \eqref{eq:zeta12=def}. Recall that for the ``$+$'' sign, it admits two real solutions $\zeta_{1,2}(\rho,s)$ with $\zeta_1 < 0 < \zeta_2$ and, due to \eqref{eq:|scF|<1 logStrip}, has no solutions in the strip $\Re(\lambda)\in(\zeta_1,\zeta_2)$. Therefore, the rightmost singularity of $\phi^+_{\text{a.c.}}(\lambda;\rho,s)$ in the left half-plane is a pole at $\lambda = \zeta_1(\rho,s)$. Similarly, the rightmost singularity of $\phi^-_{\text{a.c.}}(\lambda;\rho,s)$ is a pole at $\lambda = -\zeta_2(\rho,s)$ (see Fig.~\ref{fig:phi_lambda_anal}).

\section{Survival probability \texorpdfstring{$S_\infty(X_0)$}{S∞(X0)}}\label{sec:Survival probability Sinfty}
\par Having performed the continuation of $\phi^\pm(\lambda;\rho,s)$ in the $\lambda$-plane, the next steps are to consider the continuations in the $s$- and $\rho$-planes, as the analytic structure in these planes provides us with the asymptotic behaviors of $S_N(n\,\vert\,X_0)$ and $S_T(\tau\,\vert\,X_0)$. 
It is, however, instructive to start with a more robust quantity: the probability to survive indefinitely $S_\infty(X_0)$. The Laplace transform $\hat{S}_\infty(\lambda)$ of $S_\infty(X_0)$ is given by
\begin{equation}\label{eq:hat(Sinfty)=def 1}
    \hat{S}_\infty(\lambda) =  
    \int_{0}^{\infty} \dd X_0 \, e^{-\lambda X_0}
    \left(  
    1- \int_{0}^{\infty} \dd \tau 
    \sum_{n=0}^{\infty} \mathbb{P}[\tau,n\,\vert X_0] 
    \right)
    = \frac{1}{\lambda} - \lim_{\substack{\rho\to0 \\ s\to1}}\hat{Q}(\lambda;\rho,s).
\end{equation}
Finding the exact expression for $\hat{S}_\infty(\lambda)$ requires only computing the limit $s\to1$, $\rho\to0$ of $\hat{Q}(\lambda;\rho,s)$ rather than performing the full analytic continuation to the $\rho$- and $s$-planes.  { In the context of the toy example of Section~\ref{sec:example_of_continuation}, this limit corresponds to the situation in which one of the branch points falls onto the real line.
}

\par In this section, we compute the limit in \eqref{eq:hat(Sinfty)=def 1} and show that there are indeed two different regimes separated by the critical drift $\alpha_c$: the \textit{absorption regime}, where $S_\infty(X_0)=0$, and the \textit{survival regime}, where $S_\infty(X_0)>0$. For the survival regime, we compute the first two terms of the asymptotic expansion of $S_\infty(X_0)$ when $X_0\to0$ and the rate of exponential decay for $X_0\to\infty$. { We then verify the asymptotic expansions numerically, taking both $p(t)$ and $q(M)$ to be inverse-Gaussian distributions.}

\subsection{Explicit expression for the Laplace transform}
\par Recall that the analytic structure of integrals entering $\phi_\text{a.c.}^\pm(\lambda;\rho,s)$ in the $k$-plane described in Section~\ref{sec:Analytic structure in lambda-plane} (see Fig.~\ref{fig:logF_strip}) is valid for $s<1$ and $\rho>0$. 
In the limit $s\to1$ and $\rho\to0$ the structure changes, as one of the branch points of the logarithm falls onto the real line inducing the singularity at $k=0$.
Specifically, if $\langle M-\alpha t \rangle<0$ then $\zeta_2=0$, and if $\left\langle M-\alpha t\right\rangle>0$ then $\zeta_1=0$ (see Fig.~\ref{fig:F_imaginary_line}). 
At the same time, as $\lambda\to0$ the pole $k=-\ii\lambda$ in $\phi^+_\text{a.c.}(\lambda;\rho,s)$ also falls onto the real line at $k=0$. Recall that $\phi_\text{a.c.}^+(0;\rho,s)$ is one of the factors in the representation \eqref{eq:LTQ=generalIvanov} for $\hat{Q}(\rho,s\,\vert\,\lambda)$, hence, in the limit $\rho\to0$ and $s\to1$, the coincidence of two singularities at the origin  may induce a non-trivial behavior and needs to be dealt with.

\par Since the problem may arise due to the singularity at $k=0$, we preemptively avoid potential complications by explicitly factoring out the possibly singular behavior from the integral before taking the limit.
To simplify the notation, we focus on the case $\lambda>0$. The continuation to $\lambda<0$ can be found straightforwardly from \eqref{eq:phi_ac=cases}. 
First, we formally rewrite~\eqref{eq:phi^pm=def1} as:
\begin{multline}\label{eq:phi^pm = F2 factored out}
    \phi^{\pm}(\lambda;\rho,s)    
    = 
    \exp\left[
        -\frac{1}{2\pi} \int_{-\infty}^{\infty}
        \dd k\; \frac{1}{\lambda\pm \ii k}
        \log\left[ 
            \frac{1 - s\, c(\rho) F(k;\rho)}
                 {1 - s\, c(\rho) F_2(k;\rho)}
        \right] 
    \right]
    \\
    \times 
    \exp\left[
        -\frac{1}{2\pi}
        \int_{-\infty}^{\infty}
        \frac{\dd k}{\lambda \pm \ii k} 
        \log\left[1 - s\,  c(\rho) F_2(k;\rho)\right]
    \right],
\end{multline}
where
\begin{equation}\label{eq:F2(k;rho)=def}
    F_2(k;\rho) = 1 
        + \ii \, \mu_1(\rho) \; k 
        - \frac{1}{2} \mu_2(\rho) k^{2}.
\end{equation}
Note that \eqref{eq:phi^pm = F2 factored out} amounts to adding and subtracting the term corresponding to the expansion of the integrand in the vicinity of $k=0$.
This term can easily be computed since 
\begin{equation}\label{eq:1-scF=()()}
    1-s \, c(\rho)\, F_2(k;\rho) = 
    \frac{1}{2} s \, c(\rho)\, \mu_2(\rho)
    \big(k - \ii k_+^*(\rho,s)\big)\big(k + \ii k_-^*(\rho,s)\big),
\end{equation}
with
\begin{equation}\label{eq:k_pm=m1m2}
    k_\pm^*(\rho,s) = 
        \pm \frac{\mu_1(\rho)}{\mu_2(\rho)} + 
        \sqrt{ \left(\frac{\mu_1(\rho)}{\mu_2(\rho)}\right)^2 + \frac{2}{\mu_2(\rho)}\left(\frac{1}{s \, c(\rho)}-1\right) }.
\end{equation}
Applying \eqref{eq:1-scF=()()} to the second term in \eqref{eq:phi^pm = F2 factored out} yields an integral that, up to a constant, has the same form as the auxiliary function $\psi^+(\lambda)$ given by \eqref{eq:psi_ex=def}, with $A$ replaced by $k_{+}^*$ and $B$ replaced by $k_{-}^{*}$.
Using \eqref{eq:psi=cases} we obtain for $\Re(\lambda)>0$
\begin{multline}\label{eq:phi^pm(lambda;rho,s)=regular}
    \Re(\lambda)>0: \quad 
    \phi^{\pm}(\lambda;\rho,s) = 
    \frac{1}{\lambda+k_{\mp}^{*}(\rho,s)} \sqrt{\frac{2}{s\, c(\rho) \, \mu_2(\rho)}}
    \\ \times
    \exp\left[
        -\frac{1}{2\pi} \int_{-\infty}^{\infty}
        \dd k\; \frac{1}{\lambda\pm \ii k}
        \log\left[ 
            \frac{1 - s\, c(\rho)\, F(k;\rho)}
                 {1 - s\, c(\rho)\, F_2(k;\rho)}
        \right] 
    \right].
\end{multline}
Although this representation is obtained for $\Re(\lambda)>0$ it can be readily extended to the full $\lambda$-plane. 
Such an extension consists of combining the prefactors already found in \eqref{eq:psi_ac=cases} and \eqref{eq:phi_ac=cases}. The resulting expressions are rather cumbersome and hence we do not present them here explicitly, except for one observation. Specifically, since there is no pole at $k=0$ in the integrand when $\lambda=0$, the analytic continuation $\phi_\text{a.c.}^\pm(\lambda;\rho,s)$ evaluated at $\lambda=0$ is given by
\begin{multline}\label{eq:phi^pm(0;rho,s)=regular}
    \phi^{\pm}_\text{a.c.}(0;\rho,s) = 
    \frac{1}{ k_{\mp}^{*}(\rho,s)}
    \sqrt{\frac{2}{s\, c(\rho) \, \mu_2(\rho)}}
    \\ \times
    \exp\left[
        \mp\frac{1}{2\pi} \int_{-\infty}^{\infty}
        \dd k\; \frac{1}{\ii k}
        \log\left[ 
            \frac{1 - s\, c(\rho)\, F(k;\rho)}
                 {1 - s\, c(\rho)\, F_2(k;\rho)}
        \right] 
    \right].
\end{multline}
The main advantage of \eqref{eq:phi^pm(0;rho,s)=regular} is that the integral is regular as $s\to1$, $\rho\to0$ and the nontrivial dependence is encoded solely in the prefactor.

\par Substituting now \eqref{eq:phi^pm(lambda;rho,s)=regular} and \eqref{eq:phi^pm(0;rho,s)=regular} into \eqref{eq:LTQ=generalIvanov} we immediately obtain that the Laplace transform $\hat{Q}(\rho,s\,\vert\,\lambda)$ can be written as
\begin{multline}\label{eq:LTQ_P=general_regular_int}
\Re(\lambda)\ge0:\quad 
    \hat{Q}(\rho,s\,\vert\,\lambda)
    = \frac{1}{\lambda+\frac{\rho}{\alpha}}
    \\
    - \frac{1 - s\, c(\rho)}
           {\lambda+\frac{\rho}{\alpha}}
    \frac{2}{s \, c(\rho)\, \mu_2(\rho)}
    \frac{1}{
        k_{-}^{*}(\rho,s)
    \left( \strut \lambda+k_{+}^*(\rho,s) \right)}
    e^{ I(\lambda;\rho,s)}     ,
\end{multline}
where
\begin{equation}\label{eq:I(l,rho,s) def 1}
    I(\lambda;\rho,s) = 
    -\frac{1}{2\pi} \int_{-\infty}^{\infty}
    \dd k\,
    \left(
        \frac{1}{\ii k} + \frac{1}{\lambda+\frac{\rho}{\alpha}-\ii k}
    \right)
    \log
        \frac{1 - s \, c(\rho) \, F(k;\rho)}
             {1 - s \, c(\rho) \, F_2(k;\rho)} 
    .
\end{equation}
The representation \eqref{eq:LTQ_P=general_regular_int} is convenient to compute the limit in \eqref{eq:hat(Sinfty)=def 1} and thereby the Laplace transform $\hat{S}_\infty(\lambda)$.

\par In the limit $s\to1$, $\rho\to0$ the integral \eqref{eq:I(l,rho,s) def 1} is regular and the nontrivial behavior is contained in the prefactor. A simple calculation shows that
\begin{equation}\label{eq:lim rho->0, s->1, prefactor}
    \lim_{\rho\to0, s\to1}
    \left[
    \frac{1-s \,c(\rho)}{ k_{-}^{*}(\rho,s)}
    \frac{1}{\lambda+k_+^*(\rho,s)}
    \frac{2}{s\, c(\rho) \, \mu_2(\rho)}
    \right] = 
    \begin{cases}
        0, & \mu_1(0) \le 0,\\
        \frac{1}{1+\lambda\frac{\mu_2(0)}{2\mu_1(0)}},
           & \mu_1(0) > 0.
    \end{cases}
\end{equation}
Hence, recalling the definition \eqref{eq:mu(rho)=def<>} of the moments $\mu_\ell(\rho)$, we obtain
\begin{align}
\label{eq:S_infty=0 general}
    \left\langle M - \alpha t\right\rangle \le 0 : 
    &\qquad 
    \hat{S}_{\infty}(\lambda) = 0,
    \\
    \label{eq:S_infty>0 general}
    \left\langle M - \alpha t\right\rangle > 0 : 
    &\qquad 
    \hat{S}_{\infty}(\lambda) = \frac{1}{\lambda}\frac{1}{1+\lambda \frac{\left\langle (M-\alpha t)^2\right\rangle}{2\left\langle M-\alpha t\right\rangle}}
    e^{I(\lambda;0,1)}.
\end{align}
From \eqref{eq:S_infty=0 general} and  \eqref{eq:S_infty>0 general} the existence of \textit{survival} and \textit{absorption} regimes is evident, and so is the critical value of the drift \eqref{eq:alpha_c_def} separating these two regimes, $\alpha_c = \langle M\rangle / \langle t\rangle$.

\par The survival probability at the infinite time $S_\infty(X_0)$ in the \textit{survival regime} can in principle be obtained by inverting the Laplace transform
\begin{equation}\label{eq:hatSinfty(lambda)=exact 2}
    \Re(\lambda)\ge0 : \qquad 
    \int_{0}^{\infty}\dd X_0\, e^{-\lambda X_0} S_\infty(X_0)
    =
    \frac{1}{\lambda}
    \frac{1}{1+\lambda \frac{\left\langle (M-\alpha t)^2\right\rangle}{2\left\langle M-\alpha t\right\rangle}}
    e^{I(\lambda;0,1)}.
\end{equation}
The inversion is, however, unfeasible unless the probability distributions $p(t)$ and $q(M)$ have a specific form. Instead, we extract the asymptotic behavior of $S_\infty(X_0)$ for $X_0\to\infty$ and $X_0\to0$.

\subsection{Asymptotic behavior}
\par The behavior as $X_0\to0$ is governed by the $\lambda\to\infty$ expansion of $\hat{S}_\infty(X_0)$.
The detailed calculation is provided in Appendix~\ref{sec:app_expansions}, and here we present only the result. Specifically, substituting the expansion $I(\lambda;0,1)$ given by \eqref{eq:final-I-inf-expansion} into \eqref{eq:hatSinfty(lambda)=exact 2}, after algebraic manipulations, yields
\begin{multline}\label{eq:S_infty(lambda)=expansion}
    \hat{S}_\infty(\lambda) \underset{\lambda\to\infty}{=} 
    \frac{1}{\lambda}
    \sqrt{\frac{2\langle M-\alpha t\rangle^2}{\left\langle (M-\alpha t)^2\right\rangle}}
     \exp\left[ -\frac{1}{\pi}
        \int_{0}^{\infty} \dd k
        \Im\left[
        \frac{1}{k}
        \log \frac{1-F(k;0)}
             {1-F_2(k;0)}
        \right]
    \right]\\
    \times\left(
        1
        - \frac{1}{\pi\lambda}
        \int_{0}^{\infty}\dd k\;
        \log\left| 1 - F(k;0)\right|
    \right)
    + O\left(\frac{1}{\lambda^2}\right),
\end{multline}
where $F_2(k;\rho)$ is defined in \eqref{eq:F2(k;rho)=def}. 
According to Tauberian theorems, \eqref{eq:S_infty(lambda)=expansion} translates into
\begin{multline}\label{eq:S_infty(X)=expansion}
    S_\infty(X_0) \underset{X_0\to0}{\sim}
    \sqrt{\frac{2 \langle M-\alpha t\rangle^2}
               {\langle (M-\alpha t)^2\rangle }}
    \exp\left[ -\frac{1}{\pi}
        \int_{0}^{\infty} \dd k
        \Im\left[
        \frac{1}{k}
        \log \frac{1-F(k;0)}
             {1-F_2(k;0)}
        \right]
    \right]
    \\
    \times
    \left(
    1 - \frac{X_0}{\pi}\int_{0}^{\infty}\dd k\;
        \log\left| 1 - F(k;0)\right|
    \right).
\end{multline}
This is the expansion stated in \eqref{eq:S_infty_small_X0}.  
As a consistency check, note that if $\alpha=\alpha_c$, then $\langle M-\alpha_c t\rangle=0$ and the prefactor in~\eqref{eq:S_infty(X)=expansion} is zero, which agrees with $S_\infty(X_0)=0$ as in~\eqref{eq:S_infty=0 general}.

\par The behavior of $S_\infty(X_0)$ as $X_0\to\infty$ can be inferred from the singularity structure of $\hat{S}_\infty(\lambda)$ in the $\lambda$-plane. The representation \eqref{eq:hatSinfty(lambda)=exact 2} is regular for $\Re(\lambda)>0$, and thus this singularity lies in $\Re(\lambda)\le0$. By either keeping track of the analytic structure in \eqref{eq:phi^pm(lambda;rho,s)=regular} and combining the prefactors found in \eqref{eq:psi_ac=cases} and \eqref{eq:phi_ac=cases} or, alternatively, by repeating the same arguments as presented in Section~\ref{sec:sub:Analytic continuation of phi in lambda-plane} at the level of the representation \eqref{eq:S_infty>0 general} we find that
\begin{equation}
    \Re(\lambda)\le0:\qquad
    \hat{S}_\infty(\lambda) 
    =
    \frac{1-F_2(-\ii\lambda;0)}{1-F(-\ii\lambda;0)}
    \frac{1}{\lambda}
    \frac{1}{1+\lambda \frac{\left\langle (M-\alpha t)^2\right\rangle}{2\left\langle M-\alpha t\right\rangle}}
    e^{I(\lambda;0,1)},
\end{equation}
or, using the explicit form of $F_2(k;\rho)$ as in \eqref{eq:F2(k;rho)=def},
\begin{equation}\label{eq:Sinfty(l) l<0}
    \Re(\lambda)\le 0:\qquad
    \hat{S}_\infty(\lambda) 
    =
    - \frac{\left\langle M-\alpha t \right\rangle}{1-F(-\ii\lambda;0)}
    e^{I(\lambda;0,1)}.
\end{equation}
As we have argued before in Section~\ref{sec:Analytic structure in lambda-plane} the denominator has two real solutions at $\lambda=-\zeta_{1,2}$ (see right panel of Fig.~\ref{fig:F_imaginary_line}) and no solutions in between these two values. In the limit $\rho\to0$ and $s\to1$ we have $\zeta_1=0$, hence in \eqref{eq:Sinfty(l) l<0} there are two poles: one at $\lambda=0$, and the other at $\lambda=-\zeta_2$.

\par  The leading behavior of $S_\infty(X_0)$ as $X_0\to\infty$ is governed by the pole at $\lambda=0$. Expanding \eqref{eq:Sinfty(l) l<0} as $\lambda\to0$ we find that 
\begin{equation}\label{eq:Sinfty(l)=1/l + O(1)}
    \hat{S}_\infty(\lambda) = \frac{1}{\lambda} + O\left(1\right),\qquad \lambda\to0, 
\end{equation}
which gives the asymptotic value of $S_\infty(X_0)$ 
\begin{equation}\label{eq:Sinfty(infty)=1}
    \lim_{X_0\to\infty} S_\infty(X_0) = 1.
\end{equation}
The result \eqref{eq:Sinfty(infty)=1} is indeed very natural. 
In the survival regime $\langle M-\alpha t\rangle>0$ so that on average the process moves away from the origin. Therefore, the further away from the origin it starts, the lower is the probability of the first-passage event. In the limit $X_0\to\infty$ this probability tends to $1$. The more interesting question is how this asymptotic value is approached. 

\par The subleading corrections can be inferred from the singularities of $\hat{S}_\infty(\lambda)$. The first of them corresponds to the singularity with the largest real part. Here, this is the pole at $\lambda=-\zeta_2(0,1)$. Introducing the notation $R\equiv\zeta_2(0,1)$ we write
\begin{equation}\label{eq:Sinfty(x0)asymp}
     1 - S_\infty(X_0) \underset{X_0\to\infty}{\asymp}
     e^{- R  X_0}.
\end{equation} 
Recall that $R$ is the smallest positive solution of  $F(\ii R;0) = 1$.
\par As a final remark here, we should mention that the behavior \eqref{eq:Sinfty(x0)asymp} is actually well-known in the financial literature, in which $S_\infty(X_0)$ denotes the probability of the ultimate ruin. In this context, the condition $F(\ii R;0) = 1$ is essentially equivalent to what is known as the Lundberg fundamental equation \cite{Asmussen2010}. Here we obtain this decay rate as a byproduct, which both serves as a consistency check and demonstrates the power of the representation \eqref{eq:LTQ=generalIvanov}.

{
\subsection{Numerical verification}
The asymptotic results derived above hold for arbitrary light-tailed distributions $p(t)$ and $q(M)$ with smooth densities. To verify them, it is instructive to consider a specific example. One natural choice would be exponential distributions, which would recover the exact results from \cite{BM-25}; this is an exercise we leave to the reader. Since the present paper addresses the case of general distributions, we instead take both $p(t)$ and $q(M)$ to be inverse-Gaussian
\begin{align}
    \label{eq:p(t)=invGauss}
    &p(t) = p(t; m_t, \ell_t ) = 
    \sqrt{\frac{\ell_t }{2\pi\, t^3}}
    \exp\left[
        - \ell_t \frac{\left( t - m_t \right)^2}{2m_t^2 t}
    \right],
    \\
    \label{eq:q(M)=invGauss}
    &q(M) = q(M; m_M, \ell_M ) = 
    \sqrt{\frac{\ell_M }{2\pi\, M^3}}
    \exp\left[
        - \ell_M \frac{\left( M - m_M \right)^2}{2m_M^2 M}
    \right].
\end{align}
The critical value of the drift is given by $\alpha_c=m_M/m_t$. This choice serves two purposes: first, it clearly falls beyond the Poisson class, providing a genuine test of our general framework; second, while the integrals in \eqref{eq:S_infty(X)=expansion} still require numerical evaluation, the decay rate in \eqref{eq:Sinfty(x0)asymp} admits a simple closed form.

\begin{figure}[h]
    \includegraphics[width=\linewidth]{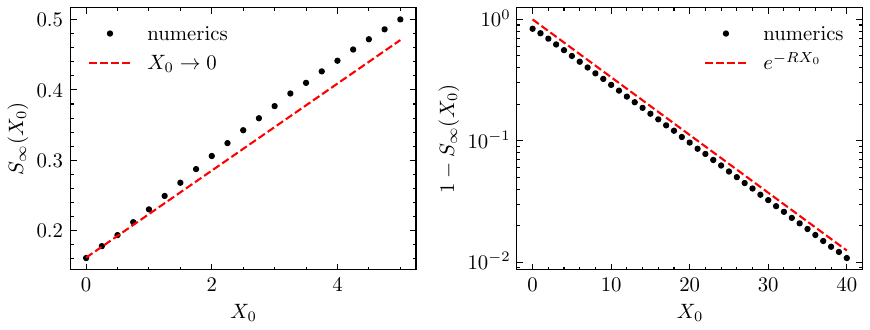}
    \caption{
    Comparison of analytical predictions (red dashed lines) with Monte Carlo simulations (black circles) using $10^6$ independent trajectory realizations for each value of $X_0$. Parameters: $m_t=2$, $\ell_t=15$, $m_M=4$, $\ell_M=10$ (so that $\alpha_c=2$), and $\alpha=1.8$. Each trajectory includes $1000$ jumps to approximate infinite-time behavior. 
    Left: asymptotic expansion \eqref{eq:S_infty(X)=expansion} with coefficients computed by numerical integration. 
    Right: exponential decay \eqref{eq:Sinfty(x0)asymp} with rate \eqref{eq:R=invGauss} (logarithmic scale).
    }\label{fig:Sinfty_numerics}
\end{figure}

\par 
For the distributions \eqref{eq:p(t)=invGauss} and \eqref{eq:q(M)=invGauss}, expressions \eqref{eq:c(rho)=def} and \eqref{eq:F(k;rho)=def} for the effective random walk reduce to
\begin{equation}\label{eq:c(rho)=invGauss}
    c(\rho) = \exp\left[
        \frac{\ell_M}{m_M}
        \left(1 - \sqrt{ 1 + 2 \rho \,\frac{m_M^2}{\alpha\, \ell_M} }\right)
    \right]
\end{equation}
and 
\begin{multline}\label{eq:F(k;rho)=invGauss}
    F(k;\rho) = 
    \frac{1}{c(\rho)}
    \exp\left[
        \frac{\ell_t}{m_t}
        \left( 
            1-\sqrt{1 + 2\ii \alpha k \frac{m_t^2}{\ell_t}}
        \right)
    \right]
    \\ 
    \times
    \exp\left[
        \frac{\ell_M}{m_M} 
        \left( 1 - 
            \sqrt{1 + 2\frac{m_M^2}{\ell_M}\left(\frac{\rho}{\alpha} - \ii k \right) }
        \right)
    \right].
\end{multline}
A straightforward calculation then yields the decay rate for the survival probability \eqref{eq:Sinfty(x0)asymp}. Specifically, solving the equation $F(\ii R; 0)=1$ for the smallest positive $R$, we find
\begin{equation}\label{eq:R=invGauss}
    R = 2 \frac{\ell_M \ell_t}{m_M m_t}
    \left(\frac{\ell_M}{m_M} + \frac{\ell_t}{m_t}\right)
    \frac{\left(m_M - \alpha m_t\right)}
         {\left(\ell_M + \alpha \ell_t \right)^2 }.
\end{equation}
Figure~\ref{fig:Sinfty_numerics} compares the analytical predictions with numerical simulations. The agreement confirms both the full asymptotic expansion \eqref{eq:S_infty(X)=expansion} (left panel) and the exponential decay \eqref{eq:Sinfty(x0)asymp} with rate \eqref{eq:R=invGauss} (right panel).

}

\par Now we proceed with the analytic continuation in the $s$- and $\rho$-planes. For simplicity and to align with the goal of studying first-passage properties, we will be performing the continuations directly for the Laplace transforms of the survival probabilities $S_N(n\,\vert\,X_0)$ and $S_T(\tau\,\vert\,X_0)$, rather than for the functions $\phi^\pm_\text{a.c.}(\lambda;\rho,s)$.

\section{Survival probability \texorpdfstring{$S_N(n\,\vert\,X_0)$}{SN(n|X0)}}\label{sec:Survival probability SnX0}
\par Let us first consider the probability of the process staying positive up to the $n$-th jump. The Laplace transform of this probability, defined as
\begin{equation}\label{eq:hatSn=}
    \hat{S}_N(s\,\vert\,\lambda) \equiv \int_{0}^{\infty}\dd X_0\, e^{-\lambda X_0} 
    \sum_{n=0}^{\infty}s^n S_N(n\,\vert\,X_0)
\end{equation}
can be represented in terms of $\hat{Q}(\rho,s\,\vert\,\lambda)$ as
\begin{equation}\label{eq:Sn(s|l)=expressionQ}
    \hat{S}_{N}(s\,\vert\,\lambda) = 
    \left. \frac{1}{1-s}\left(\frac{1}{\lambda} - \hat{Q}(\rho,s\,\vert\,\lambda)\right) \right|_{\rho=0}.
\end{equation}
Equivalently, using \eqref{eq:LTQ=generalIvanov} and replacing $\phi^\pm(\lambda;\rho,s)$ with their respective analytic continuations \eqref{eq:phi_ac=cases}, we find
\begin{equation}\label{eq:Sn(s|l)=expression}
    \hat{S}_{N}(s\,\vert\,\lambda) = \frac{1}{\lambda}
    \phi^{+}_\text{a.c.}(0;\rho,s)\;
    \phi^{-}_\text{a.c.}
        \left(\lambda+\frac{\rho}{\alpha};\rho,s\right) 
    \Big|_{\rho=0}.
\end{equation}
In contrast to Section~\ref{sec:Survival probability Sinfty}, it is more convenient to use explicit regularization for $\phi^+_\text{a.c.}(0;\rho,s)$ in \eqref{eq:phi^pm=def1}, rather than utilizing \eqref{eq:phi^pm(lambda;rho,s)=regular} or treating the integral in \eqref{eq:phi^pm=def1} in the principal value sense. Specifically, we will use the following representation:
\begin{equation}\label{eq:S(s|lambda)=limE}
    \hat{S}_{N}(s\,\vert\,\lambda) = 
    \lim_{\epsilon\to0}
    \frac{1}{\lambda}
    \phi^{+}_\text{a.c.}(\epsilon;\rho,s)\;
    \phi^{-}_\text{a.c.}
        \left(\lambda+\frac{\rho}{\alpha};\rho,s\right) 
    \Big|_{\rho=0}.
\end{equation}
The analytic structure of $\hat{S}_N(s\,\vert\,\lambda)$ in the $\lambda$-plane follows directly from that of $\phi^\pm_\text{a.c.}(\lambda;\rho,s)$ as given in \eqref{eq:phi_ac=cases}. In this section we focus on its behavior in the $s$-plane. 
{ We show that the Laplace transform \eqref{eq:S(s|lambda)=limE} can be analytically continued following the same procedure as in the toy example of Section~\ref{sec:example_of_continuation}. In short, by shifting the integration contour in the $k$-plane from the real line to a horizontal line at $\Im(k)=\zeta_*$, we can perform the continuation up to the point $s_*$, where the two branch points of the logarithm coincide. This pinch singularity determines the decay rate $\xi_n$ in \eqref{eq:xi_n xi_tau}.}

\par 
As will be shown shortly, the analytic structure of $\hat{S}_N(s\,\vert\,\lambda)$ depends on the sign of $\mu_1(0)=\langle M-\alpha t\rangle$, since $\mu_1(0)>0$ and $\mu_1(0)<0$ correspond to the \textit{survival} and \textit{absorption} regimes, respectively, and $\mu_1(0)=0$ is a \textit{critical} point.
{ Below, we study these cases separately. The resulting asymptotic expressions are then again tested against the numerical simulations with both inter-arrival times and jump amplitudes drawn from inverse-Gaussian distributions.}

\subsection{Survival regime}
We start with the \textit{survival regime}, i.e., the case where $\mu_1(0)>0$.
In this scenario, we expect a finite probability that the process stays positive indefinitely and the first-passage never occurs. 
By analyzing the analytic structure of $\hat{S}_N(s\,\vert\,\lambda)$ in the $s$-plane we show that $S_\infty(X_0)>0$ and that $S_N(n\,\vert\,X_0)$ approaches $S_\infty(X_0)$ exponentially fast with the rate given by $\xi_n(\alpha)$ as in \eqref{eq:xi_n xi_tau}.

\begin{figure}[h!]\centering
    \includegraphics{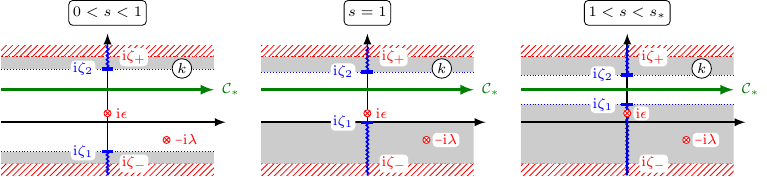}
    \caption{Schematic representation of the analytic structure of the integrand in \eqref{eq:S(s|l)=lim(int) l>0} in the complex $k$-plane within the survival regime. The integrand features two branch points at $k = \ii \zeta_{1,2}$ and two poles at $k = \ii \epsilon$ and $k = -\ii \lambda$. Upon continuation to $s > 1$, the branch points pinch the shifted contour $\mathcal{C}_*=\{k:\Im(k)=\zeta_*(0)\}$ inducing a singularity at $s=s_*$ defined by \eqref{eq:s*=condition pinch} in the $s$-plane.}\label{fig:fig_s_survival_anal}
\end{figure}

\paragraph{Analytic continuation in the $s$-plane}
Assume for the moment that $\lambda > 0$ (the case $\lambda \le 0$  will be considered shortly). Then according to \eqref{eq:phi_ac=cases} we have:
\begin{equation}
    \lambda>0:\quad \phi_\text{a.c.}^{\pm}(\lambda;0,s) = \phi^{\pm}(\lambda;0,s). 
\end{equation} 
Hence \eqref{eq:S(s|lambda)=limE} implies
\begin{equation}\label{eq:S(s|l)=lim(int) l>0}
    \hat{S}_N(s\,\vert\,\lambda) = 
    \lim_{\epsilon\to0}
    \frac{1}{\lambda}
    \exp\left[-\frac{1}{2\pi}
        \int_{-\infty}^{\infty} \dd k\,
        \left(  
            \frac{1}{\epsilon+\ii k} +
            \frac{1}{\lambda-\ii k}
        \right)
        \log\left[1 - s F(k;0)\right]
    \right].
\end{equation}
The continuation in the $s$-plane is performed by deforming the contour of integration in the $k$-plane. The analytic structure of the integrand is as follows (see Fig.~\ref{fig:fig_s_survival_anal} for the schematic illustration): The logarithm is analytic within the strip $\Im(k) \in (\zeta_1, \zeta_2)$ and has branch cuts at $k = \ii \zeta_{1,2}$, where  $\zeta_{1,2} \equiv \zeta_{1,2}(0, s)$ are the real solutions of  \eqref{eq:zeta12=def} with $\rho=0$ (see Fig.~\ref{fig:F_imaginary_line} and Fig.~\ref{fig:logF_strip}). The term in parentheses contributes two simple poles at $k = \ii \epsilon$ and $k = -\ii \lambda$. 

\par Note that $c(0) = 1$, and thus as $s \to 1$, we have  $\zeta_1(0, s) \to 0$. This implies that the branch point at  $\ii \zeta_1(0, s)$ moves onto the contour of integration at $k = 0$.  
Simultaneously, the pole at $k = \ii \epsilon$ also approaches the contour at $k = 0$ as $\epsilon\to0$.
This is exactly the coincidence of singularities studied in Section~\ref{sec:Survival probability Sinfty}.

\par To perform the continuation to the region $s>1$, we shift the contour of integration in~\eqref{eq:S(s|l)=lim(int) l>0} from the real line to a  horizontal contour $\mathcal{C}_*$ such that $\Im(k) = \zeta_*(0)$ (see Fig.~\ref{fig:fig_s_survival_anal}), where $\zeta_*(\rho)$ is the location of the minimum of  $F(k; \rho)$ along the imaginary axis (see Fig.~\ref{fig:F_imaginary_line}). Recall that if $\mu_1(0) > 0$, then $\zeta_*(0) > 0$, and when shifting the contour we must account for the residue at $k = \ii \epsilon$. The resulting expression is
\begin{multline}
    \hat{S}_N(s\,\vert\,\lambda) = 
    \lim_{\epsilon\to0}
    \Biggl\{
    \frac{1}{\lambda}
    \frac{1}{1 - s F(\ii\epsilon;0)}
    \\\times
    \exp\left[-\frac{1}{2\pi}
        \int_{\mathcal{C}_*} \dd k\,
        \left(  
            \frac{1}{\epsilon+\ii k} +
            \frac{1}{\lambda-\ii k}
        \right)
        \log\left[1 - s F(k;0)\right]
    \right]
    \Biggr\}.
\end{multline}
Taking the limit $\epsilon\to0$ is now straightforward, and we have
\begin{equation}\label{eq:S(s|lambda)=shifted_contour}
    \hat{S}_N(s\,\vert\,\lambda) = 
    \frac{1}{\lambda}
    \frac{1}{1 - s}
    \exp\left[-\frac{1}{2\pi}
        \int_{\mathcal{C}_*} \dd k\,
        \left(  
            \frac{1}{\ii k} +
            \frac{1}{\lambda-\ii k}
        \right)
        \log\left[1 - s F(k;0)\right]
    \right].
\end{equation}
As the value of $s$ increases, the branch points at $\ii\zeta_{1,2}$ approach the contour $\mathcal{C}_{*}$, finally pinching once $s$ reaches a critical value $s_*$. The pinch occurs when
\begin{equation}\label{eq:s*=condition pinch}
    s_* : \quad \zeta_1(0;s_*) = \zeta_2(0;s_*) = \zeta_*(0),
\end{equation}
or, equivalently, when \eqref{eq:zeta12=def} has one real solution, i.e., 
\begin{equation}\label{eq:s*=condition}
    \frac{1}{s_*} = F\left(\ii\zeta_*(0);0\right) = 
    \min_{k_I} F\left(\ii k_I;0\right).
\end{equation}

\par The analytic structure of \eqref{eq:S(s|lambda)=shifted_contour} in the $s$-plane is now clear. The prefactor gives rise to a simple pole at $s = 1$.  
The integral remains analytic for $\left|s\right| < s_*$, and has a singularity at $s=s_*$, which a more careful analysis reveals to be a branch point, though its precise nature is not essential here. The expression \eqref{eq:S(s|lambda)=shifted_contour}  therefore provides an analytic continuation valid for  $\left|s\right| < s_*$.
The only step remaining is to extend the above derivation to the case $\Re(\lambda)\le0$, in which according to \eqref{eq:phi_ac=cases}, we have
\begin{equation}
 \lambda<0:\quad \phi_\text{a.c.}^{\pm}(\lambda;0,s) = \frac{1}{1-s F(\pm\ii\lambda;0)}\phi^{\pm}(\lambda;0,s).
\end{equation}
The analysis proceeds in the same way. The only subtlety here is that the pole at $k=-\ii\lambda$ lies in the upper half-plane. If $\lambda\in\left(-\zeta_*,0\right)$, then while shifting the contour from the real line to $\mathcal{C}_*$ we must account for two poles: at $k=\ii\epsilon$ and at $k=-\ii\lambda$, with the contribution from the latter canceling out the prefactor.

\par Combining all the arguments above we arrive at the analytic continuation of $\hat{S}_N(s\,\vert\,\lambda)$ valid within the disk of radius $s_*$ defined by \eqref{eq:s*=condition}, namely,
\begin{equation}\label{eq:S(s|l)=ac}
\alpha<\alpha_c:\qquad 
    \hat{S}_N(s\,\vert\,\lambda) = 
    \left\{
    \begin{aligned}
    &\frac{1}{\lambda}\frac{\mathcal{G}_N(s\,\vert\,\lambda)}{1 - s} 
    ,\quad 
    \lambda>-\zeta_*(0),\\
    &
    \frac{1}{1-sF\left(-\ii\lambda;0\right)}
    \frac{1}{\lambda}\frac{\mathcal{G}_N(s\,\vert\,\lambda)}{1 - s}
    ,\quad 
    \lambda<-\zeta_*(0),
    \end{aligned}
    \right.
\end{equation}
where we use the notation
\begin{equation}\label{eq:GN(s|l)=}
    \mathcal{G}_N(s\,\vert\,\lambda) = 
    \exp\left[-\frac{1}{2\pi}
        \int_{\mathcal{C}_*} \dd k\,
        \left(  
            \frac{1}{\ii k} +
            \frac{1}{\lambda-\ii k}
        \right)
        \log\left[1 - s F(k;0)\right]
    \right].
\end{equation}
Recall that, strictly speaking, for $F(-\ii\lambda;0)$ in the second line of \eqref{eq:S(s|l)=ac} to be well-defined, we need $-\ii\lambda$ to lie within the strip of analyticity, i.e., $\lambda$ must satisfy $-\zeta_+(0)<\lambda < \zeta_{*}(0)$. Thus \eqref{eq:S(s|l)=ac} is valid for $\lambda\in(-\zeta_+(0),\infty)$.

\par
\begin{figure}[h!]\centering
\includegraphics{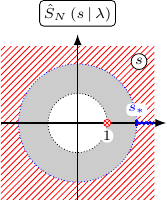}
\caption{Schematic representation of the analytic structure of the $\hat{S}_N(s\,\vert\,\lambda)$ in the $s$-plane in the survival regime for $\lambda>-\zeta_*(0)$. There is a pole at $s=1$ and a branch point at $s_*$ defined by \eqref{eq:s*=condition}. 
In the unshaded region representations \eqref{eq:S(s|l)=lim(int) l>0} and  \eqref{eq:S(s|l)=ac} are equivalent.
In the shaded region they differ, with \eqref{eq:S(s|l)=ac} providing a proper analytic continuation.}\label{fig:Sn_anal_survival}
    
\end{figure}

 To summarize, we have shown that for $\lambda>-\zeta_*(0)$ the function $\hat{S}_N(s\,\vert\,\lambda)$ in the  $s$-plane has a simple pole at $s = 1$ and a branch point at $s = s_*$ with no other singularities for $\abs{s}<s_*$. The schematic illustration is shown in Fig.~\ref{fig:Sn_anal_survival}. We emphasize that if $\lambda < -\zeta_*(0)$, then additional singularities may arise from the prefactor in \eqref{eq:S(s|l)=ac}. However, these singularities have no effect on the large-$n$ behavior of $S_N(n\,\vert\,X_0)$, as will be explained shortly.

\paragraph{Asymptotic behavior of $S_N(n\,\vert\,X_0)$} 
The standard result in the context of analytic combinatorics \cite{FlajoletSedgewick2009} is that the large-$n$ behavior of the survival probability $S_N(n\,\vert\,X_0)$ is governed by the singularities of its generating function closest to the origin. To find this generating function, we formally invert the Laplace transform with respect to $\lambda$ and obtain
\begin{equation}
    \sum_{n=0}^{\infty}s^n S_{N}(n\,\vert\,X_0)
    = \frac{1}{2\pi\ii }\int_{\gamma -\ii\infty}^{\gamma + \ii \infty} \dd \lambda 
    \, e^{\lambda X_0} \hat{S}_{N}(s\,\vert\,\lambda),
\end{equation}
where the integration is performed along the vertical contour that lies to the right of all possible singularities of $\hat{S}_{N}(s\,\vert\,\lambda)$. For example, we can choose $\gamma$ such that $ \gamma > -\zeta_*(0)$. 
With this choice, due to \eqref{eq:S(s|l)=ac}, the integrand $S_N(s\,\vert\,\lambda)$ has a pole at $s=1$ and a branch point at $s=s_*$. The pole corresponds to the asymptotic value of $S_N(n\,\vert\,X_0)$ and the branch point at $s_*$ determines the rate at which this value is approached.

\par First and foremost, the residue at $s = 1$ determines the asymptotic value of $S_N(n\,\vert\,X_0)$:
\begin{equation}
   \lim_{n\to\infty}
   \int_{0}^{\infty} \dd X_0\, e^{-\lambda X_0} 
    S_N(n\,\vert\,X_0) 
    =
    - \Res_{s=1} \left[  \hat{S}_{N}(s\,\vert\,\lambda) \right].
\end{equation}
This limit is nothing but $\hat{S}_\infty(\lambda)$, and computing the residue yields
\begin{equation}\label{eq:hatSinfty(lambda)=cases}
    \hat{S}_\infty(\lambda) =
        \left\{
    \begin{aligned}
    &\frac{1}{\lambda}\;\mathcal{G}_N(1\,\vert\,\lambda)
    ,\quad 
    \lambda>-\zeta_*(0),\\
    &
    \frac{1}{\lambda}\frac{\mathcal{G}_N(1\,\vert\,\lambda)}{1-F\left(-\ii\lambda;0\right)}
    ,\quad 
    \lambda\in(-\zeta_+(0),-\zeta_*(0)).
    \end{aligned}
    \right.
\end{equation}
Although the representation \eqref{eq:hatSinfty(lambda)=cases} is slightly different from \eqref{eq:hatSinfty(lambda)=exact 2} and \eqref{eq:Sinfty(l) l<0} obtained in Section~\ref{sec:Survival probability Sinfty}, their equivalence can be established by shifting the contour of integration in $\mathcal{G}_N(1\,\vert\,\lambda)$ as in \eqref{eq:GN(s|l)=} from $\mathcal{C}_*$ back to the real line. Since the detailed analysis of $\hat{S}_\infty(\lambda)$ and $S_\infty(X_0)$ was already performed in Section~\ref{sec:Survival probability Sinfty}, here we focus on the corrections for $S_N(n\,\vert\,X_0)$.

\par The first subleading correction is given by the singularity in the $s$-plane closest to the origin. In the present case this is a branch cut at $s_*$, hence
\begin{equation}\label{eq:LT(S-Sinfty)}
    \int_{0}^{\infty} \dd X_0\, e^{-\lambda X_0}
    \left[ S_N(n\,\vert\,X_0) - S_\infty(X_0) \right]
    \underset{n\to\infty}{\asymp}
    s_*^{-n}.
\end{equation}
Recall that $s_*$, as defined in \eqref{eq:s*=condition}, does not depend on $\lambda$, hence the $n\to\infty$  behavior of the Laplace transform \eqref{eq:LT(S-Sinfty)} translates into the $n\to\infty$ behavior of $S_{N}(n\,\vert\,X_0)$, namely, 
\begin{equation}\label{eq:Sn-Sinft=asympt}
    S_N(n\,\vert\,X_0) - S_\infty(X_0) 
    \underset{n\to\infty}{\asymp}
    \exp\left[
        - \frac{n}{\xi_n(\alpha)}
    \right],
\end{equation}
with the decay rate given by
\begin{equation}\label{eq:xi_n=def}
    \xi_n(\alpha) = \frac{1}{\log s_*},\qquad
    \frac{1}{s_*} =\min_{k_I} F(\ii k_I;0).
\end{equation}
This is the result stated in \eqref{eq:SN_ST_decays}.

\subsection{Absorption regime} We now proceed to the \textit{absorption regime} in which $\mu_1(0)<0$ and the process eventually crosses the origin, $S_\infty(X_0)=0$. The analysis is very similar to that of the survival regime. The main difference is that now $\zeta_*(0)<0$, and the contour of integration is shifted to the lower part of the $k$-plane. As a result, there is no contribution from the pole at $k=\ii\epsilon$, hence, there is no pole at $s=1$ for $\hat{S}_N(s\,\vert\,\lambda)$,  and the analytic continuation reads
\begin{equation}
    \alpha>\alpha_c:\quad
    \hat{S}_N(s\,\vert\,\lambda) = 
    \left\{
    \begin{aligned}
    &\frac{1}{\lambda}\;\mathcal{G}_N(s\,\vert\,\lambda)
    ,\quad 
    \lambda>-\zeta_*(0),\\
    &
    \frac{1}{\lambda}\frac{\mathcal{G}_N(s\,\vert\,\lambda)}{1-sF\left(-\ii\lambda;0\right)}
    ,\quad 
    \lambda\in(-\zeta_+(0),-\zeta_*(0)).
    \end{aligned}
    \right.
\end{equation}
where $\mathcal{G}_N(s\,\vert\,\lambda)$ is again given by \eqref{eq:GN(s|l)=}. 
\par Following the same procedure as in the survival regime, we find the result stated in~\eqref{eq:SN_ST_decays abs}, namely
\begin{equation}\label{eq:Sn=asympt}
    S_N(n\,\vert\,X_0) \underset{n\to\infty}{\asymp} \exp\left[-\frac{n}{\xi_n(\alpha)}\right]
\end{equation}
with the decay rate given by \eqref{eq:xi_n=def}.

\subsection{Critical point}\label{sub:critical_point_SN} 
Finally, we analyze the behavior of $S_N(n\,\vert\,X_0)$ at the critical point. Specifically, we show that as the drift approaches the critical value the  <<correlation length>> $\xi_n(\alpha)$ diverges, resulting in polynomial decay of the survival probability $S_N(n\,\vert\,X_0)$ as $n\to\infty$.  Additionally, we show that there is a nontrivial limiting behavior as $X_0\to\infty$ and $n\to\infty$ with fixed ratio $\frac{X_0}{\sqrt{n}}$ and compute the corresponding scaling function.

\subsubsection{Divergence of the <<correlation length>>}
With the notation \eqref{eq:F(k;rho)=def_<>}, the expression~\eqref{eq:xi_n=def} for the <<correlation length>> $\xi_n(\alpha)$ reads
\begin{equation}\label{eq:xi_n=def_1}
    \frac{1}{\xi_n(\alpha)}
    = - \log \left\langle e^{-(M-\alpha t)\zeta_* }\right\rangle,
\end{equation}
where $\ii \zeta_*\equiv\ii \zeta_*(0)$ is the location of the minimum of $F(k;0)$ along the imaginary axis in the $k$-plane. If $\alpha=\alpha_c$, then this minimum is located exactly at the origin, $\zeta_*|_{\alpha=\alpha_c} = 0$, which implies that $\frac{1}{\xi_n(\alpha_c)}=0$, that is, $\xi_n(\alpha)$ diverges as the strength of the drift approaches its critical value. 

\par To quantify the divergence of $\xi_n(\alpha)$ as $\alpha\to\alpha_c$ more precisely, we expand \eqref{eq:xi_n=def_1} in series using the fact that $\zeta_*|_{\alpha=\alpha_c} = 0$. The expansion yields
\begin{multline}\label{eq:1/xi_n=expansion1}
    \frac{1}{\xi_n(\alpha)} 
    \underset{\alpha\to\alpha_c}{\sim} 
    -\log\left\langle \vphantom{\frac{\strut}{\strut}} \right.
        1   
        -
        (\alpha-\alpha_c) 
        (M-\alpha_c t) 
        \left. 
        \dv{\zeta_*}{\alpha}
        \right|_{\alpha=\alpha_c}
        \\
        + \frac{(\alpha-\alpha_c)^2}{2}
        \left(
            \left.
                2 t\,  \dv{\zeta_*}{\alpha}
                + (M-\alpha_c t)^{2} 
                \left(\dv{\zeta_*}{\alpha}\right)^{2}
                - (M-\alpha_c t)\dv{^2\zeta_*}{\alpha^2}
            \right|_{\alpha=\alpha_c}
        \right)
    \left.\vphantom{\frac{\strut}{\strut}}\right\rangle.
\end{multline}
When taking the average in the logarithm, the term with the second derivative of $\zeta_*$ disappears and the expansion depends only on the first derivative. Specifically, we obtain
\begin{equation}
    \frac{1}{\xi_n(\alpha)} 
    \underset{\alpha\to\alpha_c}{\sim} 
    -\log\left[
        1   
        + \frac{(\alpha-\alpha_c)^2}{2}
        \left(
            \left.
                2 \langle t \rangle   \dv{\zeta_*}{\alpha}
                + 
                \left\langle (M-\alpha_c t)^{2}  \right\rangle
                \left(\dv{\zeta_*}{\alpha}\right)^{2}
            \right|_{\alpha=\alpha_c}
        \right)
    \right],
\end{equation}
or, after a simple calculation, 
\begin{equation}\label{eq:1/xi_n sim (a-ac)^2}
    \frac{1}{\xi_n(\alpha)} 
    \underset{\alpha\to\alpha_c}{\sim} 
    -
    \frac{(\alpha-\alpha_c)^2}{2}
        \left(
            \left.
                2 \langle t \rangle   \dv{\zeta_*}{\alpha}
                + 
                \left\langle (M-\alpha_c t)^{2}  \right\rangle
                \left(\dv{\zeta_*}{\alpha}\right)^{2}
            \right|_{\alpha=\alpha_c}
        \right).
\end{equation}

\par The only thing left is to find the derivative of $\zeta_*$ at the critical point. Recall that $\zeta_*$ is the position of the minimum, hence it satisfies
\begin{equation}\label{eq:<M-at>=0}
    \left\langle (M-\alpha t) e^{- \zeta_* (M-\alpha t)} \right\rangle = 0.
\end{equation}
By taking the derivatives with respect to $\alpha$, we can iteratively build the expansion of $\zeta_*(\alpha)$. Using that $\zeta_*|_{\alpha=\alpha_c}=0$ we obtain 
\begin{equation}
    \left\langle (M-\alpha_c t) - 
    (\alpha-\alpha_c)
        \left(t + (M-\alpha_c t)^2 \left.\dv{\zeta_*}{\alpha}\right|_{\alpha=\alpha_c} \right)
    \right\rangle 
    = 0,
\end{equation}
hence
\begin{equation}\label{eq:pd_alpha zeta*(0)=}
    \left. \dv{\zeta_*}{\alpha}\right|_{\alpha=\alpha_c} = 
        -\frac{\left\langle t\right\rangle}
            {\left\langle (M-\alpha_c t)^2\right\rangle}.
\end{equation}
Finally, substituting \eqref{eq:pd_alpha zeta*(0)=} into \eqref{eq:1/xi_n sim (a-ac)^2} we get
\begin{equation}
    \frac{1}{\xi_n(\alpha)} 
    \underset{\alpha\to\alpha_c}{\sim}
    \frac{(\alpha-\alpha_c)^2}{2}
                \frac{\left\langle t\right\rangle^2}
                     {\left\langle (M-\alpha_c t)^2\right\rangle},
    \qquad
    \xi_n(\alpha) \underset{\alpha\to\alpha_c}{\sim}
    \frac{2}{\left\langle t\right\rangle^2}
    \frac{\left\langle (M-\alpha_c t)^2\right\rangle}{\left(\alpha- \alpha_c\right)^2}.
\end{equation}
This is exactly the behavior stated in \eqref{eq:xiT xin=divergence_res}.

\subsubsection{Algebraic decay of $S_N(n\,\vert\,X_0)$}
\par If $\alpha=\alpha_c$, then $s_*=1$ and $\xi_n(\alpha_c)=\infty$, hence there is no exponential decay of $S_N(n\,\vert\,X_0)$ as $n\to\infty$. At the level of the analytic structure of $\hat{S}_N(s\,\vert\,\lambda)$ in the $s$-plane this divergence is manifested by the coincidence of the pole and the branch point, as both of them are located at $s=1$. The large-$n$ behavior of $S_N(n\,\vert\,X_0)$ is governed by the behavior of $\hat{S}_N(s\,\vert\,\lambda)$ in the vicinity of $s=1$. 

\par To construct this expansion, we again resort to the representation \eqref{eq:phi^pm(lambda;rho,s)=regular} for $\phi^\pm(\lambda;\rho,s)$. Substituting \eqref{eq:LTQ_P=general_regular_int} in \eqref{eq:Sn(s|l)=expressionQ} yields
\begin{equation}\label{eq:Sn(s|l)=critical I}
    \Re(\lambda)>0:\quad 
    \hat{S}_{N}(s\,\vert\,\lambda) = 
    \frac{1}
           {s\,\lambda}
    \frac{2}{\left\langle (M-\alpha_c t)^2\right\rangle}
    \frac{1}{
        k_{-}^{*}(0,s)
    \left( \strut \lambda+k_{+}^*(0,s) \right)}
    e^{ I(\lambda;0,s)},
\end{equation}
with $k_\pm^*(\rho,s)$ given by \eqref{eq:k_pm=m1m2}. Expanding \eqref{eq:Sn(s|l)=critical I} as $s\to1$ we obtain
\begin{equation}\label{eq:Sn(s|l)=critical expansion}
    \Re(\lambda)>0:\quad 
    \hat{S}_N(s\,\vert\,\lambda) \underset{s\to1}{\sim}
    \frac{1}{\sqrt{1-s}} \frac{1}{\lambda^2} 
        \frac{1}{\sqrt{\frac{1}{2}\langle (M-\alpha_c t)^2\rangle}}e^{I(\lambda;0,1)},
\end{equation}
where $I(\lambda;\rho,s)$ is given by \eqref{eq:I(l,rho,s) def 1}. 
The singular behavior \eqref{eq:Sn(s|l)=critical expansion} suggests that as $n\to\infty$ the survival probability $S_N(n\,\vert\,X_0)$ decays as $\frac{1}{\sqrt{n}}$. Specifically, we have
\begin{equation}\label{eq:Sn-U(X)}
    S_N(n\,\vert\,X_0) \underset{n\to\infty}{\sim}
    \frac{1}{\sqrt{\pi n}}\,  U(X_0),
\end{equation}
with the function $U(X_0)$ defined through its Laplace transform $\hat{U}(\lambda)$ as 
\begin{equation}\label{eq:U(X0)=LT}
    \Re(\lambda)\ge0:\quad 
    \hat{U}(\lambda)\equiv
    \int_{0}^{\infty} \dd X_0\, e^{-\lambda X_0} U(X_0)
    = \frac{1}{\lambda^2} \frac{1}{\sqrt{\frac{1}{2} \left\langle (M-\alpha_ct)^2\right\rangle  }}
    e^{I(\lambda;0,1)}.
\end{equation}
Now we use \eqref{eq:U(X0)=LT} to extract the asymptotic behaviors of $U(X_0)$ for $X_0\to0$ and $X_0\to\infty$ by studying the behavior of $\hat{U}(\lambda)$. 

\par The analysis is similar to the one performed in Section~\ref{sec:Survival probability Sinfty} for the survival probability $S_\infty(X_0)$. There is, however, a major difference. According to \eqref{eq:Sinfty(l)=1/l + O(1)} the Laplace transform $\hat{S}_\infty(\lambda)$ has a singularity of the order $1/\lambda$, which yields a limiting behavior $S_\infty(X_0)\to1$ as $X_0\to\infty$.  
Extracting the subleading corrections requires a proper analytic continuation to $\Re(\lambda)<0$, which eventually results in \eqref{eq:Sinfty(x0)asymp}. At the same time, the Laplace transform $\hat{U}(\lambda)$ of $U(X_0)$, as is clear from \eqref{eq:U(X0)=LT},  behaves as $1/\lambda^2$ when $\lambda\to0$. Consequently, the leading order behavior for $X_0\to\infty$ can be extracted solely from the $\lambda\to0$ expansion. While the continuation to $\Re(\lambda)<0$ can be straightforwardly performed utilizing the contour deformation in the $k$-plane, to simplify the discussion we do not present it here, especially since the resulting corrections are exponentially small as~$X_0\to\infty$.

\par The asymptotic expansions of $U(X_0)$ as $X_0\to0$ and $X_0\to\infty$ are governed by the behaviors of $\hat{U}(\lambda)$ as $\lambda\to\infty$ and $\lambda\to0$, respectively. The main technical challenge lies in expanding $I(\lambda;0,1)$. We provide the detailed derivation in Appendix~\ref{sec:app_expansions}, and present only the results here. Specifically, substituting expansion  \eqref{eq:final-I-0-expansion} into \eqref{eq:U(X0)=LT} yields, for $X_0\to\infty$,
\begin{equation}\label{eq:Un(X0->inf)}
    U(X_0)\underset{X_0\to\infty}{\sim}
    \frac{1}{\sqrt{\frac{1}{2}\langle (M-\alpha_c t)^2\rangle}}
    \left(
        X_0 
        - \frac{1}{2\pi} \int_{-\infty}^{\infty} \dd k\;
            \frac{1}{k^2}
            \log\frac{1 - F(k;0)}
                     {\frac{1}{2}k^2\langle (M-\alpha_c t)^2\rangle}
    \right) ,
\end{equation}
and similarly due to \eqref{eq:final-I-inf-expansion} for $X_0\to0$ we obtain
\begin{multline}\label{eq:Un(X0->0)}
    U(X_0)  \underset{X_0\to0}{\sim}
    \exp\left[ -\frac{1}{\pi}
        \int_{0}^{\infty} \dd k
        \Im\left[
        \frac{1}{k}
        \log \frac{1-F(k;0)}
             {\frac{1}{2}k^2\langle (M-\alpha_c t)^2\rangle}
        \right]
    \right]
    \\ \times
    \left(
    1 - \frac{X_0}{\pi}\int_{0}^{\infty}\dd k\;
        \Re\big[  \log\left[ 1 - F(k;0)\right] \big]
    \right).
\end{multline}
We stress that expressions \eqref{eq:Un(X0->inf)} and \eqref{eq:Un(X0->0)} are explicit for any given distributions $p(t)$ and $q(M)$ that have finite moments and smooth densities.

\par 
Before proceeding to the analysis of the survival probability $S_T(\tau\,\vert\,X_0)$, two remarks are in order. First, the asymptotic  expansions \eqref{eq:Un(X0->inf)} and \eqref{eq:Un(X0->0)} are derived  at the critical point $\alpha=\alpha_c$, under the assumption that the limit $n\to\infty$ is taken first. In this case, the survival probability  $S_N(n\,\vert\,X_0)$ decays algebraically as $1/\sqrt{n}$, with a  prefactor depending on the initial position $X_0$. A natural question is, then, what happens when both $n$ and $X_0$ tend to infinity simultaneously. This leads to a non-trivial scaling limit, which is essentially equivalent to the Brownian motion.  

\par 
Second, there exists a special case in which the jump distribution  $q(M)$ and the waiting-time distribution $p(t)$ are tuned so that the  effective random walk becomes symmetric. Although this situation is  restrictive, it provides a valuable benchmark, since the survival  probability of symmetric random walks on the half-line has been studied  extensively. In this case, our expressions \eqref{eq:Sn-U(X)},  \eqref{eq:U(X0)=LT}, \eqref{eq:Un(X0->inf)} and \eqref{eq:Un(X0->0)} should recover the results of \cite{MMS-17}.

\paragraph{Scaling limit}

At the critical point $\alpha=\alpha_c$, the effective random walk consists of independent identically distributed jumps with zero mean and finite variance. In this situation it is natural to expect convergence to the Brownian motion in the limit where both $X_0$ and $n$ tend to infinity with a fixed ratio $z=X_0/\sqrt{n}$. Accordingly, we anticipate that
\begin{equation}\label{eq:Sn=scalingV(z)}
    S_N(n\,\vert\,X_0) \sim V\!\left(z=\frac{X_0}{\sqrt{n}}\right),
    \qquad n\to\infty,\quad  X_0\to\infty,
\end{equation}
with a scaling function $V(z)$. Since the scaling limit corresponds to  large $n$ and large $X_0$, $V(z)$ can be obtained from the expansion of $\hat{S}_{N}(s\,\vert\,\lambda)$ as $s\to1$ and $\lambda\to0$, while  respecting the scaling $X_0\sim\sqrt{n}$.

\par To make the connection between the parameters $\lambda$ and $s$ consistent with the scaling relation between $X_0$ and $n$, we first change the integration variable in \eqref{eq:hatSn=} to $z=X_0/\sqrt{n}$. This gives
\begin{equation}\label{eq:hatSn=scaling1}
    \hat{S}_{N}(s\,\vert\,\lambda)
    =
    \sum_{n=0}^{\infty} e^{ n \log s }
    \frac{1}{\sqrt{n}}
    \int_{0}^{\infty} \dd z\, e^{-\lambda z \sqrt{n}} S_{N}\left(n\,\vert\,z \sqrt{n}\right).
\end{equation} 
Next, we introduce the variable $y=n(1-s)$. In the limit $s\to1$, the 
sum over $n$ can be replaced by an integral over $y$, while the survival 
probability $S_N(n\,\vert\,z \sqrt{n})$ can be replaced by its scaling form $V(z)$
\begin{equation}\label{eq:hatSn=scaling2}
    \hat{S}_{N}(s\,\vert\,\lambda)
    =
    \frac{1}{(1-s)^{\frac{3}{2}}}
    \int_{0}^{\infty} \dd y\,  e^{ y \frac{\log s}{1-s} }
    \frac{1}{\sqrt{y}}
    \int_{0}^{\infty} \dd z\, e^{-\frac{\lambda}{\sqrt{1-s}} z \sqrt{y}} 
    \;V(z).
\end{equation}
Representation \eqref{eq:hatSn=scaling2} clearly indicates that the correct scaling relation between $\lambda$ and $s$ is $\lambda\sim\sqrt{1-s}$. Introducing the rescaled parameter $u=\lambda/\sqrt{1-s}$, we obtain
\begin{equation}\label{eq:hatSn=scaling3}
    \hat{S}_N\left(s\,\vert\,u\, \sqrt{1-s}\right)
    \underset{s\to1}{\sim} \frac{1}{(1-s)^{\frac{3}{2}}}
    \int_{0}^{\infty} \dd y\, e^{ - y}
    \frac{1}{\sqrt{y}}
    \int_{0}^{\infty} \dd z\, e^{-u\, z \sqrt{y}} 
    \;V(z).
\end{equation}
On the other hand, the same asymptotic behavior can be obtained directly from \eqref{eq:Sn(s|l)=critical I}. Recall that $I(\lambda;0,s)$ remains regular as $s\to1$ and $\lambda\to0$, so the singularity originates solely from the prefactor. This yields
\begin{equation}\label{eq:hatSn=scaling exact}
    \hat{S}_{N}\left(s\,\vert\,u\,\sqrt{1-s}\right)
    =\frac{1}{(1-s)^{3/2}} 
    \frac{1}{u \left(u+\sqrt{\frac{1}{2}\left\langle (M-\alpha_c t)^2\right\rangle}\right)}.
\end{equation}
Comparing \eqref{eq:hatSn=scaling3} with \eqref{eq:hatSn=scaling exact} results in an integral equation for the scaling function $V(z)$:
\begin{equation}
    \int_{0}^{\infty} \dd y\, e^{ - y}
    \frac{1}{\sqrt{y}}
    \int_{0}^{\infty} \dd z\, e^{-u\, z \sqrt{y}} 
    \;V(z)
    =
    \frac{1}{u \left(u+\sqrt{\frac{1}{2}\left\langle (M-\alpha_c t)^2\right\rangle}\right)}.
\end{equation}
This is precisely the equation studied in detail in \cite{MMS-17} in the context of survival probabilities of one-dimensional random walks. Its solution reads
\begin{equation}\label{eq:Sn=erf scaling}
    S_N(n\,\vert\,X_0) \sim
    \erf\left(\frac{X_0}{\sqrt{2n\left\langle (M-\alpha_c t)^2\right\rangle}}\right),\qquad
    n\to\infty,\quad \frac{X_0}{\sqrt{n}}\text{~--- fixed}.
\end{equation}

\par It is worth emphasizing that the scaling behavior \eqref{eq:Sn=erf scaling} does not rely on the specific structure of the underlying effective random walk beyond the assumptions that it has zero mean, finite moments, and smooth density. In particular, the same functional form of the scaling limit was previously obtained in \cite{MMS-17} for any symmetric random walk with finite variance. This essentially reflects the convergence of a random walk to a Brownian motion, which is a robust property, insensitive to microscopic details.

\paragraph{Symmetric distribution}
Finally, let us consider the case where the effective random walk not only has a zero mean but is also symmetric. Since it requires fine-tuning of the distributions of the jump amplitudes $q(M)$  and inter-jump intervals $p(t)$, such a case is indeed very special. At the same time, the survival probability of symmetric random walks was studied in great detail in \cite{MMS-17}. In particular, the behavior of the survival probability of the symmetric random walk with light-tailed increment distribution was shown, in our notation, to be given by
\begin{equation}
    S_{N}\left(n\,\vert\,X_0\right) 
    \underset{n\to\infty}{\sim}
    \frac{1}{\sqrt{\pi n}} U_\text{sym}(X_0),
\end{equation}
with the function $U_\text{sym}(X_0)$ defined through the Laplace transform as
\begin{equation}\label{eq:Usym(X0)=LT}
    \int_{0}^{\infty} \dd X_0\, e^{-\lambda X_0} U_\text{sym} \left(X_0\right)
    = \frac{1}{\lambda} \exp\left[
        -\frac{\lambda}{\pi} \int_{0}^{\infty} \dd k\, \frac{1}{\lambda^2+k^2} \log\left[1 - F(k;0)\right]
    \right].
\end{equation}
At first glance, the representation \eqref{eq:Usym(X0)=LT} may seem to differ from \eqref{eq:U(X0)=LT}, but they are actually the same. 
Note that the imaginary part of the integrand in \eqref{eq:I(l,rho,s) def 1} is odd, while the real part is even, hence, by the parity argument, we can keep only the real part. Since for a symmetric random walk $F(k;0) = F(-k;0)$ we obtain
\begin{equation}\label{eq:I(l;0,1)=symm}
    I(\lambda;0,1) = - \frac{1}{\pi}\int_{0}^{\infty}\dd k\,
    \frac{\lambda}{\lambda^2 + k^2} \log \frac{1 - F(k;0)}{\frac{1}{2}k^2\langle (M-\alpha t)^2\rangle } 
\end{equation}
Then we extract the term with $\log[ 1-F_2(k;0)]$ by using the identity (which can be verified by standard residue calculus)
\begin{equation}\label{eq:int identity = symm}
    \frac{1}{\pi}\int_{0}^{\infty}\dd k\,\frac{\lambda}{\lambda^2 + k^2}
    \log\left[\frac{1}{2} \mu_2(0) k^2 \right]
    = 
    \frac{1}{2}\log\left[\frac{1}{2}\mu_2(0) \lambda^2 \right].
\end{equation}
Substituting then \eqref{eq:I(l;0,1)=symm} and \eqref{eq:int identity = symm} into \eqref{eq:U(X0)=LT} we see that it now coincides with \eqref{eq:Usym(X0)=LT}.

{
\subsection{Numerical verification}\label{sec:SN-numerical}
To test the validity of the asymptotic results for $S_N(n\,\vert\,X_0)$, we first verify that they reduce to those of \cite{BM-25} when both $p(t)$ and $q(M)$ are exponential distributions; we leave this calculation as an exercise for the reader. To confirm that the results also hold for non-Poissonian arrivals, we perform Monte Carlo simulations.

\par According to \eqref{eq:Sn-Sinft=asympt} and \eqref{eq:Sn=asympt}, if $\alpha\ne\alpha_c$, then the large-$n$ behavior of the survival probability reads
\begin{equation}\label{eq:Sn-Sinft=asympt_1_general}
\alpha\ne\alpha_c :\qquad
    S_N(n\,\vert\,X_0) - S_\infty(X_0) \underset{n\to\infty}{\asymp} 
        \exp\left[ - \frac{n}{\xi_n(\alpha)} \right],
\end{equation}
where the rate $\xi_n(\alpha)$ is given by \eqref{eq:xi_n=def}, and $S_\infty(X_0)=0$ for $\alpha>\alpha_c$. For numerical simulations, to eliminate the constant offset $S_\infty(X_0)$, we analyze the marginal first-passage distribution $\mathbb{P}_c[n\,\vert\,X_0]$ conditioned on trajectories in which first passage has occurred:
\begin{equation}\label{eq:PN=def_marginal}
    \mathbb{P}_c[n\,\vert\,X_0] \equiv \frac{1}{1-S_\infty(X_0)}
        \int_{0}^{\infty} \dd\tilde{\tau}\,
        \mathbb{P}[\tilde{\tau},n\,\vert\,X_0] 
        = \frac{S_{N}(n-1\,\vert\,X_0) - S_{N}(n\,\vert\,X_0)}{1-S_\infty(X_0)}.
\end{equation}
In the absorption regime, where the first passage is sure to happen, this is exactly the same as the marginal probability distribution of $n$. 

\par From  \eqref{eq:Sn-Sinft=asympt_1_general} we expect the exponential decay
\begin{equation}\label{eq:PN-asymp-exp}
    \mathbb{P}_c[n\,\vert\,X_0] \underset{n\to\infty}{\asymp}
    \exp\left[
        - \frac{n}{\xi_n(\alpha)}
    \right].
\end{equation}
This asymptotic behavior can be refined further. Recall that the function $F(k;0)$ has a saddle point at $k=\ii\zeta_*(0)$ (Section~\ref{sec:properties_of_phi}), which is precisely where the branch points coincide at $s=s_*$. The branch points $\zeta_{1,2}(0,s)$ are determined by the equation $1-sF(\ii\zeta;0)=0$. Expanding $F(\ii k;0)$ near the saddle point $k=\zeta_*$ yields a quadratic form, so that $\zeta_{1,2}(0,s)$ satisfy a quadratic equation in $(\zeta-\zeta_*)$. They thus exhibit square-root behavior analogous to $A(s)$ and $B(s)$ in the toy example \eqref{eq:A,B=A(s), B(s)} of Section~\ref{sec:example_of_continuation}. This reveals that the singularity at $s_*$ is a square-root branch point, implying
\begin{equation}\label{eq:Sn-Sinft=asympt_power_1_general}
    S_N(n\,\vert\,X_0) - S_\infty(X_0) 
    \underset{n\to\infty}{\sim}
    \frac{1}{\sqrt{n}}
    \exp\left[
        - \frac{n}{\xi_n(\alpha)}
    \right].
\end{equation}
At the level of the marginal probability, this gives
\begin{equation}\label{eq:PN-asymp-refined}
    \mathbb{P}_c[n\,\vert\,X_0] \underset{n\to\infty}{\sim}
    \frac{1}{n^{3/2}}
    \exp\left[
        - \frac{n}{\xi_n(\alpha)}
    \right].
\end{equation}

\par 
To verify the asymptotic behavior of $S_N(n\,\vert\,X_0)$, we compare analytical predictions with numerical simulations. As in Section~\ref{sec:Survival probability Sinfty}, we take both $p(t)$ and $q(M)$ to be inverse-Gaussian distributions given by \eqref{eq:p(t)=invGauss} and \eqref{eq:q(M)=invGauss}. In this case, the decay rate \eqref{eq:xi_n=def} can be evaluated explicitly. A straightforward computation shows that
\begin{equation}\label{eq:xi_n=InverseGaussian}
    \frac{1}{\xi_n(\alpha)} = 
    \left(\ell_t \sqrt{\alpha} + \frac{\ell_M}{\sqrt{\alpha}}\right)
    \sqrt{\frac{1}{\ell_M+\alpha\ell_t} 
         \left( \alpha \frac{\ell_M}{m_M^2} + \frac{\ell_t}{m_t^2} \right)}
    - \frac{\ell_M}{m_M} - \frac{\ell_t}{m_t}.
\end{equation}
In the numerical simulations, we chose the parameters to be
\begin{equation}\label{eq:numerical_parameters}
    m_t=1, \qquad \ell_t=5, \qquad m_M=2, \qquad \ell_M=15, \qquad \alpha_c = \frac{m_M}{m_t} = 2.
\end{equation}
With these parameters, we generated $10^9$ independent trajectories starting at $X_0=0.5$ for two values of the drift velocity: $\alpha=1.6$ and $\alpha=2.4$. 

\par Figure~\ref{fig:Sn_non_crit} confirms both the exponential decay rate \eqref{eq:xi_n=InverseGaussian} (left column) and the $n^{-3/2}$ power law (right column) through comparison with numerical simulations in both the survival (top row, $\alpha=1.6$) and absorption (bottom row, $\alpha=2.4$) regimes. 
While plotting the histogram, we retained only values of $n$ that appeared in at least $500$ configurations to ensure adequate statistics.

\begin{figure}[h]
    \includegraphics[width=\linewidth]{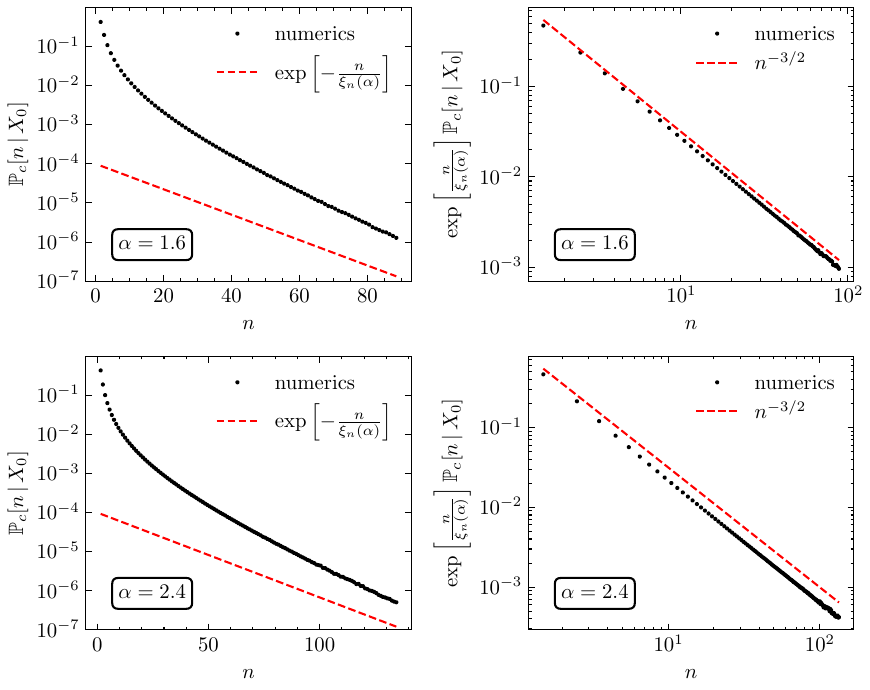}
    \caption{
    Comparison of analytical predictions (red dashed lines) with direct Monte Carlo simulations (black circles, $10^9$ trajectory realizations for $X_0=0.5$) in the survival regime (top row, $\alpha=1.6$) and absorption regime (bottom row, $\alpha=2.4$). 
    Parameters of the distribution are given by \eqref{eq:numerical_parameters}. 
    Left column: marginal first-passage probability $\mathbb{P}_c[n\,\vert\,X_0]$ versus $n$ (logarithmic scale), confirming the exponential decay \eqref{eq:PN-asymp-exp} with rate \eqref{eq:xi_n=InverseGaussian}.
    The theoretical curves are shifted vertically by a factor of $10^{-4}$ for visual comparison.
    Right column: rescaled probability $e^{n/\xi_n(\alpha)}\mathbb{P}_c[n\,\vert\,X_0]$ versus $n$ (log-log scale), confirming the $n^{-3/2}$ power law in \eqref{eq:PN-asymp-refined}.
    }
    \label{fig:Sn_non_crit}
\end{figure}

\par At the critical point $\alpha=\alpha_c$, the exponential decay \eqref{eq:Sn-Sinft=asympt_power_1_general} is replaced by the power-law behavior, which is given by \eqref{eq:Sn-U(X)}:
\begin{equation}\label{eq:Sn-U(X)_1}
    S_N(n\,\vert\,X_0) \underset{n\to\infty}{\sim} \frac{1}{\sqrt{\pi n}} U(X_0).
\end{equation}
Unlike the rates in the off-critical regimes, the asymptotic expansions \eqref{eq:Un(X0->inf)} and \eqref{eq:Un(X0->0)} of $U(X_0)$ for $X_0\to\infty$ and $X_0\to0$ require numerical integration. For the parameters \eqref{eq:numerical_parameters} we obtain
\begin{equation}\label{eq:U(X0)-numerical}
    U(X_0) \underset{X_0\to0}{\approx} 1.06 - 1.03\; X_0, \qquad
    U(X_0) \underset{X_0\to\infty}{\approx} 1.22\; X_0 + 0.90.
\end{equation}
We computed $S_N(n\,\vert\,X_0)$ for $n=100$ numerically across a range of initial positions $X_0$ to test three distinct regimes.

\begin{figure}[ht]
\includegraphics[width=\linewidth]{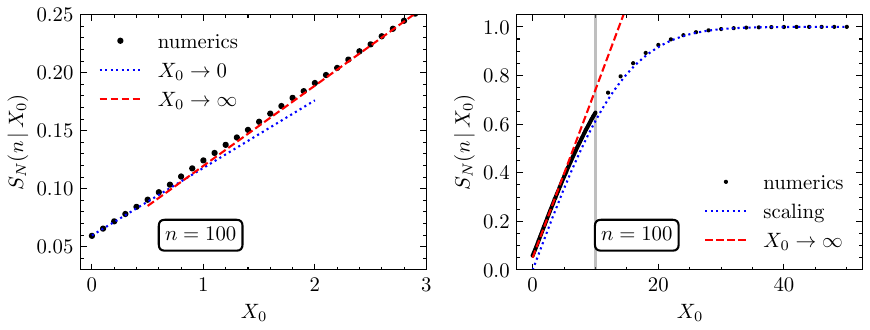}
\caption{
    Comparison of analytical predictions (dashed and dotted lines) with direct Monte Carlo simulations (black circles, $10^6$ trajectory realizations for each value of $X_0$) for $S_N(n=100\,\vert\,X_0)$ at the critical point $\alpha=\alpha_c$. 
    The parameters of the distributions are given by \eqref{eq:numerical_parameters}.
    The left panel shows moderate initial positions $X_0 \in [0,3]$, where the data are described by large-$n$ behavior as in \eqref{eq:Sn-U(X)_1}. The blue dotted and red dashed lines  represents the small-$X_0$ and large-$X_0$ expansions \eqref{eq:U(X0)-numerical} respectively.
    The right panel extends to larger $X_0$ values, displaying the crossover to the universal scaling regime. The blue dotted line shows the scaling function \eqref{eq:Sn=erf scaling}, and the red dashed line again shows the large-$X_0$ expansion. The gray vertical line at $X_0=\sqrt{n}$ indicates the crossover scale.
}\label{fig:Sn_critical}
\end{figure}

\par The results shown in Fig.~\ref{fig:Sn_critical} demonstrate agreement between analytical predictions and Monte Carlo simulations. 
The left panel focuses on moderate initial positions $X_0 \in [0,3]$ where $n=100$ is sufficiently large for the asymptotic behavior \eqref{eq:Sn-U(X)_1} to apply. 
For $X_0 \lesssim 1$, the small-$X_0$ expansion (blue dotted line) accurately captures the numerical data. 
For $X_0 \gtrsim 1$, the large-$X_0$ expansion  (red dashed line) provides a description instead. 
The right panel extends to larger initial positions. In this regime, the asymptotic limit $n\to\infty$ with fixed $X_0$ no longer applies. Instead, the data approach the universal scaling limit \eqref{eq:Sn=erf scaling} (blue dotted line), which depends only on the ratio $X_0/\sqrt{n}$. The crossover between the large-$X_0$ asymptotic and the scaling regime occurs near $X_0 \approx \sqrt{n}$ (gray vertical line), as predicted.

}

\section{Survival probability \texorpdfstring{$S_T(\tau\,\vert\,X_0)$}{ST(τ|X0)}}
\label{sec:Survival probability StX0}

Let us now study the properties of $S_T(\tau\,\vert\,X_0)$, the probability that the process remains positive up to time $\tau$. Specifically, we focus on the analytic properties of the Laplace transform of the survival probability 
\begin{equation}\label{eq:hatSt=def}
    \hat{S}_{T}(\rho\,\vert\,\lambda)
    \equiv
    \int_{0}^{\infty} \dd X_0\, e^{-\lambda X_0}
    \int_{0}^{\infty} \dd \tau\, e^{-\rho\tau} S_{T}(\tau\,\vert\,X_0)
\end{equation}
in the $\rho$-plane, from which we extract the large-$\tau$ behavior of $S_T(\tau\,\vert\,X_0)$ stated in Section~\ref{sec:main results}.

\par First, representing \eqref{eq:hatSt=def} in terms of $\hat{Q}(\rho,s\,\vert\,\lambda)$ as
\begin{equation}\label{eq:hatSt=1-Q}
    \hat{S}_{T}(\rho\,\vert\,\lambda) = 
    \frac{1}{\rho}\left.\left(\frac{1}{\lambda} - \hat{Q}\left(\rho,s\,\vert\,\lambda\right)\right)\right|_{s=1},
\end{equation}
and using the representation \eqref{eq:LTQ=generalIvanov} we obtain
\begin{equation}\label{eq:St(rho)=lim e explicit}
    \hat{S}_{T}(\rho\,\vert\,\lambda) = 
    \lim_{\epsilon\to0}
    \frac{1}{\rho}
    \left[
        \frac{1}{\lambda}
        - \frac{1}{\lambda+\frac{\rho}{\alpha}}
        + \frac{1 - c(\rho)}{\lambda+\frac{\rho}{\alpha}}\;
        \phi^-_\text{a.c} \left(\lambda+\frac{\rho}{\alpha};\rho,1\right)
        \phi^{+}_\text{a.c.} \left(\epsilon; \rho, 1\right)
    \right].
\end{equation}

\par The subsequent derivation follows along the lines of Section~\ref{sec:Survival probability SnX0}. 
That is, we examine the integrand \eqref{eq:St(rho)=lim e explicit}, move the contour of integration from the real line to $\mathcal{C}_*(\rho)$ defined by $\ii k=\zeta_*(\rho)$ and use the resulting expression for the analytic continuation. 
The main difference  from the continuation of $\hat{S}_N(s\,\vert\,\lambda)$ is that now the new contour  $\mathcal{C}_*(\rho)$ depends on $\rho$, hence, it moves when $\rho$ changes. 
This gives rise to more technically involved derivations, although the general strategy remains the same. For this reason, below we be concentrate mostly on the details that differ from $\hat{S}_N(s\,\vert\,\lambda)$.  

\par In addition, in the interest of simplicity, we perform the analytic continuation with respect to $\rho$ only along the real line, eventually finding the pole at $\rho=0$ (determining the survival probability $S_\infty(X_0)$) and a branch point at $\rho=\rho_*<0$ (determining the exponential decay of $S_T(\tau\,\vert\,X_0)$). Since $S_T(\tau\,\vert\,X_0)$ is essentially a cumulative function of the probability distribution, there are no singularities with $\Re(\rho)>\rho_*$ apart from $\rho=0$.

\par In the following, we separately analyze three cases: the \emph{survival regime} $\langle M-\alpha t\rangle>0$, the \emph{absorption regime} $\langle M-\alpha t\rangle<0$, and the \emph{critical point}  $\langle M-\alpha t\rangle=0$. 
{Finally, we compare the analytical results with the numerical simulations.}

\subsection{Survival regime}
When the drift is weak, $\langle M-\alpha t\rangle>0$, the process has a finite probability of never crossing the origin, so that $S_\infty(X_0)>0$.  By analyzing the analytic structure of $\hat{S}_T(\rho\,\vert\,\lambda)$ for $\rho<0$ we show that $S_T(\tau\,\vert\,X_0)$ approaches its asymptotic value exponentially fast with the rate given by \eqref{eq:xi_n xi_tau}.

\paragraph{Analytic continuation in the $\rho$-plane}
\par Let us first assume that $\Re\!\left(\lambda+\tfrac{\rho}{\alpha}\right)>0$. In this case, according to \eqref{eq:phi_ac=cases}, we have
\begin{equation}\label{eq:re(l+r/a)>0 phi_ac=}
    \Re\left(\lambda+\frac{\rho}{\alpha}\right)>0:\qquad 
    \phi_\text{a.c.}^{\pm}\left(\lambda+\frac{\rho}{\alpha};\rho,1\right) = \phi^{\pm}\left(\lambda+\frac{\rho}{\alpha};\rho,s\right),
\end{equation}
and hence from \eqref{eq:St(rho)=lim e explicit} we obtain
\begin{multline}\label{eq:StX=int Re}
    \hat{S}_T(\rho\,\vert\,\lambda) = 
    \lim_{\epsilon \to 0}
        \frac{1}{\rho}
        \left\{\vphantom{\frac{\strut}{\strut}} \right.
            \frac{1}{\lambda}
            - \frac{1}{\lambda+\frac{\rho}{\alpha}}
            + \frac{1-c(\rho)}{\lambda+\frac{\rho}{\alpha}}
        \\
        \times 
        \exp\left[
            -\frac{1}{2\pi}
                \int_{-\infty}^{\infty} \dd k\,
                \left(
                    \frac{1}{\epsilon + \ii k}
                    +
                    \frac{1}{\lambda+\frac{\rho}{\alpha}-\ii k}
                \right)
                \log\left[1 - c(\rho) F(k;\rho) \right]
        \right]
        \left.\vphantom{\frac{\strut}{\strut}} \right\}.
\end{multline}
The nontrivial analytic structure is entirely contained in the exponential factor. By this stage, the reader has likely developed enough intuition from the previous analysis and may anticipate the next steps. As before, we deform the contour of integration from the real axis to a horizontal contour $\mathcal{C}_*(\rho)$ with $\Im(k)=\zeta_*(\rho)$, which provides the analytic continuation required in the survival regime.

\par Now we need to determine whether the deformed contour $\mathcal{C}_*(\rho)$ lies above or below the real axis, or, equivalently, to establish the sign of $\zeta_*(\rho)$. 
As argued in Section~\ref{sec:properties_of_phi}, this is fixed by the sign of $\mu_1(\rho)$. Recall from  \eqref{eq:mu(rho)=def} that
\begin{equation}\label{eq:mu1=def 2}
    \mu_1(\rho) = \frac{1}{c(\rho)}
    \int_{0}^{\infty} \dd M\, q(M)\,
    e^{-\rho \frac{M}{\alpha}}\,
    \bigl(M - \alpha\langle t\rangle\bigr).
\end{equation}
If $\mu_1(0)>0$, then for $\rho<0$ the sign remains unchanged, so that $\zeta_*(\rho)>0$ and the contour $\mathcal{C}_*(\rho)$, although moving with $\rho$, remains in the upper half-plane.

\par
\begin{figure}[h!]\centering
    \includegraphics{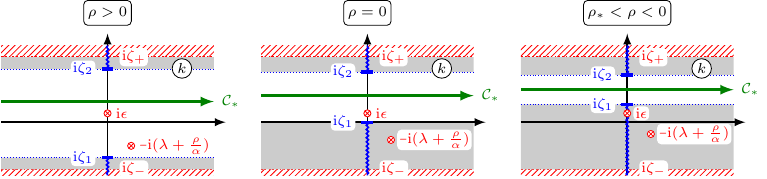}
    \caption{Schematic representation of the analytic structure of the integrand in \eqref{eq:StX=int Re} in the complex $k$-plane within the survival regime. The integrand features two branch points at $k = \ii \zeta_{1,2}(\rho)$ and two poles at $k = \ii \epsilon$ and $k = -\ii (\lambda+\frac{\rho}{\alpha})$. Upon continuation to $\rho < 0$, the branch points pinch the shifted contour $\mathcal{C}_*(\rho)$ inducing a singularity at $\rho=\rho_*$ defined by \eqref{eq:rho*=condition pinch} in the $\rho$-plane.}\label{fig:fig_T_survival_anal}
\end{figure}

 The analytic structure of the integrand in the $k$-plane is similar to that of \eqref{eq:S(s|l)=lim(int) l>0}; the logarithm is analytic within the strip $\Im(k)\in(\zeta_1,\zeta_2)$ and has branch cuts emanating from $k=\ii\zeta_{1,2}(\rho)$, where $\zeta_{1,2}(\rho)\equiv\zeta_{1,2}(\rho,1)$ are real solutions of \eqref{eq:zeta12=def} with $s=1$ (see Figs.~\ref{fig:F_imaginary_line} and \ref{fig:logF_strip}). The term in parentheses provides two simple poles at $k = \ii \epsilon$ and $k = -\ii (\lambda+\frac{\rho}{\alpha})$. The schematic representation of the analytic structure is given in Fig.~\ref{fig:fig_T_survival_anal}.

\par Therefore, when shifting the contour, the residue at $k=\ii\epsilon$ must be accounted for.  Specifically, the residue induces a factor $(1-c(\rho)F(\ii\epsilon;\rho))^{-1}$. Once the contour is shifted, taking the limit $\epsilon\to0$ is straightforward, yielding
\begin{multline}\label{eq:hatS(rho|l)=int C*}
    \hat{S}_T(\rho\,\vert\,\lambda) = 
        \frac{1}{\rho}
        \left\{\vphantom{\frac{\strut}{\strut}} \right.
            \frac{1}{\lambda}
            - \frac{1}{\lambda+\frac{\rho}{\alpha}}
            + \frac{1}{\lambda+\frac{\rho}{\alpha}}
        \\
        \times 
        \exp\left[
            -\frac{1}{2\pi}
                \int_{\mathcal{C}_*(\rho)} \dd k\,
                \left(
                    \frac{1}{\ii k}
                    +
                    \frac{1}{\lambda+\frac{\rho}{\alpha}-\ii k}
                \right)
                \log\left[1 - c(\rho) F(k;\rho) \right]
        \right]
        \left.\vphantom{\frac{\strut}{\strut}} \right\}.
\end{multline}

\par Note, however, that if $\rho$ is extended to the positive half of the real line, there exists a value $\tilde{\rho}$ such that, for $\rho>\tilde{\rho}$ we have $\mu_1(\rho)<0$. For this reason, the procedure of shifting the contour to $\mathcal{C}_*(\rho)$ should not be applied for $\rho>\tilde{\rho}$. In other words, the representation \eqref{eq:hatS(rho|l)=int C*} can be used as a definition only for $\rho\in(0,\tilde{\rho})$, whereas for $\rho>\tilde{\rho}$ one must instead rely on \eqref{eq:StX=int Re}. In the region $\rho\in(0,\tilde{\rho})$, these two representations coincide.

\par We now continue \eqref{eq:hatS(rho|l)=int C*} to $\rho<0$. To determine the rightmost singularity, and thereby the asymptotic behavior for large $\tau$, note that \eqref{eq:hatS(rho|l)=int C*} can be used as an analytic continuation up to the value $\rho_*<0$ at which $\zeta_1(\rho)$ and $\zeta_2(\rho)$ pinch the contour of integration, i.e.,
\begin{equation}\label{eq:rho*=condition pinch}
    \rho_*:\quad 
    \zeta_1(\rho_*;1) = \zeta_2(\rho_*;1) = \zeta_*(\rho_*).
\end{equation}
This is essentially the same condition as \eqref{eq:s*=condition pinch}. The coincidence of the branch points implies that at $\rho=\rho_*$, equation \eqref{eq:zeta12=def} has only one real solution, which gives an implicit expression for $\rho_*$:
\begin{equation}\label{eq:rho*=condition}
    \rho_*:\quad 
    \frac{1}{c(\rho_*)} = \min_{k_I} F\left(\ii k_I;\rho_*\right).
\end{equation}

\par The case $\Re(\lambda+\frac{\rho}{\alpha})<0$ is treated similarly. The differences are: (i) the analytic continuation in \eqref{eq:re(l+r/a)>0 phi_ac=} includes a prefactor; (ii) the contour shift from the real line to $\mathcal{C}_*(\rho)$ picks up a residue from the pole at $k = -\ii(\lambda+\frac{\rho}{\alpha})$. These two contributions cancel, so that the analytic structure in the $\rho$-plane remains unchanged. 

\par Finally, at some point $\rho_0$ defined by $\Re(\lambda + \frac{\rho_0}{\alpha}) = -\zeta_*(\rho_0)$, the pole at $k = -\ii (\lambda+\frac{\rho}{\alpha})$ traverses the contour $\mathcal{C}_*(\rho)$ giving rise to an additional prefactor in front of the integral. 
As a result, for  $\rho_* < \rho < \tilde{\rho}$ the analytic continuation of $\hat{S}_T(\rho\,\vert\,\lambda)$ reads
\begin{equation}
    \hat{S}_{T}(\rho\,\vert\,\lambda) = 
    \left\{\begin{aligned}
    & \frac{1}{\rho}\left(
            \frac{1}{\lambda} 
            -
            \frac{1-\mathcal{G}_T(\rho\,\vert\,\lambda)}{\lambda+\frac{\rho}{\alpha}}
        \right),
        \quad
            \lambda+\frac{\rho}{\alpha} > - \zeta_*(\rho),\\
    &\frac{1}{\rho}\left(
            \frac{1}{\lambda} 
            -
            \frac{1}{\lambda+\frac{\rho}{\alpha}}
            \frac{1-\mathcal{G}_T(\rho\,\vert\,\lambda)}{1-c(\rho)F\left(-\ii\left(\lambda+\frac{\rho}{\alpha}\right);\rho\right)}
        \right),
        \quad
            \lambda+\frac{\rho}{\alpha} < - \zeta_*(\rho),\\
    \end{aligned}\right.
\end{equation}
with 
\begin{equation}\label{eq:GT(rho|l)=}
    \mathcal{G}_T(\rho\,\vert\,\lambda) = 
    \exp\left[
            -\frac{1}{2\pi}
                \int_{\mathcal{C}_*(\rho)} \dd k\,
                \left(
                    \frac{1}{\ii k}
                    +
                    \frac{1}{\lambda+\frac{\rho}{\alpha}-\ii k}
                \right)
                \log\left[1 - c(\rho) F(k;\rho) \right]
        \right].
\end{equation}
Recall the analogous expressions \eqref{eq:S(s|l)=ac} and \eqref{eq:GN(s|l)=} for $\hat{S}_{N}(s\,\vert\,\lambda)$.

\paragraph{Asymptotic behavior of $S_T(\tau\,\vert\,X_0)$}
Formally inverting the Laplace transform with respect to $\lambda$, we obtain
\begin{equation}
    \int_{0}^{\infty} \dd \tau e^{-\rho\tau} S_{T}(\tau\,\vert\,X_0)
    = 
    \frac{1}{2\pi\ii} \int_{\gamma-\ii\infty}^{\gamma+\ii\infty} \dd \lambda
    \, e^{\lambda X_0} \hat{S}_{T}(\rho\,\vert\,\lambda).
\end{equation}
Choosing $\gamma$ to be sufficiently large, say $\gamma > -\zeta_*(\rho_*) -\frac{\rho_*}{\alpha}$, we see that the singularities of $\hat{S}_{T}(\rho\,\vert\,\lambda)$ in the $\rho$-plane are a pole at $\rho=0$ and the branch point at $\rho=\rho_*$. 

\par The pole at $\rho=0$ corresponds to the survival probability $S_\infty(X_0)$. Computing the residue at $\rho=0$ yields
\begin{equation}
\Re(\lambda)>0:\quad 
    \hat{S}_{\infty}(\lambda) = 
            \frac{1}{\lambda}
            \exp\left[
            -\frac{1}{2\pi}
                \int_{\mathcal{C}_*(0)} \! \dd k
                \left(
                    \frac{1}{\ii k}
                    +
                    \frac{1}{\lambda-\ii k}
                \right)
                \log\left[1 - F(k;0) \right]
        \right].
\end{equation}
This coincides with \eqref{eq:hatSinfty(lambda)=cases}, which originates from the pole at $s=1$ of $\hat{S}_N(s\,\vert\,\lambda)$. We have already analyzed this expression in detail; here it is treated merely as a consistency check, and we need not analyze it further.

\par The branch point at $\rho=\rho_*$ determines the leading order correction, and implies that the survival probability decays exponentially as
\begin{equation}\label{eq:S(T)-Sin=decay}
    \alpha<\alpha_c:\quad 
    S_{T}(\tau\,\vert\,X_0) - S_{\infty}(X_0) 
    \underset{\tau\to\infty}{\asymp}
    \exp\left[ -\frac{\tau}{\xi_\tau(\alpha)} \right],
\end{equation}
where according to \eqref{eq:rho*=condition}, the value of $\rho_*$ is determined by solving
\begin{equation}\label{eq:xiT(alpha)=}
    \xi_\tau(\alpha) = - \frac{1}{\rho_*},\qquad
    \frac{1}{c(\rho_*)} = \min_{k_I} F\left(\ii k_I;\rho_*\right).
\end{equation}
This is the result stated in \eqref{eq:xiT xin=divergence_res} for the \textit{survival regime}.

\subsection{Absorption regime}
If the drift is strong and $\langle M-\alpha t\rangle<0$, then the process is expected to eventually cross the origin.

\par At the level of analytic structure in the $\rho$-plane, the difference from the survival regime is that $\mu_1(0) < 0$, and hence $\mathcal{C}_*(0)$ lies in the lower part of the $k$-plane. Consequently, when shifting the contour, there is no contribution from the pole at $k=\ii\epsilon$, and thus
\begin{multline}\label{eq:hatS(rho|l)=int C* absorption}
    \hat{S}_T(\rho\,\vert\,\lambda) = 
        \frac{1}{\rho}
        \left\{\vphantom{\frac{\strut}{\strut}} \right.
            \frac{1}{\lambda}
            - \frac{1}{\lambda+\frac{\rho}{\alpha}}
            + \frac{1-c(\rho)}{\lambda+\frac{\rho}{\alpha}}
        \\
        \times 
        \exp\left[
            -\frac{1}{2\pi}
                \int_{\mathcal{C}_*(\rho)} \dd k\,
                \left(
                    \frac{1}{\ii k}
                    +
                    \frac{1}{\lambda+\frac{\rho}{\alpha}-\ii k}
                \right)
                \log\left[1 - c(\rho) F(k;\rho) \right]
        \right]
        \left.\vphantom{\frac{\strut}{\strut}} \right\}.
\end{multline}
Notably, the expression above has no pole at $\rho=0$, since 
$c(0)=1$.

\par To determine the asymptotic behavior of the survival probability, we analyze the pinching of the contour $\mathcal{C}_*(\rho)$ by branch points at $k = \zeta_{1,2}(\rho)$ in the $k$-plane. From~\eqref{eq:mu1=def 2} it follows that at some $\rho_c < 0$, the function $\mu_1(\rho)$ changes sign, causing the contour $\mathcal{C}_*(\rho)$ to shift from the lower to the upper half of the $k$-plane. This could, in principle, alter the asymptotic behavior by changing the dominant singularities, but, in fact, it does not. 

\par If $\rho = \rho_c$, then $\mu_1(\rho_c) = 0$, and hence the function $F(\ii k; \rho_c)$ reaches its minimum at $k = 0$, with $F(0; \rho_c) = 1$. Since $\rho_c$ is negative, we have $c(\rho_c) > 1$, and thus the equation \eqref{eq:zeta12=def} has no real solutions. 
This confirms that when continuing $\rho$ to the negative half-line, the contour pinching occurs before any potential complications caused by the movement of the shifted contour $\mathcal{C}_*(\rho)$. In other words, $\rho_c<\rho_*$, ensuring that the decay rate of the survival probability is indeed determined by $\rho_*$.

\par Repeating now the same arguments as in the survival regime we find that 
\begin{equation}\label{eq:St-asymp-exp}
\alpha>\alpha_c:
    \qquad
    S_{T}(\tau\,\vert\,X_0)
    \underset{\tau\to\infty}{\asymp}
    \exp\left[ -\frac{\tau}{\xi_\tau(\alpha)} \right],
\end{equation}
with the rate given by \eqref{eq:xiT(alpha)=}. This is the result stated in \eqref{eq:xiT xin=divergence_res} for the \textit{absorption regime}.

\subsection{Critical point}
We conclude the analysis of the survival probability  $S_T(\tau\,\vert\,X_0)$ by examining the critical point where $\langle M-\alpha t\rangle = 0$. The analysis closely follows the approach presented in Section~\ref{sub:critical_point_SN} for 
$S_N(n\,\vert\,X_0)$. Therefore, we outline only the main steps and present the key results.

\subsubsection{Divergence of the <<correlation length>>} 
We first demonstrate that as the drift strength $\alpha$ approaches the critical value $\alpha_c$, the correlation length $\xi_\tau(\alpha)$ defined in \eqref{eq:xiT(alpha)=} diverges.

Using the notation \eqref{eq:F(k;rho)=def_<>}, the equation \eqref{eq:xiT(alpha)=} for $\rho_*$ can be written as
\begin{equation}\label{eq:r* condition average critical}
    \left\langle \exp\left[ - \rho_* \frac{M}{\alpha} 
    - \zeta_*(\rho_*)(M-\alpha t)\right]\right\rangle = 1.
\end{equation}
At the critical point, we have $\mu_1(0) = 0$, which implies 
\begin{equation}\label{eq:rho=0, zeta=0}
 \rho_*\big|_{\alpha=\alpha_c}=0, \qquad   \zeta_*\big|_{\alpha=\alpha_c} = 0.
\end{equation}

\par Treating both $\rho_*$ and $\zeta_*$ as functions of $\alpha$ and expanding \eqref{eq:r* condition average critical} to first order as $\alpha\to\alpha_c$, we obtain
\begin{equation}
    1 = 
    \left\langle 1 -
    (\alpha-\alpha_c)
    \left. \left[
    (M-\alpha_c t) \dv{\zeta_*}{\alpha}
    - \frac{M}{\alpha_c} \dv{\rho_*}{\alpha}
    \right]\right|_{\alpha=\alpha_c}
    \right\rangle
    + O\left( (\alpha-\alpha_c)^2 \right).
\end{equation}
Taking the average and using $\langle M-\alpha_c t\rangle=0$, we find
\begin{equation}\label{eq:drho/da=0}
    \left. \dv{\rho_*}{\alpha} \right|_{\alpha=\alpha_c} = 0.
\end{equation}
Further expanding \eqref{eq:r* condition average critical} to the second order and examining the coefficient of $(\alpha-\alpha_c)^2$, we obtain the consistency condition
\begin{equation}
    \left\langle 
    \left. \dv{\zeta_*}{\alpha}
        \left( 2t + (M-\alpha_c t)^2 \dv{\zeta_*}{\alpha} \right)
    - \left(M-\alpha_c t\right)\dv{^2\zeta_*}{\alpha^2}
    - \dv{^2\rho_*}{\alpha^2} \frac{M}{\alpha_c}
    \right|_{\alpha=\alpha_c}
    \right\rangle 
    = 0,
\end{equation}
and hence
\begin{equation}\label{eq:d2rho/da2=}
\left. \dv{^2\rho_*}{\alpha^2} \right|_{\alpha=\alpha_c}= 
    \frac{\alpha_c}{\langle M\rangle}
    \left. \dv{\zeta_*}{\alpha}
        \left( 2\langle t \rangle  + \left\langle(M-\alpha_c t)^2\right\rangle \dv{\zeta_*}{\alpha} \right)
    \right|_{\alpha=\alpha_c}. 
\end{equation}

\par 
To complete the calculation, we need the derivative of $\zeta_*$ at the critical point. Since $\zeta_*$ is the location of the minimum of $F(\ii k_I;\rho_*)$, it satisfies 
\begin{equation}\label{eq:<M-at>rho=0}
    \left\langle 
        (M-\alpha t)
        \exp\left[ -\rho_* \frac{M}{\alpha} 
        - \zeta_*(M-\alpha t) \right]
    \right\rangle = 0.
\end{equation}
Expanding as $\alpha\to\alpha_c$ gives
\begin{equation}\label{eq:dzeta/da=}
    \left. \dv{\zeta_*}{\alpha}\right|_{\alpha=\alpha_c} = 
    -\frac{\left\langle t\right\rangle}
            {\left\langle (M-\alpha_c t)^2\right\rangle}.
\end{equation}
Note that due to \eqref{eq:rho=0, zeta=0} and \eqref{eq:drho/da=0}, in the first order, \eqref{eq:<M-at>rho=0} is equivalent to \eqref{eq:<M-at>=0}, and hence \eqref{eq:dzeta/da=} coincides with \eqref{eq:pd_alpha zeta*(0)=}. Substituting now \eqref{eq:dzeta/da=} into \eqref{eq:d2rho/da2=} yields
\begin{equation}
    \left. \dv{^2\rho_*}{\alpha^2} \right|_{\alpha=\alpha_c} = -
    \frac{\alpha_c}{\left\langle M\right\rangle}
    \frac{\langle t\rangle^2 }{\left\langle (M-\alpha_c t)^2\right\rangle}
\end{equation}
which, according to \eqref{eq:xiT(alpha)=}, translates into the divergence
\begin{equation}\label{eq:xiT=divergence}
    \xi_\tau(\alpha) \underset{\alpha\to\alpha_c}{\sim}
    \frac{\langle M\rangle}{\alpha_c}
    \frac{2}{\langle t\rangle^2}
    \frac{\langle (M-\alpha_c t)^2\rangle}{(\alpha-\alpha_c)^2}.
\end{equation}
Recalling that $\alpha_c=\frac{\langle M\rangle}{\langle t\rangle}$, we obtain the result stated in \eqref{eq:xiT xin=divergence_res}.

\subsubsection{Algebraic decay of $S_T(\tau\,\vert\,X_0)$} 
From \eqref{eq:xiT=divergence} it is clear that $\xi_\tau(\alpha_c)=\infty$ and hence there is no exponential decay of the survival probability $S_T(\tau\,\vert\,X_0)$; algebraic behavior is expected instead. 
At the level of the analytic structure of $\hat{S}_T(\rho\,\vert\,\lambda)$ in the $\rho$-plane, it corresponds to the coincidence of the branch point and the pole (recall that at the critical point $\rho_*=0$). The large $\tau$ behavior of $S_T(\tau\,\vert\,X_0)$ is therefore governed by the expansion of $\hat{S}_T(\rho\,\vert\,\lambda)$ in the vicinity of $\rho=0$.

\par Substituting the representation \eqref{eq:LTQ_P=general_regular_int} into \eqref{eq:hatSt=1-Q} and expanding the resulting expression for $\rho\to0$ we obtain
\begin{equation}\label{eq:ST(rho|l)=critical expansion}
    \Re(\lambda)>0:\quad 
    \hat{S}_T(\rho\,\vert\,\lambda)
    \underset{\rho\to0}{\sim}
    \frac{1}{\sqrt{\rho}} \frac{1}{\lambda^2}
    \sqrt{
    \frac{\langle t\rangle}{ \frac{1}{2}\left\langle (M-\alpha_c t)^2\right\rangle}}
    e^{I(\lambda;0,1)},
\end{equation}
where we used 
\begin{equation}
    c(\rho) = 1 - \frac{\left\langle M \right\rangle }{\alpha}\rho + O(\rho^2) 
    = 1 - \langle t\rangle \rho + O(\rho^2).
\end{equation}
Note that \eqref{eq:ST(rho|l)=critical expansion} is very similar to \eqref{eq:Sn(s|l)=critical expansion}, and the square root divergence implies that the large-$\tau$ behavior is
\begin{equation}\label{eq:St-U(X)}
    S_T(\tau\,\vert\,X_0) \underset{\tau\to\infty}{\sim}
    \sqrt{\frac{\langle t\rangle}{\pi \tau}}  U(X_0),
\end{equation}
where $U(X_0)$ is the same function as for the survival probability $S_N(n\,\vert\,X_0)$. Recall that it is defined through the Laplace transform as in \eqref{eq:U(X0)=LT}, and the asymptotic behaviors for $X_0\to\infty$ and $X_0\to0$ are given by \eqref{eq:Un(X0->inf)} and \eqref{eq:Un(X0->0)}, respectively.

\par As argued in Section~\ref{sub:critical_point_SN}, at the critical point, after a proper rescaling the process converges to a Brownian motion, and the survival probability $S_N(n\,\vert\,X_0)$ admits a scaling form defined by \eqref{eq:Sn=erf scaling}. Repeating essentially the same derivation, we obtain
\begin{equation}\label{eq:St=erf scaling}
    S_T(\tau\,\vert\,X_0) \sim 
    \erf\left(
        \frac{X_0}{\sqrt{2\frac{\tau}{\langle t\rangle} \langle (M-\alpha_ct)^2\rangle}}
    \right),\qquad 
    \tau\to\infty, \quad 
    \frac{X_0}{\sqrt{\tau}}\text{~--- fixed}.
\end{equation}
With this we conclude the analysis of the survival probabilities.

\begin{figure}[h]
    \includegraphics[width=\linewidth]{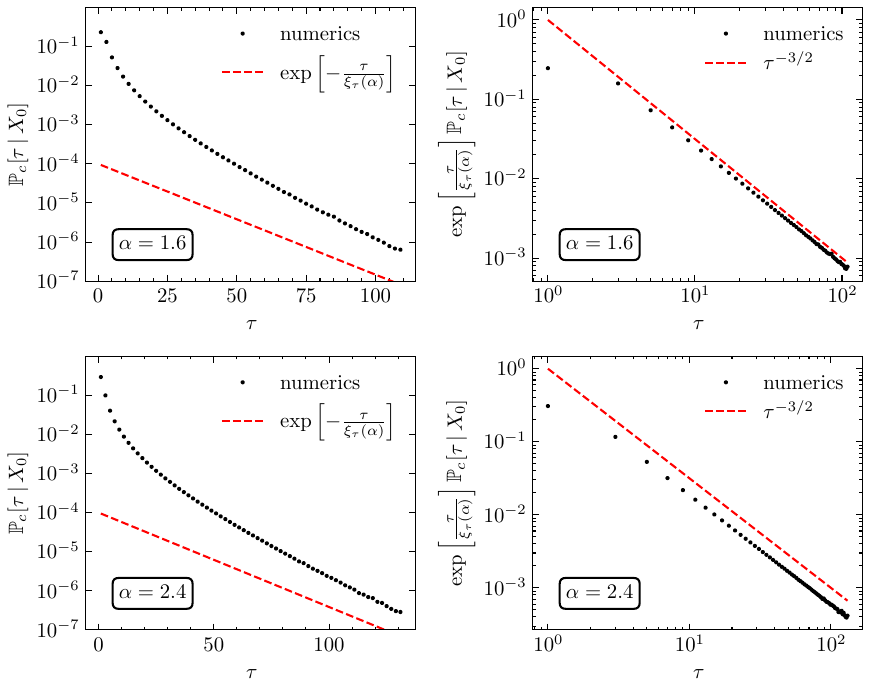}
    \caption{
    Comparison of analytical predictions (red dashed lines) with direct Monte Carlo simulations (black circles, $10^9$ trajectory realizations for $X_0=0.5$) in the survival regime (top row, $\alpha=1.6$) and absorption regime (bottom row, $\alpha=2.4$). 
    Parameters of the distribution are given by \eqref{eq:numerical_parameters}. 
    Left column: marginal first-passage probability $\mathbb{P}_c[\tau\,\vert\,X_0]$ versus $\tau$ (logarithmic scale), confirming the exponential decay \eqref{eq:Pt-asymp-refined} with rate \eqref{eq:xi_t=InverseGaussian}.
    The theoretical curves are shifted vertically by a factor of $10^{-4}$ for visual comparison.
    Right column: rescaled probability $e^{\tau/\xi_\tau(\alpha)}\mathbb{P}_c[\tau\,\vert\,X_0]$ versus $\tau$ (log-log scale), confirming the $\tau^{-3/2}$ power law in \eqref{eq:Pt-asymp-refined}.
    }
    \label{fig:St_non_crit}
\end{figure}

{
\subsection{Numerical verification}
To verify the asymptotic behavior of the survival probability $S_T(\tau\,\vert\,X_0)$, we use the same simulations as in Section~\ref{sec:SN-numerical} for $S_N(n\,\vert\,X_0)$. 
Both distributions $p(t)$ and $q(M)$ are inverse-Gaussian as given by \eqref{eq:p(t)=invGauss} and \eqref{eq:q(M)=invGauss} with parameters \eqref{eq:numerical_parameters}. The only difference is that now we concentrate on the marginal first-passage time probability distribution $\mathbb{P}_c[\tau\,\vert\,X_0]$ conditioned on the trajectories in which the first passage occurs:
\begin{equation}
    \mathbb{P}_c[\tau\,\vert\,X_0] = 
    -\frac{1}{1-S_\infty(X_0)} \pdv{}{\tau}\Big[ S_T(\tau\,\vert\,X_0) \Big].
\end{equation}
Repeating the arguments of Section~\ref{sec:SN-numerical} we expect the right tail of the distribution to be given by
\begin{equation}\label{eq:Pt-asymp-refined}
    \alpha\ne\alpha_c:\qquad \mathbb{P}_{c}\left[\tau\,\vert\,X_0\right]
    \underset{\tau\to\infty}{\sim}
    \frac{1}{\tau^{3/2}} \exp\left[ - \frac{\tau}{\xi_\tau(\alpha)} \right]
\end{equation}
Substituting \eqref{eq:F(k;rho)=invGauss} and \eqref{eq:c(rho)=invGauss} into \eqref{eq:xiT(alpha)=}, a straightforward computation shows that
\begin{equation}\label{eq:xi_t=InverseGaussian}
    \frac{1}{\xi_\tau(\alpha)} = \frac{1}{2}
    \frac{\ell_M \ell_t}{\ell_M+\alpha\ell_t}
    \left(
        \frac{1}{m_t} - \frac{\alpha}{m_M}
    \right)^2.
\end{equation}
Figure~\ref{fig:St_non_crit} demonstrates that the behavior \eqref{eq:Pt-asymp-refined} is correct.  The results of the simulations at the critical point $\alpha=\alpha_c$ are shown in Fig.~\ref{fig:St_critical}.

\begin{figure}[h]
\includegraphics[width=\linewidth]{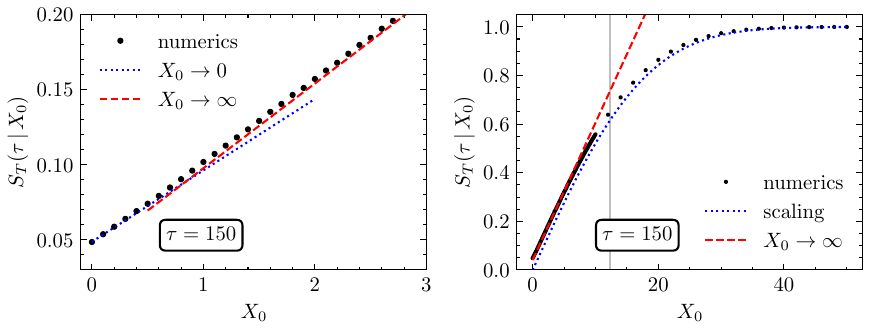}
\caption{
    Comparison of analytical predictions (dashed and dotted lines) with direct Monte Carlo simulations (black circles, $10^6$ trajectory realizations for each value of $X_0$) for $S_T(\tau=150\,\vert\,X_0)$ at the critical point $\alpha=\alpha_c$. 
    The parameters of the distributions are given by \eqref{eq:numerical_parameters}.
    The left panel shows moderate initial positions $X_0 \in [0,3]$, where the data are described by large-$\tau$ behavior as in \eqref{eq:St-U(X)}. The blue dotted and red dashed lines  represents the small-$X_0$ and large-$X_0$ expansions \eqref{eq:U(X0)-numerical}, respectively.
    The right panel extends to larger $X_0$ values, displaying the crossover to the universal scaling regime. The blue dotted line shows the scaling function \eqref{eq:St=erf scaling}, and the red dashed line again shows the large-$X_0$ expansion. The gray vertical line at $X_0=\sqrt{\tau}$ indicates the crossover scale.
}\label{fig:St_critical}
\end{figure}

}

\section{Moments of \texorpdfstring{$\tau$ and $n$}{τ and n} }
\label{sec:moments}
Having established and thoroughly quantified the large-$\tau$ and  large-$n$ behavior of the survival probabilities using the representation \eqref{eq:LTQ=generalIvanov}, the next natural step is to consider other quantities of interest. In the absorption regime, an important question concerns the moments of the first-passage time $\tau$ and the number of jumps $n$ before the first passage. In this section, we address this question and compute the first two leading terms in the $X_0\to\infty$ and $X_0\to0$ expansions for the means and the variances of $\tau$ and $n$.

\par The Laplace transform of the moments can be obtained by differentiating the function $\hat{Q}(\rho,s\,\vert\,\lambda)$. For example, the mean values are
\begin{align}
    & \int_{0}^{\infty} \dd X_0\, e^{-\lambda X_0}
        \mathbb{E}\left[\tau\,\vert\,X_0\right] 
    = - \left.\pdv{}{\rho}\hat{Q}(\rho,s\,\vert\,\lambda)\right|_{\rho=0,s=1},\\
   & \int_{0}^{\infty} \dd X_0\, e^{-\lambda X_0}
        \mathbb{E}\left[n\,\vert\,X_0\right] 
    = \left.\pdv{}{s}\hat{Q}(\rho,s\,\vert\,\lambda)\right|_{\rho=0,s=1},
\end{align}
or, using the representation \eqref{eq:LTQ_P=general_regular_int},
\begin{align}
\label{eq:E[n|X]=LT}
    & \int_{0}^{\infty} \dd X_0\, e^{-\lambda X_0} \mathbb{E}\left[n\,\vert\,X_0\right] = 
    -\frac{1}{\lambda^2} \frac{1}{\langle M-\alpha t\rangle} e^{I(\lambda;0,1)},
    \\
\label{eq:E[t|X]=LT}    
    & \int_{0}^{\infty}\dd X_0\, e^{-\lambda X_0} \mathbb{E}\left[\tau\,\vert\,X_0\right] = \frac{1}{\alpha \lambda^2} - \frac{1}{\alpha\lambda^2} 
    \frac{\langle M\rangle}{\langle M-\alpha t\rangle} e^{I(\lambda;0,1)},
\end{align}
where $I(\lambda;\rho,s)$ is given by
\begin{equation}\label{eq:I(l,rho,s) def 2}
    I(\lambda;\rho,s) = 
    -\frac{1}{2\pi} \int_{-\infty}^{\infty}
    \dd k\,
    \left(
        \frac{1}{\ii k} + \frac{1}{\lambda+\frac{\rho}{\alpha}-\ii k}
    \right)
    \log
        \frac{1 - s \, c(\rho) \, F(k;\rho)}
             {1 - s \, c(\rho) \, F_2(k;\rho)} 
    .
\end{equation}
The representations \eqref{eq:E[n|X]=LT} and  \eqref{eq:E[t|X]=LT}, however, still involve rather complicated integrals, and their practical utility might not be immediately apparent. This is especially true since we are dealing with a very general class of distributions, and it is unrealistic to expect that the Laplace transforms can be inverted analytically for arbitrary $p(t)$ and $q(M)$.

\par 
What we can do instead is to extract the asymptotic behavior of the mean values as $X_0\to\infty$ and $X_0\to 0$ from the expansions of the Laplace transforms as $\lambda\to 0$ and $\lambda\to\infty$. 
Typically such expansions can be constructed by expanding the integrand in $I(\lambda;0,1)$ term by term, and the rest becomes a technical exercise. In our case, however, this naive approach results in divergent integrals, necessitating a more careful treatment.

\par 
The resolution lies in recognizing that $I(\lambda;\rho,s)$ can be expressed as a Mellin convolution integral, for which powerful asymptotic techniques exist. In this section, we illustrate how such asymptotic expansions can be constructed via Mellin transform techniques, using the method presented in \cite{L-08}. We compute the first two terms in the expansion of both the mean and the variance for $\tau$ and $n$ as $X_0\to\infty$ and $X_0\to 0$ (16 terms in total). This demonstrates that the mapping to an effective random walk, together with the representation \eqref{eq:LTQ=generalIvanov}, provides a powerful tool.

\par 
Before proceeding to the technical details, let us make a useful observation. From \eqref{eq:E[n|X]=LT} and \eqref{eq:E[t|X]=LT}, it is clear that the means of $\tau$ and $n$ are related via
\begin{equation}\label{eq:mean(tau) mean(n)}
    \mathbb{E}\left[\tau\,\vert\,X_0\right] = \frac{X_0}{\alpha} 
    + \frac{\langle M\rangle}{\alpha} \mathbb{E}\left[n\,\vert\,X_0\right].
\end{equation}
This relation can readily be used to recover the asymptotic expansions for the mean value of $\tau$ from those of the mean value of $n$. In other words, instead of 16 terms, we need to compute only 12.

\paragraph{Asymptotic expansions of $I(\lambda;\rho,s)$}
Computing the $X_0\to\infty$ and $X_0\to 0$ asymptotic behaviors of the first two moments involves asymptotic expansions of $I(\lambda;\rho,s)$ (for means) and its first derivatives with respect to $\rho$ and $s$ (for second moments of $\tau$ and $n$) as $\lambda\to 0$ and $\lambda\to\infty$. As previously mentioned, simply expanding the integrand in series with respect to $\lambda$ results in divergent integrals. We leave verification of this fact to the reader and instead present an alternative approach based on Mellin transform techniques.

\par 
We first simplify the integral representation. Note that the imaginary part of the integrand in \eqref{eq:I(l,rho,s) def 2} is odd, whereas the real part is even. By parity, we can write
\begin{equation}\label{eq:I(l|rho,s)=int Re}
    I(\lambda;\rho,s) = 
    -\frac{1}{\pi} \int_{-\infty}^{\infty}
    \dd k\Re\left\{
    \left(
        \frac{1}{\ii k} + \frac{1}{\lambda+\frac{\rho}{\alpha}-\ii k}
    \right)
    \log
        \frac{1 - s \, c(\rho) \, F(k;\rho)}
             {1 - s \, c(\rho) \, F_2(k;\rho)} 
    \right\}
    .
\end{equation} 
Let us now focus on $I(\lambda;0,1)$. Setting $\rho=0$ and $s=1$ in \eqref{eq:I(l|rho,s)=int Re}, rescaling the integration variable $k\mapsto\lambda k$, and separating the real and imaginary parts explicitly, we obtain
\begin{multline}\label{eq:I(lambda;0,1)-sep-example}
    I(\lambda;0,1) = - \frac{1}{\pi}
    \int_{0}^{\infty} \dd k\;
    \frac{1}{k^2 + 1}
    \log \left| \frac{1 - F(\lambda k;0)}
                     {1 - F_2(\lambda k;0)} \right|
    \\ - \frac{1}{\pi}
    \int_{0}^{\infty} \dd k\;
    \frac{1}{k(k^2+1)} 
      \arg \frac{1 - F(\lambda k;0)}
              {1 - F_2(\lambda k;0)}.
\end{multline}
Each integral is now of the form
\begin{equation}\label{eq:I_Mellin}
    I(\lambda) = \int_{0}^{\infty} \dd k  \, f(k) \, h(\lambda k),
\end{equation}
which are Mellin convolution integrals. There exists a well-developed body of theory regarding the asymptotic analysis of such integrals. Here we employ the method from \cite{L-08} for its simplicity and directness.

\par The key result is as follows. Suppose we aim to find the $\lambda\to 0$ behavior of \eqref{eq:I_Mellin}, where the functions $f(k)$ and $h(k)$ have asymptotic expansions
\begin{equation}\label{eq:f,h=expansions}
    f(k) \underset{k\to\infty}{=}
    \sum_{\ell=0}^{n-1} \frac{a_\ell}{k^{\alpha_\ell}}
    + O\left(k^{-\alpha_n}\right),
    \qquad
    h(k) \underset{k\to0}{=} 
    \sum_{\ell=0}^{n-1}b_\ell k^{\beta_\ell}
    + O(k^{\beta_n})
\end{equation}
with $\alpha_\ell - \beta_j \ne 1$ for all pairs $(\ell,j)$. Then the asymptotic behavior of \eqref{eq:I_Mellin} reads
\begin{equation}\label{eq:fh answer_general}
    I(\lambda)
    =
    \sum_{\ell = 0}^{n-1} 
    a_\ell 
    \lambda^{\alpha_\ell-1}
    \mathcal{M}[h; 1-\alpha_\ell]
    +
    \sum_{j=0}^{m-1} b_j \lambda^{\beta_j} \mathcal{M}[f;\beta_j+1]
    + 
    O\left(\lambda^{\beta_m} + \lambda^{\alpha_n-1}\right),
\end{equation}
where $\mathcal{M}[g;z]$ is the Mellin transform defined as the integral
\begin{equation}\label{eq:M[g;z]=def}
    \mathcal{M}[g;z] = \int_0^{\infty} k^{z-1} g(k) \dd k,
\end{equation}
or, when integral diverges, its analytic continuation.

\par 
If there exist pairs $(\ell, j)$ such that $\alpha_\ell - \beta_j = 1$, the asymptotic behavior in \eqref{eq:fh answer_general} must be modified. Although such situations do not arise in our context, we present the necessary modification for completeness. Specifically, for each such pair $(\ell,j)$, the sum of terms
\begin{equation}\label{eq:fh canceling_terms}
    a_\ell \lambda^{\alpha_\ell-1} \mathcal{M}[h;1-\alpha_\ell]
    +
    b_j \lambda^{\beta_j} \mathcal{M}[f;\beta_j+1]
\end{equation}
is replaced by
\begin{equation}\label{eq:fh logarithmic_terms}
    \lambda^{\beta_j}
    \lim_{z\to0}\left\{\vphantom{\frac{}{}}
    a_\ell \mathcal{M}[h; z+1-\alpha_\ell] 
    + b_j \mathcal{M}[f; z+1+\beta_j]
    \right\}-  a_\ell b_j \; \lambda^{\beta_j} \log \lambda.
\end{equation}
Furthermore, if $\alpha_n-\beta_m=1$, then the error term $O\left(\lambda^{\beta_m} + \lambda^{\alpha_n-1}\right)$ becomes $O\left(\lambda^{\beta_m}\log\lambda\right)$. 

\par The beauty of \eqref{eq:fh answer_general} together with \eqref{eq:fh logarithmic_terms} is that it systematically handles the divergences that would  appear from naive expansion of \eqref{eq:I(l,rho,s) def 2}, converting them into well-defined Mellin transforms.

\par The final step is to apply \eqref{eq:fh answer_general} to $I(\lambda;0,1)$ and to its derivatives $\partial_s I(\lambda;0,1)$ and $\partial_\rho I(\lambda;0,1)$. The $\lambda\to 0$ expansions yield the $X_0\to\infty$ behavior, while the $\lambda\to \infty$ expansions (obtained by reversing the roles of $f$ and $h$) give the $X_0\to 0$ behavior. Although the calculations involve considerable effort, they reduce to systematic application of Taylor expansions and Mellin transform evaluations.

\paragraph{Asymptotic expansions for the mean and the variance}
\par Now that we have established a method to compute the asymptotic expansions of $I(\lambda;0,1)$ and its derivatives, the remaining task is computational. Rather than burden the reader with extensive algebraic details, we defer the computational part to Appendix~\ref{sec:app_expansions} and proceed directly to presenting the final results in the form of asymptotic expansions
\begin{gather}
\label{eq:E[n]-X0}
    \mathbb{E}[n\,\vert\, X_0] 
    \underset{X_0\to\infty}{\sim}
    X_0 A_{1}
    + A_0,
    \qquad
    \mathbb{E}[n\,\vert\, X_0] 
    \underset{X_0\to0}{\sim}
    a_0
    + a_1 X_0,
    \\
    \label{eq:E[t]-X0}
    \mathbb{E}[\tau\,\vert\, X_0] 
    \underset{X_0\to\infty}{\sim}
    X_0 \tilde{A}_{1}
    + \tilde{A}_0,
    \qquad
    \mathbb{E}[\tau\,\vert\, X_0] 
    \underset{X_0\to0}{\sim}
    \tilde{a}_0
    + \tilde{a}_1 X_0,
    \\
    \label{eq:Var[n]-X0}
    \mathrm{Var}[n\,\vert\,X_0] \underset{X_0\to\infty}{\sim}
    X_0 B_{1} + B_{0},
    \qquad
    \mathrm{Var}[n\,\vert\,X_0] \underset{X_0\to0}{\sim}
    b_0 + b_1 X_0,
    \\
    \label{eq:Var[t]-X0}
    \mathrm{Var}[\tau\,\vert\,X_0] \underset{X_0\to\infty}{\sim}
    X_0 C_{1} + C_{0},
    \qquad
    \mathrm{Var}[\tau\,\vert\,X_0] \underset{X_0\to0}{\sim}
    c_0 + c_1 X_0.
\end{gather}
Note that the asymptotics of $\mathbb{E}[\tau\,\vert\,X_0]$ are recovered from \eqref{eq:E[n]-X0} using the relation \eqref{eq:mean(tau) mean(n)}.

\par The coefficients in \eqref{eq:E[n]-X0}, \eqref{eq:Var[n]-X0}, and \eqref{eq:Var[t]-X0} are expressed in terms of the following auxiliary integrals:
{\allowdisplaybreaks
\begin{gather}
    J_1 =  \frac{1}{\pi}
        \int_{0}^{\infty} \dd k\; 
        \Im\left[ 
         \frac{1}{k} \log \frac{1- F(k;0)}{1-F_2(k;0)}
        \right],
    \\
    J_2 =  
    \frac{1}{\pi}
    \int_{0}^{\infty} \dd k \,
    \Re \left[ \frac{1}{k^2} \log 
        \frac{1- F(k;0)}{1-F_2(k;0)} 
        \right],\\
    J_3 =  \frac{1}{\pi}
        \int_{0}^{\infty} \dd k\; 
        \Im\left[ 
         \frac{1}{k^3} \log \frac{1- F(k;0)}{1-F_2(k;0)}
        \right],
    \\
    K_1 = \frac{1}{\pi}
    \int_{0}^{\infty} \dd k \,
    \frac{1}{k}\Im \left[
        \frac{1}{1-F(k;0)} - \frac{1}{1-F_2(k;0)}
    \right],
    \\
    K_2 = \frac{1}{\pi}
    \int_{0}^{\infty} \dd k \,
    \frac{1}{k^2}\Re \left[
        \frac{1}{1-F(k;0)} - \frac{1}{1-F_2(k;0)}
    \right],
    \\
    D_1 =
    \frac{1}{\pi}
    \int_{0}^{\infty} \dd k \,
    \left. \frac{1}{k}\Im \left[
        \frac{\partial_\rho c(\rho) F(k;\rho)}{1-F(k;0)} 
        - \frac{\partial_\rho c(\rho) F_2(k;\rho)}{1-F_2(k;0)}
    \right]\right|_{\rho=0},
    \\
    D_2 =
    \frac{1}{\pi}
    \int_{0}^{\infty} \dd k \,
    \left. \frac{1}{k^2}\Re \left[
        \frac{\partial_\rho c(\rho) F(k;\rho)}{1-F(k;0)} 
        - \frac{\partial_\rho c(\rho) F_2(k;\rho)}{1-F_2(k;0)}
    \right]\right|_{\rho=0},
    \\
    L_1 = \frac{1}{\pi}\int_{0}^{\infty}\dd k
    \Re\left[\strut \log (1 -F(k;0)) \right],
    \\
    L_2 = \frac{1}{\pi}\int_{0}^{\infty} \dd k
    \Re\left[ \frac{F(k;0)}{1-F(k;0)} \right],
    \\
    L_3 = \frac{1}{\pi}\int_{0}^{\infty} \dd k
    \Re\left[
        \frac{\partial_\rho c(\rho) F(k;\rho)}{1-F(k;0)} 
    \right].
\end{gather}
}
While these integrals appear complex, their key virtue is that they provide explicit expressions valid for arbitrary distributions $p(t)$ and $q(M)$. Given specific distributions, these integrals can be evaluated numerically with standard techniques.

\par
The advantage of the effective random walk approach becomes particularly evident when compared to alternative methods. Traditional renewal equation techniques, while elegant for special cases, face significant challenges when dealing with general distributions and higher-order moments. 
In contrast, the triple Laplace transform representation \eqref{eq:LTQ=generalIvanov} provides a systematic computational framework with no conceptual barriers. The method extends naturally to higher moments: computing asymptotic expansions for third or fourth moments would require evaluating second or third derivatives of $I(\lambda;\rho,s)$, leading to additional auxiliary integrals similar to those above. 
The computational complexity would indeed grow, but the calculation remains tractable through systematic application of Taylor expansions and Mellin transform techniques.

\par The explicit coefficient expressions presented below serve primarily as a concrete demonstration of this computational power. They illustrate that the effective random walk mapping, combined with our integral representation, transforms what would otherwise be barely tractable asymptotic problems into algebraically intensive, yet very systematic, calculations.

\par We now give the explicit coefficient expressions. For the mean number of jumps \eqref{eq:E[n]-X0}, the $X_0\to\infty$ coefficients are
\begin{equation}\label{eq:A1A0=}
    A_{1} = -\frac{1}{\langle M -\alpha t\rangle}
    ,\qquad
    A_{0} = \frac{J_2}{\langle M-\alpha t \rangle},
\end{equation}
while the $X_0\to 0$ coefficients are
\begin{equation}\label{eq:a1a0=}
    a_0 = -\frac{\sqrt{\frac{1}{2}\langle (M-\alpha t)^2 \rangle} }
    {\langle M-\alpha t\rangle} e^{-J_1}
    ,\qquad
    a_1 = - a_0 \; L_1.
\end{equation}
For the variance of $n$ the coefficients in the expansion \eqref{eq:Var[n]-X0} for $X_0\to\infty$ are
\begin{equation}
\begin{aligned}\label{eq:B1B0=}
    B_1 & = - 
    \frac{\langle M^2\rangle - \langle M\rangle^2 
        + \langle (\alpha t)^2\rangle - \langle \alpha t \rangle^2 }
         {\langle M - \alpha t\rangle^3},\\
    B_0 & = - B_1 J_2 - 
    \frac{\langle (M-\alpha t)^3\rangle}
         {3\langle M-\alpha t\rangle^3}
         -\frac{2J_3}{\langle M-\alpha t\rangle^2}
    - \frac{2K_2}{ \langle M-\alpha t \rangle},
\end{aligned}
\end{equation}
and, similarly, for $X_0\to0$ we have:
\begin{equation}\label{eq:b1b0=}
\begin{aligned}
    & b_0  = a_0\left[2 K_1 -a_0
    + \frac{\langle (M-\alpha t)^2\rangle}
             {\langle M-\alpha t\rangle^2} \right],
    \\
    & b_1  = L_1\left(a_0^2 - b_0\right) 
        - \frac{a_0}{\langle M-\alpha t\rangle}
        + 2 a_0 L_2.
\end{aligned}
\end{equation}
Finally, for the variance of $\tau$ given in \eqref{eq:Var[t]-X0}, the $X_0\to\infty$ coefficients are
\begin{equation}\label{eq:C1C0=}
\begin{aligned}
    C_1 & = 
    - \frac{
        \langle M\rangle^2 
        \left[ \langle t^2\rangle-\langle t\rangle^2 \right]
        -
        \langle t\rangle^2
        \left[ \langle M^2\rangle - \langle M\rangle^2 \right]}
        {\left\langle M-\alpha t\right\rangle^3},
    \\
    C_0 & = -C_1 J_2 
    - \frac{\langle M\rangle^2}{3 \alpha^2}
      \frac{\langle (M-\alpha t)^3\rangle}
           {\langle M-\alpha t\rangle^3}
    - \frac{\langle M\rangle^2}{\alpha^2}
      \frac{2J_3}{\langle M-\alpha t\rangle^2}
    + \frac{\langle M\rangle}{\alpha} 
      \frac{2 D_2}{\langle M-\alpha t\rangle},
\end{aligned}
\end{equation}
and the $X_0\to 0$ coefficients are
\begin{equation}\label{eq:c1c0=}
\begin{aligned}
    c_0 = & - \frac{\langle M\rangle^2 }{\alpha^2} a_0^2
    - 2\frac{\langle M\rangle}{\alpha} a_0 D_1
    + a_0\frac{\langle M\rangle \langle M(M-\alpha t)^2\rangle}
             {\alpha^2 \langle (M-\alpha t)^2\rangle}
    \\ &
    \qquad + a_0 
        \frac{
        \langle M\rangle^2 \left[\langle t^2\rangle-\langle t\rangle^2\right]
        +
        \langle t\rangle^2 \left[\langle M^2\rangle-\langle M\rangle^2\right]
        }
        {\langle M-\alpha t\rangle^2},
    \\
    c_1 = & L_1 \left(a_0^2 \frac{\langle M\rangle^2}{\alpha^2} -c_0\right)
    -\frac{\langle M\rangle^2}{\alpha^2}
    \frac{a_0}{\langle M-\alpha t\rangle}
    - 2 a_0 \, L_3 \frac{\langle M\rangle}{\alpha}.
\end{aligned}
\end{equation}

\paragraph{Numerical verification}
Having derived rather cumbersome expressions for the asymptotic coefficients, we must test their validity. The first check is to compare them  against the exactly solvable case where both $p(t)$ and $q(M)$ are exponentially distributed, for which the analytical expressions were obtained in \cite{BM-25}. A motivated reader can verify that the asymptotic expressions do coincide with these exact results.
\begin{figure}[h!]
    \includegraphics[width=\linewidth]{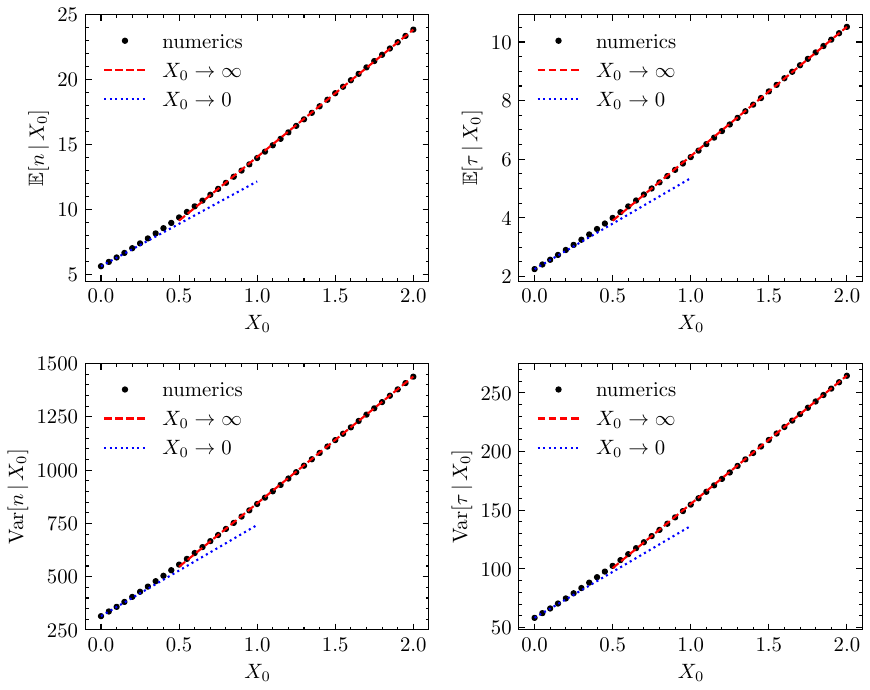}
    \caption{Comparison between direct Monte Carlo simulations (black circles, $10^8$ independent trajectory realizations for each value of $X_0$)  and the asymptotic expansions \eqref{eq:E[n]-X0},\eqref{eq:E[t]-X0}, \eqref{eq:Var[n]-X0}, and \eqref{eq:Var[t]-X0} for the means and the variances of $\tau$ and $n$. The simulations use distributions  \eqref{eq:p,q uni half Gaussian}: uniform inter-arrival times and half-Gaussian jump amplitudes. Parameters: $\alpha = 2$, $\gamma = 1$, $\ell = 0{.}9$. Red dashed lines show the large-$X_0$ expansions; blue dotted lines show the small-$X_0$ expansions.}
    \label{fig:meanVar_numerics}
\end{figure}

\par 
However, testing the asymptotics only against this special case would not be prudent. In the double exponential case, the mean and variance are linear functions of $X_0$, which means that the asymptotic behaviors for $X_0\to\infty$ and $X_0\to 0$ are essentially identical. More importantly, this might mislead the reader into believing that such linear behavior is generic. This suggests that we need to test the asymptotic expansions against a more general case of the distributions $p(t)$ and $q(M)$, making numerical simulations the natural choice.

\par 
{Unlike the previous sections, where we chose inverse-Gaussian distributions for their explicit decay rates, here the auxiliary integrals must be evaluated numerically in any case. We therefore take this opportunity to test against a different family of distributions, further demonstrating the generality of the approach.} Specifically, we perform Monte Carlo simulations for a process where the time intervals are drawn from a half-Gaussian distribution and the jumps follow a uniform distribution:
\begin{equation}\label{eq:p,q uni half Gaussian}
    p(t) = \begin{cases}
        \frac{1}{\ell}, &  0<t\le\ell, \\
        0, & t>\ell,
    \end{cases}    
    \qquad
    q(M) = \sqrt{\frac{2}{\pi}} \gamma\;  
           e^{-\frac{1}{2}\gamma^2 M^2},
\end{equation} 
For this choice we have
\begin{equation}
    \left\langle M - \alpha t \right\rangle
    = \frac{1}{\gamma} \sqrt{\frac{2}{\pi}} - \frac{\alpha\ell}{2},
    \qquad
    \alpha_c = \frac{2}{\gamma\ell} \sqrt{\frac{2}{\pi}}.
\end{equation}
The choice \eqref{eq:p,q uni half Gaussian} illustrates two key features: (i) the support of $p(t)$ is bounded rather than infinite; (ii) this combination appears to have no analytical solution~--- neither the auxiliary integrals nor the corresponding renewal equations have known closed forms. 

\par By generating trajectories for different initial positions, we computed the sample means and variances of $\tau$ and $n$ and compared them with our theoretical predictions. The comparison in Fig.~\ref{fig:meanVar_numerics} demonstrates perfect agreement and confirms the predicted asymptotic behavior.

\section{Conclusion}
\label{sec:conclusion}

\par In this paper, we applied the mapping of the jump process with drift to an effective random walk for the case where both inter-arrival times and jump amplitudes follow arbitrary light-tailed distributions with smooth densities. Through detailed analysis of the Laplace transform of the joint probability distribution of $\tau$ and $n$, we obtained exact asymptotic results for general distributions.

\par Our main result is confirming that the intuitive picture obtained in \cite{BM-25} for exactly solvable cases extends to the general setting. We identified three regimes: the \textit{survival regime} (weak drift) where the process has finite probability of remaining positive indefinitely, the \textit{absorption regime} (strong drift) where first-passage occurs with certainty, and the \textit{critical point} separating these regimes. In both survival and absorption regimes, survival probabilities approach their asymptotic values exponentially fast, with explicitly computable decay rates. At the critical point, exponential decay is replaced by algebraic decay with universal scaling behavior described by error functions.
\par Beyond the decay rates, we derived asymptotic expansions for the mean and variance of both $\tau$ and $n$ in the limits $X_0\to\infty$ and $X_0\to0$. All results are expressed in terms of the Fourier transform of the effective random walk, making them applicable to arbitrary distributions. This demonstrates the power of the random walk mapping approach for analyzing first-passage properties beyond exactly solvable cases.

\par A natural question concerns the case where our assumptions are violated. As mentioned earlier, the smooth density requirement is largely technical and can likely be relaxed without substantial modifications to the analysis. The case of heavy-tailed distributions, where the second moment of the effective random walk is infinite, on the other hand, presents a challenge. Allowing the distributions to be heavy-tailed would alter the behavior of $F(k;\rho)$ in ways that could invalidate our asymptotic analysis. The heavy-tailed case therefore merits detailed investigation. Another important assumption is that $t_i$ and $M_j$ are independent. It would be interesting to study whether the proposed formalism can be generalized to some extent to situations where correlations are present, as was done for the Sparre-Andersen theorem in \cite{ACEK-14}.

\par Another direction for future work is extending this analysis to multi-particle systems and studying extreme first-passage times \cite{WSL-83,L-20,GMO-20,HCC-24,HDCC-24,GMO-25}. One could also study the fraction of particles that remain positive up to a fixed time and examine how initial conditions affect this quantity. Such systems would likely exhibit memory effects similar to those observed in related models \cites{BM-07,DG-09,KM-12,BMRS-20,BJC-22,BMR-23,BMR-24,SD-24,DMS-23,JRR-23,JRR-23b,BHMMRS-23,HMS-24,MDS-24,SM-24}.

\subsection*{Acknowledgement}
The author is indebted to S.\,N. Majumdar for his guidance and valuable discussions throughout this work.

\subsection*{Data availability statement} No new data were created or analyzed in this study.

\addtocontents{toc}{\protect\setcounter{tocdepth}{1}}

\begin{appendix} 

\section{Asymptotic expansions for the means and variances}
\label{sec:app_expansions}
This appendix is devoted to the computation of the asymptotic expansions of the means and variances of $n$ and $\tau$. First we extract the asymptotic behavior of $I(\lambda;\rho,s)$ and its two first derivatives using the approach of \cite{L-08}. A key ingredient is the expression \eqref{eq:fh answer_general} that allows us to construct the asymptotic expansion for the Mellin convolution integrals \eqref{eq:I_Mellin} from the asymptotic behavior of the convoluted functions \eqref{eq:f,h=expansions}. These results are then used to construct the expansions of the mean and variances given by \eqref{eq:E[n]-X0}, \eqref{eq:Var[n]-X0}, and \eqref{eq:Var[t]-X0}.

\subsection{Expansion of \texorpdfstring{$I(\lambda;0,1)$}{I(λ; 0,1)}}
We begin with the expansions of $I(\lambda;0,1)$, which appears in the expressions for both the means and variances of $\tau$ and $n$. 

\paragraph{Expansion $\lambda\to0$.} 
We first construct the $\lambda\to0$ expansion of $I(\lambda;0,1)$. Recall that according to \eqref{eq:I(lambda;0,1)-sep-example} this integral is given by
\begin{equation}\label{eq:I(lambda;0,1)-sep}
    I(\lambda;0,1) = - \frac{1}{\pi}
    \int_{0}^{\infty} \dd k\,
    f_1(k) h_1(\lambda k)
    - \frac{1}{\pi}
    \int_{0}^{\infty} \dd k\,          
    f_2(k) h_2(\lambda k).
\end{equation}
We recognize that \eqref{eq:I(lambda;0,1)-sep} is decomposed into two separate Mellin convolution integrals of the form \eqref{eq:I_Mellin}, with the functions $f_{1,2}(k)$ are given by
\begin{equation}\label{eq:f1-f2=def}
    f_1(k) = \frac{1}{k^2+1},
    \qquad 
    f_2(k) = \frac{1}{k(k^2+1)},
\end{equation}
and for the functions $h_{1,2}(k)$ we have
\begin{equation}\label{eq:h1-h2=def}
    h_1(k) = \Re \left[  \log \frac{1 - F(k;0)}
                     {1 - F_2(k;0)} \right],
    \quad 
    h_2(k) = \Im \left[ \log \frac{1 - F(k;0)}
                         {1 - F_2(k;0)} \right].
\end{equation}
To apply \eqref{eq:fh answer_general} to the first term we need the asymptotic expansions of $f_1(k)$ as $k\to\infty$ and of $h_1(k)$ as $k\to0$. The former is straightforward and the latter follows from expansion \eqref{eq:F(k;rho) expansion k->0} of the Fourier transform $F(k;\rho)$ for $k\to0$, namely
\begin{equation}\label{eq:f1-h1-expansion}
    f_1(k) 
        \underset{k\to\infty}{=} \frac{1}{k^2} + O\left(\frac{1}{k^4}\right) ,
    \qquad
    h_1(k) 
    \underset{k\to0}{=} 
                - \frac{1}{6} 
                \frac{\mu_3(0)}{\mu_1(0)}\, k^2
                + O\left(k^3\right) .
\end{equation}
Hence in the notation \eqref{eq:f,h=expansions} we have
\begin{equation}\label{eq:a0b0alpha0beta0}
    a_0 = 1, \quad \alpha_0 = 2,  \quad 
    \alpha_1 = 4,\quad 
    b_0 = - \frac{1}{6} 
                \frac{\mu_3(0)}{\mu_1(0)},\quad 
    \beta_0 = 2,\quad 
    \beta_1 = 3.
\end{equation}
Then we apply the general formula \eqref{eq:fh answer_general}. Substituting \eqref{eq:a0b0alpha0beta0} into this general result yields
\begin{equation}\label{eq:f1h1-integral-expansion-Mellin}
    \int_{0}^{\infty} \dd k\, f_1(k) h_1(\lambda k)
    \underset{\lambda\to0}{=} 
    \lambda \; \mathcal{M}[h_1;-1]
    - \lambda^2 \, \frac{1}{6} 
        \frac{\mu_3(0)}{\mu_1(0)} 
        \mathcal{M}[f_1;-1]
        + O(\lambda^3).
\end{equation}
To proceed further, we evaluate the Mellin transform of $f_1(k)$, which can be computed using standard complex analysis techniques,
\begin{equation}\label{eq:M[f1]-Mellin}
    \mathcal{M}\left[f_1; z\right] = 
    \int_{0}^{\infty} \dd k\; k^{z-1}
    \frac{1}{k^2+1}
    =
    \frac{\pi}{2} \frac{1}{\sin \frac{\pi z}{2}},
    \qquad 
    \mathcal{M}[f_1; -1] = -\frac{\pi}{2}.
\end{equation}
Since $h_1$ explicitly depends on $F(k;0)$ and hence is determined by the exact form of the distributions $p(t)$ and $q(M)$ we keep its Mellin transform in the integral form. The contribution from the first term in \eqref{eq:I(lambda;0,1)-sep} then becomes
\begin{equation}\label{eq:f1h1-integral-expansion}
    \frac{1}{\pi}
    \int_{0}^{\infty} \dd k\,
    f_1(k) h_1(\lambda k)
    \underset{\lambda\to0}{\sim}
        \frac{\lambda}{\pi} 
        \int_{0}^{\infty} \dd k\;
            \frac{1}{k^2}
            \log \left| \frac{1 - F(k;0)}{1 - F_2(k;0)} \right|
         + \frac{\lambda^2}{12} \frac{\mu_3(0)}{\mu_1(0)}.
\end{equation}

\par
We now turn to the second term in \eqref{eq:I(lambda;0,1)-sep}, for which the convoluted functions are $f_2(k)$ and $h_2(k)$. The asymptotic analysis proceeds similarly, though the function $h_2$ has a higher-order vanishing at $k = 0$,
\begin{equation}\label{eq:f2-h2-expansion}
    f_2(k) \underset{k\to\infty}{=} \frac{1}{k^3}  + O\left(\frac{1}{k^5}\right) ,
    \qquad
    h_2(k)  \underset{k\to0}{=}  O\left( k^3 \right).
\end{equation}
In the notation \eqref{eq:f,h=expansions} we have
\begin{equation}
    a_0 = 1,\quad \alpha_0 = 3,\quad \alpha_1 = 5, \quad \beta_0 = 3,
\end{equation}
and hence the asymptotic behavior of the second term reads
\begin{equation}
    \int_{0}^{\infty} \dd k\,
         f_2(k) h_2(\lambda k) 
    \underset{\lambda\to0}{=} 
    \lambda^2\mathcal{M}[h_2; -2] + O\left(\lambda^3\right).
\end{equation}
Keeping the Mellin transform of $h_2$ in the integral form, we obtain the expansion for the second term, i.e.,
\begin{equation}\label{eq:f2h2-integral-expansion}
    \frac{1}{\pi}
    \int_{0}^{\infty} \dd k\,
         f_2(k) h_2(\lambda k)
    \underset{\lambda\to0}{\sim}
    \frac{\lambda^2}{\pi} 
        \int_{0}^{\infty}\dd k\;
        \frac{1}{k^3} \arg \frac{1 - F(k;0)}{1-F_2(k;0)} 
    .
\end{equation}
Combining \eqref{eq:f1h1-integral-expansion} with \eqref{eq:f2h2-integral-expansion} gives the expansion for $I(\lambda;0,1)$ 
\begin{equation}\label{eq:final-I-0-expansion}
    I(\lambda;0,1) \underset{\lambda\to0}{=}
    \mathcal{A}_0^{(1)} \lambda + \mathcal{A}_0^{(2)} \lambda^2  
    +O(\lambda^3).
\end{equation}
Recalling the definition \eqref{eq:mu(rho)=def} of the moments $\mu_\ell(\rho)$, the coefficients can be expressed as
\begin{align}
    \label{eq:A_0^1=int}
    \mathcal{A}_0^{(1)} & = -\frac{1}{\pi}
    \int_{0}^{\infty} \dd k\;
    \Re\left[ \frac{1}{k^2}\log \frac{1 - F(k;0)}{1-F_2(k;0)}\right],
    \\ 
    \label{eq:A_0^2=int}
    \mathcal{A}_0^{(2)} & =   
    -
    \frac{1}{12} 
    \frac{\left\langle (M-\alpha t)^3\right\rangle}
         {\left\langle M-\alpha t\right\rangle}
    - \frac{1}{\pi} 
    \int_{0}^{\infty} \dd k\; 
    \frac{1}{k^3}\Im\left[\log \frac{1 - F(k;0)}{1-F_2(k;0)}\right].
\end{align}
Having established the $\lambda\to0$ behavior, we now turn to the expansion of $I(\lambda;0,1)$ as $\lambda\to\infty$.

\paragraph{Expansion $\lambda\to\infty$}
In principle, to obtain the expansion of $I(\lambda;0,1)$ as $\lambda\to\infty$, we could rescale the integration variable $\lambda k\mapsto k$ and apply \eqref{eq:fh answer_general} with the roles of $f$ and $h$ reversed. However, this procedure would generate the logarithmic terms \eqref{eq:fh logarithmic_terms}. A more convenient strategy is to first extract the logarithmic behavior explicitly by rewriting \eqref{eq:I(lambda;0,1)-sep} as
\begin{multline}\label{eq:I(lambda;0,1)-sep-inf}
    I(\lambda;0,1) = 
    \frac{1}{2}
        \log \left[ \frac{\mu_2(0)}{2}\right]
    + \log\left[ \strut \lambda + k_+^{*}(0,1) \right]
    \\
    -\frac{1}{\pi}
    \int_{0}^{\infty} \dd k\, 
    \Im\left[
        \frac{1}{k}
        \log
            \frac{1 - F(k;0)}{1 - F_2(k;0)}
    \right]
    - \frac{1}{\pi}
    \int_{0}^{\infty} \Re\left[
        \frac{\log\left[ 1 - F(k;0)\right]}
             {\lambda-\ii k}
    \right],
\end{multline}
where $k_+^*(0,1)$ is given by \eqref{eq:k_pm=m1m2}.
To obtain \eqref{eq:I(lambda;0,1)-sep-inf} from \eqref{eq:I(lambda;0,1)-sep} we essentially went from the regularized representation \eqref{eq:phi^pm(lambda;rho,s)=regular} of $\phi_\text{a.c.}^-(\lambda;\rho,s)$ back to the original one \eqref{eq:phi_ac=cases}. 

\par 
The nontrivial  $\lambda$-dependence is now contained in the last term. Separating the real and imaginary parts explicitly we rewrite it as
\begin{multline}\label{eq:I(l;0,1)=ltoinfInt}
    \frac{1}{\pi}\int_{0}^{\infty}
    \dd k\;
    \Re\left[
        \frac{\log[1 - F(k;0)]}{\lambda-\ii k}
    \right]
    = \frac{1}{\pi\lambda}
    \int_{0}^{\infty}\dd k\;
    \frac{1}{1 + (k/\lambda)^2}\Re\Big[ \log \left[ 1 - F(k;0)\right]\Big]
\\    
-
    \frac{1}{\pi\lambda}
    \int_{0}^{\infty}\dd k\;
    \frac{(k/\lambda)}{1 + (k/\lambda)^2} \Im \Big[\log\left[ 1 - F(k;0)\right]\Big]
    .
\end{multline}
The asymptotic behavior of \eqref{eq:I(l;0,1)=ltoinfInt} can now be found from \eqref{eq:fh answer_general}. However, for our purposes it is enough to extract the leading term, which is already clear from \eqref{eq:I(l;0,1)=ltoinfInt}. Specifically, the leading order behavior is governed solely by the first term, and we have
\begin{equation}
    \frac{1}{\pi}\int_{0}^{\infty}
    \dd k\;
    \Re\left[
        \frac{\log[1 - F(k;0)]}{\lambda-\ii k}
    \right]
    \underset{\lambda\to\infty}{\sim}
    \frac{1}{\pi\lambda}
    \int_{0}^{\infty}\dd k\;
    \Re\Big[ \log \left[ 1 - F(k;0)\right]\Big].
\end{equation}
Hence the expansion of $I(\lambda;0,1)$ as $\lambda\to\infty$
\begin{equation}\label{eq:final-I-inf-expansion}
    I(\lambda;0,1) \underset{\lambda\to\infty}{=}
    \log\left[ \lambda + k_{+}^{*}(0,1) \right]
    + \mathcal{A}_{\infty}^{(0)}
    + \frac{1}{\lambda}\mathcal{A}_{\infty}^{(1)}
    +O\left(\frac{1}{\lambda^2}\right),
\end{equation} 
with
\begin{align}
    \label{eq:A_inf^0=int}
    \mathcal{A}_{\infty}^{(0)}
    & = \frac{1}{2}\log\left[
        \frac{\mu_2(0)}{2}
    \right]
    -\frac{1}{\pi}
        \int_{0}^{\infty} \dd k\;
        \Im\left[
        \frac{1}{k}
        \log \frac{1-F(k;0)}{1-F_2(k;0)}
        \right],
    \\
    \label{eq:A_inf^1=int}
    \mathcal{A}_{\infty}^{(1)}
    & =
    -\frac{1}{\pi}
        \int_{0}^{\infty}\dd k\;
        \Re\Big[ \log \left[ 1 - F(k;0)\right]\Big].
\end{align}

\par 
As a final remark, note that the first term in \eqref{eq:final-I-inf-expansion} actually differs depending on whether the system is in \textit{survival} or \textit{absorption} regime,
\begin{equation}
    \log\left[\lambda + k_{+}^{*}(0,1)\right]
    = 
    \begin{cases}
        \log \lambda, \quad & \langle M-\alpha t\rangle<0
        \\[10pt]
        \log \left[ \lambda + 2\frac{\langle M-\alpha t\rangle}
                                    {\langle (M-\alpha t)^2\rangle} \right],
             \quad & \langle M-\alpha t\rangle\ge0
    \end{cases}
\end{equation}

\subsection{Expansion of \texorpdfstring{$\partial_s I(\lambda;0,s)$}{I(λ; 0,s)}}
The next step in our asymptotic analysis is to build the expansions for the derivative $\partial_s I(\lambda;0,s)|_{s=1}$, which will appear in the representation for the variance of $n$.
\par From \eqref{eq:I(l|rho,s)=int Re} it immediately follows that
\begin{equation}\label{eq:pd_s_I-expansion}
    \partial_s I(\lambda;0,s)\big|_{s=1} = 
    -\frac{1}{\pi}
    \int_{0}^{\infty} \dd k
    \Re\left[
    \left( \frac{1}{\ii k} + \frac{1}{\lambda-\ii k} \right)
    \left( \frac{F_2(k;0)}{1-F_2(k,0)}
    -
    \frac{F(k;0)}{1-F(k;0)}\right)
    \right].
\end{equation}
Separating real and imaginary parts and rescaling the integration  variable yields decomposition of \eqref{eq:pd_s_I-expansion} into a sum of two Mellin convolution integrals:
\begin{equation}\label{eq:pd_s_I-separated}
    \partial_s I(\lambda;0,s)\Big|_{s=1} = 
    -
    \frac{1}{\pi}
    \int_{0}^{\infty} \dd k\;
    f_1(k) g_1(\lambda k)
    -
    \frac{1}{\pi}
    \int_{0}^{\infty} \dd k\;
    f_2(k) g_2(\lambda k),
\end{equation}
where $f_{1,2}(k)$ are given by \eqref{eq:f1-f2=def} and we introduced two new functions given by
\begin{equation}\label{eq:g1-g2=def}
\begin{aligned}
    g_1(k) & = \Re\left[
    \left( \frac{1}{1-F_2(k,0)}
    -
    \frac{1}{1-F(k;0)}\right)
    \right],
    \\
    g_2(k) & = \Im\left[
        \left( \frac{1}{1-F_2(k,0)}
        -
        \frac{1}{1-F(k;0)}\right)
        \right].
    \end{aligned}
\end{equation}
We now apply the general result \eqref{eq:fh answer_general} to each term in \eqref{eq:pd_s_I-separated} separately and obtain $\lambda\to0$ and $\lambda\to\infty$ expansions.

\paragraph{Expansion $\lambda\to0$} 
For the first term in \eqref{eq:pd_s_I-separated}, the required behaviors are
\begin{equation}\label{eq:pd_s f1-h1-expansion}
    f_1(k) \underset{k\to\infty}{=} \frac{1}{k^2} + O\left(\frac{1}{k^4}\right) ,
    \quad
    g_1(k)  \underset{k\to0}{=} O\left( k^2 \right).
\end{equation}
Hence in the notation \eqref{eq:f,h=expansions} we have
\begin{equation}
    a_0 =1,\quad \alpha_0=2, \quad \alpha_1 = 4,\quad  \beta_0 = 2,
\end{equation}
and therefore the asymptotic expansion reads
\begin{equation}
    \int_{0}^{\infty} \dd k\; f_1(k) g_1(\lambda k) 
    \underset{\lambda\to0}{=}
    \lambda \; \mathcal{M}[g_1; -1]
    + O\left(\lambda^2\right), 
\end{equation}
or in the explicit way
\begin{equation}\label{eq:pd_s f1g1-integral-expansion}
    \frac{1}{\pi}\int_{0}^{\infty} \dd k\; f_1(k) g_1(\lambda k) 
    \underset{\lambda\to0}{\sim} 
    \frac{\lambda}{\pi}
        \int_{0}^{\infty} \dd k\; 
        \frac{1}{k^2} 
        \Re\left[
        \frac{1}{1-F_2(k,0)}
        -
        \frac{1}{1-F(k;0)}
        \right].
\end{equation}

Similarly for the convoluted functions  $f_2(k)$ and $g_2(k)$  in the second term in \eqref{eq:pd_s_I-separated} the required expansions are
\begin{equation}\label{eq:pd_s f2-g2-expansion}
    f_2(k) 
    \underset{k\to\infty}{\sim}O\left(\frac{1}{k^3}\right) ,
    \quad
    g_2(k)
    \underset{k\to0}{=}
    - \frac{1}{6} \frac{\mu_3(0)}{\mu_1(0)^2} k + O(k^2),
\end{equation}
so that
\begin{equation}
    \alpha_0 = 3, \quad b_0 =  - \frac{1}{6} \frac{\mu_3(0)}{\mu_1(0)^2},\quad 
    \beta_0=1, \quad \beta_1 = 2,
\end{equation}
and hence the behavior
\begin{equation}\label{eq:pd_s f2g2-integral-expansion-Mellin}
    \int_{0}^{\infty} \dd k\; f_2(k) g_2(\lambda k) 
    \underset{\lambda\to0}{=} 
     - \frac{1}{6} \frac{\mu_3(0)}{\mu_1(0)^2} \lambda \;  \mathcal{M}[f_2;2]
     + O(\lambda^2).  
\end{equation}
The Mellin transform of $f_2(k)$  can be computed explicitly resulting in
\begin{equation}\label{eq:pd_s M[f2]-Mellin}
    \mathcal{M}[f_2;z]
    = \int_{0}^{\infty} \dd k\;
    k^{z-1}
    \frac{1}{k(k^2+1)} = -\frac{\pi}{2} \frac{1}{\cos \frac{\pi z}{2}}, \qquad
    \mathcal{M}[f_2;2] = \frac{\pi}{2}.
\end{equation}
Substituting  now \eqref{eq:pd_s M[f2]-Mellin} into \eqref{eq:pd_s f2g2-integral-expansion-Mellin}  we obtain the asymptotic behavior of the second term in \eqref{eq:pd_s_I-separated}
\begin{equation}\label{eq:pd_s f2g2-integral-expansion}
    - \frac{1}{\pi}
    \int_0^\infty \dd k\; 
        f_2(k) g_2(\lambda k) \underset{\lambda\to0}{\sim}
     \frac{1}{12} \frac{\mu_3(0)}{\mu_1(0)^2} \lambda
\end{equation}

Combining \eqref{eq:pd_s f2g2-integral-expansion} with \eqref{eq:pd_s f1g1-integral-expansion} we arrive at the asymptotic expansion of $\partial_s I(\lambda;0,s)|_{s=1}$ as $\lambda\to0$. Specifically, recalling the definition of the moments \eqref{eq:mu(rho)=def} we obtain
\begin{equation}\label{eq:pd_s final-I-expansion}
    \partial_s I(\lambda;0,s)\Big|_{s=1} 
    \underset{\lambda\to0}{=} 
    \mathcal{B}_{0}^{(1)} \lambda
    + O(\lambda^2) 
\end{equation}
where the coefficient is given by
\begin{equation}\label{eq:B_0^1=int}
    \mathcal{B}_0^{(1)} = 
    \frac{1}{12} 
        \frac{\langle (M-\alpha t)^3\rangle}
             {\langle M-\alpha t\rangle^2}
        -\frac{1}{\pi}
        \int_{0}^{\infty} \dd k\; 
        \frac{1}{k^2} 
        \Re\left[
        \frac{1}{1-F_2(k,0)}
        -
        \frac{1}{1-F(k;0)}
        \right].
\end{equation}

\paragraph{Expansion $\lambda\to\infty$} To find the expansion for $\lambda\to\infty$, we rescale the variable of integration $k\mapsto k/\lambda$ in \eqref{eq:pd_s_I-separated}, arriving at 
\begin{equation}\label{eq:pd_s_I-separated-rescaled}
    \partial_s I(\lambda;0,s)\Big|_{s=1} = 
    -\frac{1}{\pi \lambda} 
    \int_{0}^{\infty} \dd k
    \left(\strut 
        f_1(k/\lambda) g_1(k) + f_2(k/\lambda) g_2(k) \right).
\end{equation}
This is again a sum of two Mellin convolution integrals with roles of the $f$ and $g$ reversed. Now, we shall use the asymptotic behaviors of $f_1$ as $k\to0$ and $g_1$ as $k\to\infty$, i.e., 
\begin{equation}\label{eq:pd_s f1-g1-expansions-inf}
    g_1(k) \underset{k\to\infty}{=} - 1 +  O\left( \frac{1}{k^2} \right)
    ,\quad
    f_1(k) \underset{k\to0}{=} 1 +  O(k^2).
\end{equation}
Note that the next term in the expansion of $g_1(k)$ may contain oscillating components when the distributions $p(t)$ or $q(M)$ have bounded support, but this affects only the terms in the expansion of \eqref{eq:pd_s_I-separated-rescaled} of order $O(1/\lambda^2)$ and higher.

\par Expansions \eqref{eq:pd_s f1-g1-expansions-inf} when combined with the general result \eqref{eq:fh answer_general} imply that 
\begin{equation}
    a_0=-1,\quad \alpha_0=0, \quad \alpha_1 = 2,
    \quad 
    b_0 =1, \quad \beta_0 = 0, \quad \beta_1 = 2,
\end{equation}
and hence
\begin{equation}\label{eq:pd_s f1g1-integral-expansion-inf1}
    \int_{0}^{\infty} \dd k\;
    f_1(k/\lambda) g_1(k) \underset{\lambda\to\infty}{=} 
    -\lambda \;  \mathcal{M}[f_1; 1]
    +\mathcal{M}[g_1;1] + O\left(\frac{1}{\lambda}\right).
\end{equation}
Now we need to compute the Mellin transforms. The Mellin transform of $f_1$ can be computed explicitly as in \eqref{eq:M[f1]-Mellin}, i.e.,
\begin{equation}\label{eq:pd_s M[f1]-Mellin}
    \mathcal{M}\left[f_1; z\right] = 
    \int_{0}^{\infty} \dd k\; k^{z-1}
    \frac{1}{k^2+1}
    =
    \frac{\pi}{2} \frac{1}{\sin \frac{\pi z}{2}},
    \qquad 
    \mathcal{M}[f_1; 1] = \frac{\pi}{2}.
\end{equation}
The transform of $g_1$ is however more subtle, as the integral \eqref{eq:M[g;z]=def} diverges, and hence the Mellin transform should be understood in the analytical continuation sense. The procedure is explained in \cite{L-08}; we state only the result here
\begin{equation}\label{eq:pd_s M[g1]-Mellin inf}
    \mathcal{M}[g_1;1] =
    \int_{0}^{\infty}\dd k\;
    \Re\left[
    1+
    \left( \frac{1}{1-F_2(k,0)}
    -
    \frac{1}{1-F(k;0)}\right)
    \right].
\end{equation}
Substituting \eqref{eq:pd_s M[g1]-Mellin inf} and \eqref{eq:pd_s M[f1]-Mellin} into \eqref{eq:pd_s f1g1-integral-expansion-inf1} after simplification yields
\begin{equation}\label{eq:pd_s f1g1-integral-expansion-inf}
    \int_{0}^{\infty} \dd k\;
    f_1(k/\lambda) g_1(k)
    \underset{\lambda\to\infty}{\sim}
    -\frac{\pi \lambda}{2} 
    +
    \frac{\pi}{2\abs{\mu_1(0)}}
    -
    \int_{0}^{\infty}\dd k\;
    \Re\left[
    \frac{F(k;0)}{1-F(k;0)}\right].
\end{equation}

\par Similarly, for the second integral in \eqref{eq:pd_s_I-separated-rescaled} we shall utilize the expansions of $g_2(k)$ as $k\to\infty$ and $f_2(k)$ as $k\to0$ that are easily extracted from \eqref{eq:f1-f2=def} and \eqref{eq:g1-g2=def},
\begin{equation}\label{eq:pd_s f2-h2-expansions-inf}
    g_2(k) \underset{k\to\infty}{=}  O\left( \frac{1}{k^2} \right) ,
    \quad
    f_2(k) \underset{k\to0}{=} \frac{1}{k} + O\left(k\right) .
\end{equation}
In the notation of \eqref{eq:fh answer_general} this corresponds to
\begin{equation}
    \alpha_0 = 2,\quad 
    b_0 = 1,\quad  \beta_0 = -1, \quad
    \beta_1 = 1,
\end{equation}
hence the asymptotic expansion 
\begin{equation}
    \int_{0}^{\infty} \dd k\;
    f_2(k/\lambda) g_2(k) \underset{\lambda\to\infty}{=} 
    \lambda \; \mathcal{M}[g_2; 0] + O\left(\frac{1}{\lambda}\right),
\end{equation}
or with the explicit form of the Mellin transform 
\begin{equation}\label{eq:pd_s f2g2-integral-expansion-inf}
    \int_{0}^{\infty} \dd k\;
    f_2(k/\lambda) g_2(k)
    \underset{\lambda\to\infty}{\sim}
    \lambda
    \int_{0}^{\infty}\dd k\;
    \frac{1}{k}
    \Im\left[
    \frac{1}{1-F_2(k;0)} -  \frac{1}{1-F(k;0)}\right].
\end{equation}
Substituting \eqref{eq:pd_s f1g1-integral-expansion-inf} and \eqref{eq:pd_s f2g2-integral-expansion-inf} into \eqref{eq:pd_s_I-separated-rescaled} yields the asymptotic expansion of $\partial_s I(\lambda;0,s)|_{s=1} $ as $\lambda\to\infty$
\begin{equation}\label{eq:pd_s final-I-expansion-inf}
    \partial_s I(\lambda;0,s)\Big|_{s=1} 
    \underset{\lambda \to\infty}{=}
    \mathcal{B}_{\infty}^{(0)} 
    + \frac{1}{\lambda}\mathcal{B}_\infty^{(1)}
    + O\left(\frac{1}{\lambda^2}\right),
\end{equation}
with
\begin{align}
    \mathcal{B}_{\infty}^{(0)} & = 
    \frac{1}{2} 
    - \frac{1}{\pi}\int_{0}^{\infty}\dd k\;
    \frac{1}{k}
    \Im\left[
    \frac{1}{1-F_2(k;0)} -  \frac{1}{1-F(k;0)}\right],
    \\
    \mathcal{B}_{\infty}^{(1)} & = -
    \frac{1}{2 \abs{\left\langle M-\alpha t\right\rangle}}
    +\frac{1}{\pi}
    \int_{0}^{\infty}\dd k\;
    \Re\left[
    \frac{F(k;0)}{1-F(k;0)}\right]
    .
\end{align}

\subsection{Expansion of \texorpdfstring{$\partial_\rho I(\lambda;\rho,1)$}{I(λ; 0,1)}}
Finally we proceed to the asymptotic expansions of $\partial_\rho I(\lambda;\rho,1)|_{\rho=0}$. Taking the derivative of \eqref{eq:I(l|rho,s)=int Re} we arrive at
\begin{multline}\label{eq:pd_rho I}
    \partial_\rho I(\lambda;\rho,1)\Big|_{\rho=0}
    =
    -\frac{1}{\pi\alpha} \int_{0}^{\infty}\dd k\;
    \Re\left[
        \frac{1}{(k + \ii \lambda)^2}
        \log \frac{1 - F(k;\rho)}{1-F_2(k;\rho)}
    \right]
    \\
    -\frac{1}{\pi} \int_{0}^{\infty}\dd k\;
    \Re\left[
        \frac{\lambda}{k(k+\ii\lambda)}
        \left.\left(
             \frac{\partial_\rho c(\rho)F_2( k;\rho)}
                 {1-F_2( k;\rho)}
            -
            \frac{\partial_\rho c(\rho) F( k;\rho)}
                 {1-F( k;\rho)}
        \right)\right|_{\rho=0}
    \right].
\end{multline}
Separating real and imaginary parts yields a decomposition into a sum of four Mellin convolution integrals, as
\begin{multline}\label{eq:pd_rho I-separated}
    \partial_\rho I(\lambda;\rho,1)\Big|_{\rho=0}=
     - \frac{1}{\pi} \int_{0}^{\infty} \dd k\; f_1(k) r_1(\lambda k)
    -\frac{1}{\pi} \int_{0}^{\infty} \dd k\; f_2(k) r_2(\lambda k)
    \\
    - \frac{1}{\pi\lambda\alpha} \int_{0}^{\infty} \dd k\; f_3(k) h_1(\lambda k)
    - \frac{1}{\pi\lambda\alpha} \int_{0}^{\infty} \dd k\; f_4(k) h_2(\lambda k)
\end{multline} 
with the functions $f_{1,2}(k)$, $h_{1,2}(k)$ given by \eqref{eq:f1-f2=def} and \eqref{eq:h1-h2=def}, and four additional auxiliary functions defined as
\begin{equation}\label{eq:f3-f4-def}
    f_3(k) = \frac{k^2-1}{\left(k^2+1\right)^2}, \qquad
    f_4(k) = \frac{2 k }{\left(k^2+1\right)^2},
\end{equation}
\begin{equation}\label{eq:r1-r2-def}
\begin{aligned}
    r_1(k) & = \Re\left.\left[ 
            \frac{\partial_\rho c(\rho)F_2(\lambda k;\rho)}
                 {1-F_2(\lambda k;\rho)}
            -
            \frac{\partial_\rho c(\rho) F(\lambda k;\rho)}
                 {1-F(\lambda k;\rho)}
        \right]\right|_{\rho=0}, \\
    r_2(k) & = \Im\left.\left[ 
                \frac{\partial_\rho c(\rho)F_2(\lambda k;\rho)}
                     {1-F_2(\lambda k;\rho)}
                -
                \frac{\partial_\rho c(\rho) F(\lambda k;\rho)}
                     {1-F(\lambda k;\rho)}
            \right]\right|_{\rho=0}. 
\end{aligned}
\end{equation}
The technique however remains the same, that is we apply \eqref{eq:fh answer_general} to each of four integrals separately.

\paragraph{Expansion $\lambda\to0$} The functions $r_1(k)$ and $r_2(k)$ contain the factor $\partial_\rho c(\rho)F(k;\rho)$. Let us first take a closer look at it. From \eqref{eq:c(rho)=def}, \eqref{eq:F(k;rho)=def} it follows that
\begin{multline}\label{eq:pd_rho cF-expansion}
    \partial_\rho c(\rho) F(k;\rho)\Big|_{\rho=0}
    \underset{k\to0}{\sim}
    - \frac{\left\langle M \right\rangle}
           {\alpha} 
    - \ii k \;
    \frac{
        \left\langle M(M-\alpha t) \right\rangle}{\alpha}
    \\
    + k^2\;
    \frac{
        \left\langle M(M-\alpha t)^2 \right\rangle
    }{2\alpha}
    +
    \ii k^3\;
    \frac{
        \left\langle M(M-\alpha t)^3 \right\rangle
    }{6\alpha},
\end{multline}
and similarly from \eqref{eq:F2(k;rho)=def} we obtain
\begin{equation}\label{eq:pd_rho cF2}
    \partial_\rho c(\rho) F_2(k;\rho)\Big|_{\rho=0}
    =
    - \frac{\left\langle M \right\rangle}
           {\alpha} 
    - \ii k \;
    \frac{
        \left\langle M(M-\alpha t) \right\rangle}{\alpha}
    + k^2\;
    \frac{
        \left\langle M(M-\alpha t)^2 \right\rangle
    }{2\alpha}.
\end{equation}
Therefore the relevant asymptotic expansions for the first term are
\begin{equation}\label{eq:pd_rho f1-r1-expansions}
    f_1(k) \underset{k\to\infty}{=} \frac{1}{k^2} + O(1) ,\quad
    r_1(k) \underset{k\to0}{=} O\left(k^2\right)
\end{equation}
which in notations \eqref{eq:fh answer_general} corresponds to
\begin{equation}
    a_0 = 1,\quad \alpha_0 = 2, \quad \alpha_1=0,\quad
    \beta_0=2,
\end{equation}
hence the asymptotic behavior
\begin{equation}
    \int_{0}^{\infty}\dd k\, f_1(k) r_1(\lambda k) 
    = \lambda \, \mathcal{M}\left[r_1; -1 \right] + O(\lambda^2).
\end{equation}
In the explicit form it reads 
\begin{equation}\label{eq:pd_rho f1r1-integral-expansion}
    \int_{0}^{\infty} f_1(k) r_1(\lambda k) 
    \underset{\lambda\to0}{\sim}
    \lambda
        \int_{0}^{\infty} \dd k\;
        \frac{1}{k^2}
        \Re\left.\left[ 
            \frac{\partial_\rho c(\rho)F_2(k;\rho)}
                 {1-F_2(k;\rho)}
            -
            \frac{\partial_\rho c(\rho) F(k;\rho)}
                 {1-F(k;\rho)}
        \right]\right|_{\rho=0}.
\end{equation}

For the second term in \eqref{eq:pd_s_I-separated} the asymptotic behaviors are
\begin{equation}\label{eq:pd_rho f2-r2-expansions}
    f_2(k) \underset{k\to\infty}{=} \frac{1}{k^3} + O\left(\frac{1}{k}\right)
    ,
    \quad
    r_2(k) \underset{k\to0}{=} \frac{\left\langle M\right\rangle}{6\alpha}
        \frac{\mu_3(0)}{\mu_1(0)^2} k + O\left(k^2\right)
\end{equation}
which implies
\begin{equation}
    a_0 = 1,\quad \alpha_0=3, \quad \alpha_1 = 1,\quad
    b_0 =  \frac{\left\langle M\right\rangle}{6\alpha}
        \frac{\mu_3(0)}{\mu_1(0)^2},\quad 
        \beta_0 = 1,
    \quad \beta_1=2.
\end{equation}
This leads to
\begin{equation}
    \int_{0}^{\infty}\dd k\, f_2(k) r_2(\lambda k) 
    \underset{\lambda\to0}{=}
    \frac{\left\langle M\right\rangle}{6\alpha}
        \frac{\mu_3(0)}{\mu_1(0)^2}\, 
        \lambda\, 
        \mathcal{M}\left[f_2; 2 \right] 
    + O(\lambda^2).
\end{equation}
Computing the Mellin transform of $f_2$ as in \eqref{eq:pd_s M[f2]-Mellin} and using the explicit representation for the moments \eqref{eq:mu(rho)=def} we obtain 
\begin{equation}\label{eq:pd_rho f2r2-integral-expansion}
    \int_{0}^{\infty} \dd k\;
    f_2(k) r_2(\lambda k) \underset{\lambda\to0}{\sim}
    \lambda \, \frac{\pi}{12} \frac{\left\langle M \right\rangle}{\alpha}
    \frac{\left\langle (M-\alpha t)^3\right\rangle  }
         {\left\langle M-\alpha t\right\rangle ^2}.
\end{equation}

\par The third term in \eqref{eq:pd_rho I-separated} requires the expansions
\begin{equation}\label{eq:pd_rho f3-h1-expansions}
    f_3(k) \underset{k\to\infty}{=}
        \frac{1}{k^2} + O\left(\frac{1}{k^4}\right),
    \quad
    h_1(k) \underset{k\to0}{=} 
    - \frac{1}{6} \frac{\mu_3(0)}{\mu_1(0)} k^2 + O\left(k^3\right) 
\end{equation}
which means
\begin{equation}
    a_0 = 1, \quad \alpha_0 = 2, \quad \alpha_1 = 4,\quad
    b_0 = - \frac{1}{6} \frac{\mu_3(0)}{\mu_1(0)}, \quad \beta_0 = 2, \quad \beta_1=3.
\end{equation}
Hence the asymptotic expansion 
\begin{equation}
    \int_{0}^{\infty}\dd k\, f_2(k) h_1(\lambda k) 
    \underset{\lambda\to0}{=} 
    \lambda\, \mathcal{M}[h_1; -1]
    - \frac{1}{6} \frac{\mu_3(0)}{\mu_1(0)} \lambda^2\, \mathcal{M}\left[f_3; 3\right]
    + O\left(\lambda^3\right) .
\end{equation}
The Mellin transform of $f_3$ can be computed straightforwardly and reads
\begin{equation}\label{eq:pd_rho M[f3]-Mellin}
    \mathcal{M}[f_3; z] = 
    \int_{0}^{\infty} \dd k\, k^{z-1} 
    \frac{k^2 - 1}{(k^2+1)^2}
    = \frac{\pi}{2} \frac{z-1}{\sin \frac{\pi}{2} z},
    \qquad
    \mathcal{M}[f_3; 3] = -\pi.
\end{equation}
The asymptotic expansion for the third term now reads
\begin{equation}\label{eq:pd_rho f3h1-integral-expansion}
    \int_{0}^{\infty}\dd k\,
    f_3(k) h_1(\lambda k)
    \underset{\lambda\to0}{\sim}
    \lambda
    \int_{0}^{\infty} \dd k\;
        \frac{1}{k^2} 
            \Re\left[ \log \frac{1-F(k;0)}{1-F_2(k,0)} \right]
    +
    \lambda^2 \, \frac{\pi }{6} \frac{\mu_3(0)}{\mu_1(0)}.
\end{equation}

For the last term we have
\begin{equation}\label{eq:pd_rho f4-h2-expansions}
    f_4(k)  \underset{k\to\infty}{=} \frac{2}{k^3} + O\left(\frac{1}{k^5}\right) ,
    \quad
    h_2(k) \underset{k\to0}{=} O(k^3),
\end{equation}
hence
\begin{equation}
    a_0 = 2,\quad \alpha_0 = 3, \quad \alpha_1=5, \quad
    \beta_0=3,
\end{equation}
and the asymptotic expansion of the integral is
\begin{equation}
    \int_{0}^{\infty}\dd k\, f_4(k) h_2(\lambda k) 
    \underset{\lambda\to0}{=} 
    2 \lambda^{2} \mathcal{M}[h_2, -2]
    + O\left(\lambda^3 \right),
\end{equation}
or explicitly
\begin{equation}\label{eq:pd_rho f4h2-integral-expansion}
    \int_{0}^{\infty}
    f_4(k) h_2(\lambda k)
    \underset{\lambda\to0}{\sim}
    2\lambda^2
    \int_{0}^{\infty} \dd k\;
        \frac{1}{k^3} 
            \Im\left[ \log  \frac{1-F(k;0)}{1-F_2(k,0)} \right].
\end{equation}

Finally, we substitute \eqref{eq:pd_rho f1r1-integral-expansion}, \eqref{eq:pd_rho f2r2-integral-expansion}, \eqref{eq:pd_rho f3h1-integral-expansion}, and \eqref{eq:pd_rho f4h2-integral-expansion} into \eqref{eq:pd_rho I-separated} to obtain the $\lambda\to0$ expansion of $\partial_\rho I(\lambda;\rho,1)|_{\rho=0}$, namely
\begin{equation}\label{eq:pd_rho I-final-0}
    \partial_\rho I(\lambda;\rho,1)\Big|_{\rho=0}
    \underset{\lambda\to0}{=}
    \mathcal{C}_{0}^{(0)}
    +  
    \mathcal{C}_{0}^{(1)} \lambda
    + O(\lambda^2) ,
\end{equation}
where the coefficients are given by
\begin{align}
    \mathcal{C}_0^{(0)} & 
    =
    -\frac{1}{\pi\alpha}
    \int_{0}^{\infty} \dd k\;
        \frac{1}{k^2}\Re\left[ \log  \frac{1-F(k;0)}{1-F_2(k,0)} \right]
    =
    \frac{1}{\alpha} \mathcal{A}_{0}^{(1)}
    \\\nonumber
    \mathcal{C}_{0}^{(1)} &= 
     -\frac{1}{6\alpha} 
      \frac{\langle (M-\alpha t)^3\rangle}
           {\langle M-\alpha t\rangle} 
      \left(
      1 + \frac{\left\langle M \right\rangle}{2\langle M-\alpha t\rangle}
      \right)     
    \\\nonumber
    &\qquad
    - \frac{2}{\alpha\pi} 
    \int_{0}^{\infty} 
        \dd k\, 
        \frac{1}{k^3} 
        \Im\left[ \log  \frac{1-F(k;0)}{1-F_2(k,0)} \right]
    \\&\qquad
    - \frac{1}{\pi}
        \int_{0}^{\infty}
        \dd k\, \frac{1}{k^2}
        \Re\left.\left[ {}
            \frac{\partial_\rho c(\rho)F_2(k;\rho)}
                 {1-F_2(k;\rho)}
            -
            \frac{\partial_\rho c(\rho) F(k;\rho)}
                 {1-F(k;\rho)}
        \right]\right|_{\rho=0}.
\end{align}

\paragraph{Expansion $\lambda\to\infty$}
To shorten the calculation a little bit, let us look at the second term in \eqref{eq:pd_rho I}. Using an integral identity
\begin{equation}\label{eq:pd_rho int F2}
    \int_{0}^{\infty}\dd k\;
    \Re\left[
        \frac{\log \left[ 1 - F_2(k;\rho)\right]}
             {(k + \ii \lambda)^2}
    \right]
    =\frac{\pi}{\lambda}
\end{equation}
which can be verified by direct calculation  we can simplify it as
\begin{multline}\label{eq:pd_rho I-second-term-alt}
    \int_{0}^{\infty}\dd k
    \Re\left[
        \frac{1}{(k + \ii \lambda)^2}
        \log \frac{1 - F(k;\rho)}{1-F_2(k;\rho)}
    \right]
    =
    -\frac{\pi}{\lambda}
    \\+
    \frac{1}{\lambda^2} \int_{0}^{\infty}\dd k
    \Re\left[
        \frac{\log [ 1 - F(k;\rho)] }
             {\left((k/\lambda)^2 + \ii \right)^2}
    \right].
\end{multline}
The second term impacts the expansion as $\lambda\to\infty$ only at the order higher than $\lambda^{-2}$. Therefore after rescaling \eqref{eq:pd_rho I-separated} transforms into 
\begin{multline}\label{eq:pd_rho I-inf}
    \partial_\rho I(\lambda;\rho,1)\Big|_{\rho=0}
    \underset{\lambda\to\infty}{=}
    \frac{1}{\alpha\lambda}
    -\frac{1}{\pi \lambda}\int_{0}^{\infty} \dd k
        f_1\left(k/\lambda\right) r_1(k)
    \\ 
    -\frac{1}{\pi \lambda}\int_{0}^{\infty} \dd k
        f_2\left(k/\lambda\right) r_2(k)
    +O\left(\frac{1}{\lambda^2}\right),
\end{multline}
where the functions $f_{1,2}(k)$ and $r_{1,2}(k)$ are given by \eqref{eq:f1-f2=def} and \eqref{eq:r1-r2-def} respectively. Thus, we need to apply \eqref{eq:fh answer_general} only to two terms out of four in \eqref{eq:pd_rho I-separated}.

\par Now we need the expansions of $r_{1,2}(k)$ for $k\to\infty$. A simple calculation shows that due to \eqref{eq:F(k;rho)=k->infty} we have
\begin{equation}\label{eq:c(rho)F(k;rho)/1-F=asympt}
    \left. \frac{\pd_\rho c(\rho) F(k;\rho)}{1-F(k;\rho)}\right|_{\rho=0} 
    \underset{k\to\infty}{=} O\left(\frac{1}{k^2}\right),
\end{equation}
and using the explicit form \eqref{eq:F2(k;rho)=def} of $F_2(k;\rho)$ we obtain 
\begin{multline}\label{eq:c(rho)F2(k;rho)/1-F2=asympt}
    \left. \frac{\pd_\rho c(\rho) F(k;\rho)}{1-F(k;\rho)}\right|_{\rho=0} 
    \underset{k\to\infty}{=} 
    - \frac{1}{\alpha} 
        \frac{\left\langle M(M-\alpha t)^2\right\rangle}
             {\left\langle (M-\alpha t)^2\right\rangle}
    \\
    + \frac{2\ii}{k}\left(
        \frac{\left\langle M (M-\alpha t)\right\rangle}
             {\left\langle (M-\alpha t)^2\right\rangle}
        -
        \frac{\left\langle M-\alpha t\right\rangle
              \left\langle M (M-\alpha t)^2\right\rangle }
             {\big[\left\langle (M-\alpha t)^2\right\rangle\big]^2}
    \right)
    + O\left(\frac{1}{k^2}\right).
\end{multline}
Therefore, only the term involving $F_2(k;\rho)$ contributes to the asymptotic behavior of $r_{1,2}(k)$ for $k\to\infty$. Note, however, that this does not mean that  $\lambda\to\infty$ expansion does not depend on $F(k;\rho)$.

\par Now we apply \eqref{eq:fh answer_general} to the first integral term in \eqref{eq:pd_rho I-inf}. The relevant expansions are
\begin{equation}\label{eq:pd_rho f3-h3-expansion-inf}
    r_1(k) 
    \underset{k\to\infty}{=} 
    \frac{1}{\alpha}
    \frac{\left\langle M(M-\alpha t)^2 \right\rangle}
         {\left\langle ( M-\alpha t)^2 \right\rangle}
    +O\left(\frac{1}{k^2}\right),
    \qquad
    f_1(k) 
    \underset{k\to0}{=}
        1+ O\left(k^2\right)
    ,
\end{equation}
hence the coefficients in \eqref{eq:fh answer_general} 
\begin{equation}
    a_0 = \frac{1}{\alpha}
    \frac{\left\langle M(M-\alpha t)^2 \right\rangle}
         {\left\langle ( M-\alpha t)^2 \right\rangle}
    ,\quad \alpha_0=0,
    \quad \alpha_1 = 1,\quad
    b_0 = 1,\quad 
    \beta_0 = 0,\quad 
    \beta_1 = 2
\end{equation}
which implies
\begin{equation}
    \int_{0}^{\infty} \dd k
        f_1\left(k/\lambda\right) r_1(k)
    \underset{\lambda\to\infty}{=}
    \frac{1}{\alpha}
    \frac{\left\langle M(M-\alpha t)^2 \right\rangle}
         {\left\langle ( M-\alpha t)^2 \right\rangle}
    \,\lambda\, \mathcal{M}[f_1; 1]
    + \mathcal{M}\left[r_1;1\right]
    + O\left(\frac{1}{\lambda}\right).
\end{equation}
The Mellin transform of $f_1$ we computed in \eqref{eq:M[f1]-Mellin}, and it reads
\begin{equation}\label{eq:M[f1]-Mellin-1}
    \mathcal{M}\left[f_1; z\right] = 
    \int_{0}^{\infty} \dd k\; k^{z-1}
    \frac{1}{k^2+1}
    =
    \frac{\pi}{2} \frac{1}{\sin \frac{\pi z}{2}},
    \qquad 
    \mathcal{M}[f_1; 1] = \frac{\pi}{2}.
\end{equation}
As for the Mellin transform of $r_1$, we face the same problem as for $g_1$ in \eqref{eq:pd_s M[g1]-Mellin inf}. Specifically, the integral \eqref{eq:M[g;z]=def} diverges and we need to perform the analytic continuation as prescribed in \cite{L-08}. The resulting expression is 
\begin{equation}\label{eq:pd_rho M[r1]-Mellin-0}
    \mathcal{M}[r_1; 1] =\int_{0}^{\infty}\dd k\, 
    \Re\left.\left[ 
            \frac{\partial_\rho c(\rho)F_2(k;\rho)}
                 {1-F_2( k;\rho)}
            -
            \frac{\partial_\rho c(\rho) F(k;\rho)}
                 {1-F(k;\rho)}
    - \frac{1}{\alpha}
    \frac{\left\langle M(M-\alpha t)^2 \right\rangle}
         {\left\langle ( M-\alpha t)^2 \right\rangle}
        \right]\right|_{\rho=0}.
\end{equation}
This integral can be simplified if we compute the contribution of the term with $F_2(k,\rho)$, specifically
\begin{multline}\label{eq:pd_rho M[h3]-Mellin-F2-term}
    \int_{0}^{\infty}\dd k\, 
    \Re\left.\left[ 
            \frac{\partial_\rho c(\rho)F_2(k;\rho)}
                 {1-F_2(k;\rho)}
    - \frac{1}{\alpha}
    \frac{\left\langle M(M-\alpha t)^2 \right\rangle}
         {\left\langle ( M-\alpha t)^2 \right\rangle}
        \right]\right|_{\rho=0}
    \\
    = \frac{\pi}{2\alpha}
    \frac{\left\langle M \right\rangle}
         {\left\langle M-\alpha t\right\rangle}
    + \frac{\pi}{\alpha}
    \frac{
        \left\langle M-\alpha t\right\rangle
        \left\langle M(M-\alpha t)^2\right\rangle
    }
        {[ 
            \left\langle ( M-\alpha t)^2  \right\rangle
        ]^2}
    - \frac{\pi}{\alpha} 
    \frac{\left\langle M(M-\alpha t)\right\rangle}
         {\left\langle (M-\alpha t)^2\right\rangle},
\end{multline}
and hence the expansion
\begin{multline}\label{eq:pd_rho f1r1-integral-expansion-inf}
     -\frac{1}{\pi\lambda}\int_{0}^{\infty}\dd k \, 
    f_1(k/\lambda) r_1(k)
    \underset{\lambda\to\infty}{\sim}
    -\frac{1}{2\alpha}
    \frac{\left\langle M(M-\alpha t)^2 \right\rangle}
         {\left\langle ( M-\alpha t)^2 \right\rangle}
    \\ - 
    \frac{1}{\alpha\lambda}
    \left\{
    \frac{1}{2}
    \frac{\left\langle M \right\rangle}
         {\left\langle M-\alpha t\right\rangle}
    + 
    \frac{
        \left\langle M-\alpha t\right\rangle
        \left\langle M(M-\alpha t)^2\right\rangle
    }
        {[ 
            \left\langle ( M-\alpha t)^2  \right\rangle
        ]^2}
    -
    \frac{\left\langle M(M-\alpha t)\right\rangle}
         {\left\langle (M-\alpha t)^2\right\rangle}
    \right\}
    \\ +\frac{1}{\pi \lambda}
        \int_{0}^{\infty}\dd k\,
    \Re\left.\left[
    \frac{\partial_\rho c(\rho) F(k;\rho)}
                 {1-F( k;\rho)}
    \right]\right|_{\rho=0}.
\end{multline}

\par Finally, for the second integral term in \eqref{eq:pd_rho I-inf} the relevant expansions are
\begin{equation}\label{eq:pd_rho f2-r2-expansion-inf}
    r_2(k) 
    \underset{k\to\infty}{=} \frac{a_0}{k}
    + O\left(\frac{1}{k^3}\right)
    ,\quad
    f_2(k) 
    \underset{k\to0}{=} 
    \frac{1}{k} + O\left(k\right),
\end{equation}
where  
\begin{equation}
    a_0 = 2  
    \frac{
        \left\langle M-\alpha t\right\rangle
        \left\langle M(M-\alpha t)^2\right\rangle
    }
        {\big[ \left\langle ( M-\alpha t)^2  \right\rangle\big]^2}
    -
    2
    \frac{\left\langle M(M-\alpha t)\right\rangle}
         {\left\langle (M-\alpha t)^2\right\rangle}.
\end{equation}
Then according to \eqref{eq:fh answer_general} we have
\begin{equation}
    \alpha_0 = 1, \quad \alpha_1 = 3,\quad
    b_0 = 1,\quad \beta_0 = -1, \quad \beta_1 = 1,
\end{equation}
hence
\begin{equation}
    \int_{0}^{\infty} \dd k
        f_2\left(k/\lambda\right) r_2(k)
    \underset{\lambda\to\infty}{=}
    \lambda \mathcal{M}[r_2;0] 
    + 
    a_0 \mathcal{M}\left[f_2; 0\right].
\end{equation}
The Mellin transform of $f_2$ is given by \eqref{eq:pd_s M[f2]-Mellin} and it reads
\begin{equation}\label{eq:pd_s M[f2]-Mellin-1}
    \mathcal{M}[f_2;z]
    = \int_{0}^{\infty} \dd k\;
    k^{z-1}
    \frac{1}{k(k^2+1)} = -\frac{\pi}{2} \frac{1}{\cos \frac{\pi z}{2}}, \qquad
    \mathcal{M}[f_2;0] = -\frac{\pi}{2},
\end{equation}
hence the expansion 
\begin{multline}\label{eq:pd_rho f2r2-integral-expansion-inf}
     -\frac{1}{\pi\lambda}\int_{0}^{\infty}\dd k \, 
    f_2(k/\lambda) r_2(k)
    \underset{\lambda\to\infty}{\sim}
    \\
    -\frac{1}{\pi} \int_{0}^{\infty} \dd k\; 
    \frac{1}{k}
    \Im\left.\left[ 
            \frac{\partial_\rho c(\rho)F_2(k;\rho)}
                 {1-F_2( k;\rho)}
            -
            \frac{\partial_\rho c(\rho) F( k;\rho)}
                 {1-F( k;\rho)}
    \right]\right|_{\rho=0}
    \\
    + \frac{1}{\lambda}\left\{
    \frac{
        \left\langle M-\alpha t\right\rangle
        \left\langle M(M-\alpha t)^2\right\rangle
    }
        {[ 
            \left\langle ( M-\alpha t)^2  \right\rangle
        ]^2}
    -
    \frac{\left\langle M(M-\alpha t)\right\rangle}
         {\left\langle (M-\alpha t)^2\right\rangle}
    \right\}.
\end{multline}
Combining now \eqref{eq:pd_rho f2r2-integral-expansion-inf} and \eqref{eq:pd_rho f1r1-integral-expansion-inf} with \eqref{eq:pd_rho I-inf} we arrive at the expansion
\begin{equation}\label{eq:pd_rho I-final-inf}
    \partial_\rho I(\lambda;\rho,1)\Big|_{\rho=0}
    \underset{\lambda\to\infty}{=}
    \mathcal{C}_{\infty}^{(0)} + \frac{1}{\lambda} \mathcal{C}_{\infty}^{(1)} + O\left(\frac{1}{\lambda^2}\right)
\end{equation}
with 
\begin{align}
    \nonumber
    \mathcal{C}_\infty^{(0)} & =    - \frac{1}{2\alpha}
    \frac{\left\langle M(M-\alpha t)^2 \right\rangle}
         {\left\langle ( M-\alpha t)^2 \right\rangle}\\
    \label{eq:C_infty^0}
    & \quad
    -\frac{1}{\pi} \int_{0}^{\infty} \dd k\; 
    \frac{1}{k}
    \Im\left.\left[ 
            \frac{\partial_\rho c(\rho)F_2(k;\rho)}
                 {1-F_2( k;\rho)}
            -
            \frac{\partial_\rho c(\rho) F( k;\rho)}
                 {1-F( k;\rho)}
    \right]\right|_{\rho=0}
    \\
    \label{eq:C_infty^1}
    \mathcal{C}_\infty^{(1)} &=
        \frac{1}{\alpha}
        -
        \frac{1}{2\alpha}
        \frac{\left\langle M \right\rangle}
        {\left\langle M-\alpha t\right\rangle}
    +\frac{1}{\pi}
    \int_{0}^{\infty}\dd k\,
    \Re\left.\left[
    \frac{\partial_\rho c(\rho) F(k;\rho)}
                 {1-F(k;\rho)}
    \right]\right|_{\rho=0}.
\end{align}
Note that in \eqref{eq:C_infty^0}, we can actually extract the term with $F_2(k;\rho)$ by explicit integration. However, this provides no simplification, so we keep \eqref{eq:C_infty^0} in the present form.

\subsection{Asymptotic expansion for the moments}
Having computed all necessary expansions of $I(\lambda;0,s)$ and its first two derivatives, we now proceed to constructing asymptotic expansions of the mean and variances of $\tau$ and $n$. We provide the detailed derivation for the mean of $n$ as an illustration. The variance of $n$ and the moments of $\tau$ follow by the same procedure, and we present only the main steps.

\paragraph{Mean values}
Taking the limit $\rho\to0$ in \eqref{eq:LTQ_P=general_regular_int} and expanding it in series we find that
\begin{equation}
    \label{eq:hat Q = pd_s 1}
    \pdv{}{s}\hat{Q}(0,s\,\vert\,\lambda)\Big|_{s=1}
        = - \frac{1}{\lambda^2}
        \frac{e^{I(\lambda;0,1)}}
        {\left\langle M-\alpha t\right\rangle}.
\end{equation}
For the expansion as $\lambda\to0$, we substitute \eqref{eq:final-I-0-expansion} into \eqref{eq:hat Q = pd_s 1} to obtain
\begin{equation}\label{eq:LT n = l->0 expansion}
    \pdv{}{s}\hat{Q}(0,s\,\vert\,\lambda)\Big|_{s=1}
    \underset{\lambda\to0}{=}
    -\frac{1}{\lambda^2} 
     \frac{1}{\left\langle M-\alpha t\right\rangle}
    -
    \frac{1}{\lambda}
    \frac{\mathcal{A}_{0}^{(1)}}{\left\langle M-\alpha t\right\rangle}
    +O(1).
\end{equation}
This suggests a linear behavior of the mean value, namely
\begin{equation}\label{eq:E[n|X]=ansatz}
    \mathbb{E}[n\,\vert\,X_0] 
    \underset{X_0\to\infty}{\sim} A_1 X_0 + A_0.
\end{equation}
Comparing the Laplace transform of the ansatz \eqref{eq:E[n|X]=ansatz} with \eqref{eq:LT n = l->0 expansion}, we fix the coefficients as
\begin{equation}
    A_1 = -\frac{1}{\langle M-\alpha t\rangle},
    \qquad
    A_0 = - \frac{\mathcal{A}_{0}^{(1)}}
               {\langle M-\alpha t\rangle}.
\end{equation}
Recalling the definition \eqref{eq:A_0^1=int} of 
$\mathcal{A}_0^{(1)}$, we immediately obtain the representation 
\eqref{eq:A1A0=}.

\par For the expansion as $\lambda\to\infty$, we substitute \eqref{eq:final-I-inf-expansion} into \eqref{eq:hat Q = pd_s 1}. In the absorption regime, $\log[\lambda + k_+^*(0,1)] = \log\lambda$. Expanding the exponential then yields
\begin{equation}\label{eq:LT n = l->inf expansion}
    \left.\pdv{}{s}\hat{Q}(0,s\,\vert\,\lambda)\right|_{s=1}
    \underset{\lambda\to\infty}{=} 
    - \frac{e^{\mathcal{A}_\infty^{(0)}}}{\left\langle M - \alpha t \right\rangle}\left(
    \frac{1}{\lambda} + \frac{\mathcal{A}_\infty^{(1)}}{\lambda^2}
    \right)
    + O\left(\frac{1}{\lambda^3}\right).
\end{equation}
This suggests the behavior
\begin{equation}\label{eq:E[n|X]=ansatz-inf}
    \mathbb{E}[n\,\vert\,X_0] 
    \underset{X_0\to0}{\sim} a_0 + a_1 X_0,
\end{equation}
with coefficients
\begin{equation}
    a_0 = - \frac{e^{\mathcal{A}_\infty^{(0)}}}
               {\left\langle M - \alpha t \right\rangle},
    \qquad
    a_1 = - \frac{e^{\mathcal{A}_\infty^{(0)}}}
               {\left\langle M - \alpha t \right\rangle}
          \mathcal{A}_\infty^{(1)} .
\end{equation}
Using the explicit forms of $\mathcal{A}_\infty^{(0)}$ and $\mathcal{A}_{\infty}^{(1)}$ given in \eqref{eq:A_inf^0=int} and \eqref{eq:A_inf^1=int}, we recover the coefficients \eqref{eq:a1a0=}. This completes the analysis for the mean of $n$.

\par The asymptotic behavior of the mean value of $\tau$ for both $X_0\to0$ and $X_0\to\infty$ is readily restored from \eqref{eq:E[n|X]=ansatz} and \eqref{eq:E[n|X]=ansatz-inf} since due to \eqref{eq:mean(tau) mean(n)} the mean values of $\tau$ and $n$ are related via
\begin{equation}
    \mathbb{E}\left[\tau\,\vert\,X_0\right] 
    = \frac{X_0}{\alpha} + \frac{\left\langle M\right\rangle}{\alpha}
    \mathbb{E}\left[n\,\vert\,X_0\right].
\end{equation}

\paragraph{Variances} To compute the variance of $n$ we first construct the asymptotic expansion for the second moment, and then combine the result with \eqref{eq:E[n|X]=ansatz} and \eqref{eq:E[n|X]=ansatz-inf} we have obtained for the mean value.

\par The Laplace transform of the second moment in terms of $\hat{Q}(\lambda;\rho,s)$ is given by
\begin{equation}\label{eq:E[n2]=pdv s Q}
    \int_{0}^{\infty} \dd X_0\, e^{-\lambda X_0} \mathbb{E}\left[n^2\,\vert\,X_0\right] = 
    \left. \left[\pdv{^2}{s^2} + \pdv{}{s} \right]
    \hat{Q}(0,s\,\vert\,\lambda) \right|_{s=1}.
\end{equation}
Therefore we need to compute the asymptotic expansions of the second derivative. A straightforward computation yields
\begin{multline}\label{eq:hat Q = pd_s 2}
    \pdv{^2}{s^2}\hat{Q}(0,s\,\vert\,\lambda)\Big|_{s=1}
        = 
        \frac{2}{\lambda^3}
        \frac{e^{I(\lambda;0,1)}}
             {\left\langle M-\alpha t\right\rangle^2}
        \\
        +\frac{1}{\lambda^2}
        \frac{e^{I(\lambda;0,1)}}
             {\left\langle M-\alpha t\right\rangle}
        \left[
        \frac{\left\langle (M-\alpha t)^2\right\rangle}
             {\left\langle M-\alpha t\right\rangle^2}
        + 2\left( \pdv{}{s} I(\lambda;0,s)\Big|_{s=1} - 1\right) 
        \right].
\end{multline}
The behavior as $X_0\to\infty$ is governed by the $\lambda\to0$ expansion of \eqref{eq:hat Q = pd_s 2}, which can be constructed straightforwardly using the expansions \eqref{eq:final-I-0-expansion} for $I(\lambda;0,1)$ and \eqref{eq:pd_s final-I-expansion} for $\pd_{s}I(\lambda;0,s)|_{s=1}$.  Substituting the resulting expansions into \eqref{eq:E[n2]=pdv s Q}, we find quadratic behavior for the second moment in the form:
\begin{equation}    
    \mathbb{E}\left[n^2\,\vert\,X_0\right] 
    \underset{X_0\to\infty}{\sim} \sigma_2 X_0^2 + \sigma_1 X_0 + \sigma_0,
\end{equation} 
where the coefficients $\sigma_i$ are represented in terms of $\mathcal{A}_{0}^{(1)}$, $\mathcal{A}_0^{(2)}$ and $\mathcal{B}_{0}^{(1)}$ given by \eqref{eq:A_0^1=int}, \eqref{eq:A_0^2=int} and \eqref{eq:B_0^1=int}. Then using the definition of the variance,
\begin{equation}
    \mathrm{Var}\left[n\,\vert\,X_0\right] =
    \mathbb{E}\left[n^2\,\vert\,X_0\right] - 
    \big(\mathbb{E}\left[n\,\vert\,X_0\right]\big)^2,
\end{equation}
we obtain the asymptotic behavior 
\begin{equation}
    \mathrm{Var}\left[n\,\vert\,X_0\right] 
    \underset{X_0\to\infty}{\sim}
    B_1 X_0 + B_0,
\end{equation}
with coefficients \eqref{eq:B1B0=}.

\par Repeating the same procedure for $\lambda\to\infty$, substituting the expansions \eqref{eq:final-I-inf-expansion} and \eqref{eq:pd_s final-I-expansion-inf} into \eqref{eq:hat Q = pd_s 2}, after elementary algebraic manipulations we reconstruct the behavior of the variance for $X_0\to0$, namely 
\begin{equation}
    \mathrm{Var}\left[n\,\vert\,X_0\right] 
    \underset{X_0\to0}{\sim} 
    b_0 + b_1 X_0,
\end{equation}
with the coefficients \eqref{eq:b1b0=}.

\par The variance of $\tau$ is computed in the same way. We start with the Laplace transform of the second moment 
\begin{equation}\label{eq:E[tau2]=pdv rhp Q}
    \int_{0}^{\infty} \dd X_0\, e^{-\lambda X_0} \mathbb{E}\left[\tau^2\,\vert\,X_0\right] = 
    \left. \pdv{^2}{\rho^2} 
    \hat{Q}(\rho,1\,\vert\,\lambda) \right|_{\rho=0},
\end{equation}
which is given by
\begin{multline}\label{eq:hat Q = pd_rho 2}
    \left. \pdv{^2}{\rho^2} \hat{Q}(\rho,1\,\vert\,\lambda) \right|_{\rho=0}
    = \frac{2}{\alpha^2\lambda^3}
    \left(
        1 - \frac{\langle M \rangle}{\langle M-\alpha t\rangle}
        \left[ 
        1 - \frac{\langle M\rangle}{\langle M-\alpha t\rangle} 
        \right]
        e^{I(\lambda;0,1)}
    \right)
    \\
    \frac{1}{\lambda^2}
    \frac{\langle t\rangle^2\big[\left\langle M^2\right\rangle - \langle M\rangle^2\big]
        +\langle M\rangle^2\big[\left\langle t^2\right\rangle-\left\langle t\right\rangle^2\big]}
        {\left\langle M-\alpha t\right\rangle^3} e^{I(\lambda;0,1)}
    \\
    +\frac{2}{\alpha \lambda^2}
    \frac{\langle M\rangle}{\langle M-\alpha t\rangle}
    e^{I(\lambda;0,1)}
            \left. \pdv{}{\rho} I(\lambda;\rho,1)\right|_{\rho=0}
    .
\end{multline}
The $X_0\to\infty$ behavior is obtained by substituting \eqref{eq:final-I-0-expansion} and \eqref{eq:pd_rho I-final-0} into \eqref{eq:hat Q = pd_rho 2}, which yields a quadratic form for the second moment. Combining this with the mean value from \eqref{eq:E[n|X]=ansatz} and the relation \eqref{eq:mean(tau) mean(n)} between $\tau$ and $n$, we find
\begin{equation}
    \mathrm{Var}\left[\tau\,\vert\,X_0\right] 
    \underset{X_0\to\infty}{\sim}
    C_1 X_0 + C_0,
\end{equation}
with coefficients given by \eqref{eq:C1C0=}.

Similarly, for the $X_0\to0$ behavior, we substitute \eqref{eq:final-I-inf-expansion} and \eqref{eq:pd_rho I-final-inf} into \eqref{eq:hat Q = pd_rho 2}. 
Following the same procedure, we obtain
\begin{equation}
    \mathrm{Var}\left[\tau\,\vert\,X_0\right] 
    \underset{X_0\to0}{\sim} c_0 + c_1 X_0,
\end{equation}
with coefficients \eqref{eq:c1c0=}.

\end{appendix}

\sloppy
\printbibliography

\end{document}